\documentclass[prx, twocolumn, superscriptaddress]{revtex4-2}
\usepackage{bm, amsmath, amssymb, amsfonts, latexsym, mathtools, braket}
\usepackage{subfigure, multirow}
\usepackage{graphicx}
\usepackage{float, color}

\usepackage[
pagebackref=false,
colorlinks=true,
linkcolor=blue,
urlcolor=blue,
filecolor=black,
citecolor=red,
pdfstartview=FitV,
pdftitle={},
pdfauthor={},
pdfsubject={},
pdfkeywords={},
pdfpagemode=None,
bookmarksopen=true
]{hyperref}

\begin{document}

\newcommand{\ii}{\text{i}}
\newcommand{\currentdensity}{i}
\newcommand{\disorderedhopping}{\Delta J_{n, n+1}}

\title{Nonlinear Landauer formula:\\ 
Nonlinear response theory of disordered and topological materials}

\author{Kohei Kawabata}
\email{kohei.kawabata@princeton.edu}
\affiliation{Department of Physics, University of Tokyo, 7-3-1 Hongo, Bunkyo-ku, Tokyo 113-0033, Japan}
\affiliation{Department of Physics, Princeton University, Princeton, New Jersey, 08540, USA}

\author{Masahito Ueda}
\email{ueda@phys.s.u-tokyo.ac.jp}
\affiliation{Department of Physics, University of Tokyo, 7-3-1 Hongo, Bunkyo-ku, Tokyo 113-0033, Japan}
\affiliation{RIKEN Center for Emergent Matter Science (CEMS), Wako, Saitama 351-0198, Japan}
\affiliation{Institute for Physics of Intelligence, University of Tokyo, 7-3-1 Hongo, Bunkyo-ku, Tokyo 113-0033, Japan}

\date{\today}

\begin{abstract}
The Landauer formula provides a general scattering formulation of electrical conduction.
Despite its utility, it has been mainly applied to the linear-response regime, and a scattering theory of nonlinear response has yet to be fully developed.
Here, we extend the Landauer formula to the nonlinear-response regime.
We show that while the linear conductance is directly related to the transmission probability, the nonlinear conductance is given by its derivatives with respect to energy.
This sensitivity to the energy derivatives is shown to produce unique nonlinear transport phenomena of mesoscopic systems including disordered and topological materials.
By way of illustration, we investigate nonlinear conductance of disordered chains and identify their universal behavior according to symmetry.
In particular, we find large singular nonlinear conductance for zero modes, including Majorana zero modes in topological superconductors.
We also show the critical behavior of nonlinear response around the mobility edges due to the Anderson transitions.
Moreover, we study nonlinear response of graphene as a prime example of topological materials featuring quantum anomaly.
Furthermore, considering the geometry of electronic wave functions, we develop a scattering theory of the nonlinear 
Hall effect.
We establish a new connection between the nonlinear 
Hall response and the nonequilibrium quantum fluctuations. 
We also discuss the influence of disorder and Anderson localization on the nonlinear 
Hall effect.
Our work opens a new avenue in quantum physics beyond the linear-response regime.
\end{abstract}

\maketitle

\section{Introduction}

Electronic transport plays a central role in condensed matter physics and gives insights into underlying electronic properties of a rich variety of materials.
As well as the practical relevance, it is of fundamental significance to develop a theory of electrical conduction in statistical physics.
Based on kinetic equations for the distribution of particles, the Boltzmann transport theory describes semiclassical transport phenomena~\cite{AM-textbook, Abrikosov-textbook}.
A fully quantum description of electrical conduction is given by the linear response theory that treats an external field as a perturbation to the system near thermal equilibrium~\cite{Kubo-57, Kubo-textbook, AS-textbook}.
Its successful applications include the quantum Hall effect, where topology of the wave functions plays a key role~\cite{Klitzing-80, Laughlin-81, Halperin-82, TKNN-82, Buttiker-88-QHE, Haldane-88, QHE-review}.

The Landauer formula provides yet another general formulation of electrical conduction from a different perspective~\cite{Landauer-57, *Landauer-70, Fisher-81, Buttiker-86, *Buttiker-88, Stone-88,  Datta-textbook, Imry-textbook}.
It relies on the scattering formulation of electronic transport and addresses experimental situations in which a system is attached to electrodes. 
A crucial advantage of the Landauer formula is its wide utility to quantum transport.
It provides a clear understanding about mesoscopic quantum experiments of point contacts~\cite{vanWees-88, Wharam-88}, wires~\cite{Tarucha-95}, and carbon nanotubes~\cite{Frank-98}.
The Landauer formula also describes Anderson localization~\cite{Anderson-58, Thouless-74, Abrahams-79, Lee-review, Evers-review}, which is the disorder-induced localization as a consequence of the wave nature of electrons.
Furthermore, it explains electronic transport in topological materials~\cite{Kane-review, Zhang-review, Ryu-review, Vishwanath-review} including graphene~\cite{graphene-NG-review, graphene-DasSarma-review}.
The scattering formulation is also instrumental in understanding statistical behavior of transport properties such as the universal conductance fluctuations~\cite{Stone-85, *Lee-85, *Lee-86, Altshuler-85, Imry-86, Altshuler-86, Washburn-86, Beenakker-review, *Beenakker-review-sc}.
Another decisive advantage of the Landauer formula is its applicability to the far-from-equilibrium regime.
While the linear response theory is applicable only close to thermal equilibrium, the Landauer formula describes far-from-equilibrium phenomena such as shot noise~\cite{Khlus-87, Lesovik-89, Buttiker-90, *Buttiker-92, Martin-92, Buttiker-review} and dissipative transport in open systems~\cite{Mostafazadeh-09, Schomerus-10, Chong-10, *Chong-11, Longhi-10, Lin-11, McDonald-18, Dogra-19, *Corman-19, Kawabata-21}.

Recently, nonlinear response of electron systems has attracted growing interest~\cite{Oka-09, Sodemann-15, Morimoto-16, Wu-17, deJuan-17, Ma-19, Kang-19, He-19, Isobe-20, Du-19PRL, *Du-19, *Du-21, Nandy-19, Xiao-19, Singh-20, Watanabe-20, *Watanabe-20JStatPhys, *Takasan-21, *Liu-21, Ahn-20, Watanabe-Yanase-21, Kumar-21, Michishita-21, Zhang-20, Resta-21, Tanikawa-21, *Tanikawa-21v2, Fava-21, Bhalla-21}.
Nonlinearity of electrical conduction gives rise to new transport phenomena unseen in the linear regime, such as the quantized circular photogalvanic effect~\cite{deJuan-17} and the high-frequency rectification~\cite{Isobe-20}. 
Moreover, several fundamental relations in thermodynamics and statistical physics, including the Onsager reciprocal relations~\cite{Onsager-31a, *Onsager-31b}, require reconsideration in the nonlinear regime.
A prime example is the nonlinear 
Hall effect, which arises even in the presence of time-reversal invariance~\cite{Sodemann-15}.
The nonlinear 
Hall conductance also offers new pieces of information about topological band structures of materials such as the Berry curvature dipole. 
Experimentally, the nonlinear 
Hall effect was observed in WTe$_{\text 2}$ with time-reversal invariance~\cite{Ma-19, Kang-19}.
Nonlinear electronic transport opens a new avenue in condensed matter physics and statistical physics, awaiting further theoretical and experimental advances.

In view of the considerable recent interest in nonlinear transport phenomena, there seems to be an urgent need to develop a scattering formulation of nonlinear response.
Several previous works were based on nonlinear optical approaches~\cite{Shen-textbook, Boyd-textbook}, which are likely to be valid in the high-frequency regime.
For electrical conduction in the low-frequency regime, by contrast, the validity of such an optical approach is unclear, and the scattering approach should be more relevant.
While some previous works studied nonlinear transport of mesoscopic conductors~\cite{Christen-96, Sheng-98, Sanchez-04, Leturcq-06, Lewenkopf-07, Lewenkopf-09, Sanchez-13, Texier-18}, 
nonlinear transport of disordered and topological materials has been largely unexplored.
Furthermore, the nonlinear 
Hall effect has been studied solely by the semiclassical Boltzmann transport theory~\cite{Sodemann-15}.
As a result, the fully quantum nature of the nonlinear Hall effect has yet to be revealed.

In this work, we develop a nonlinear response theory based on the scattering formulation and use it to explore nonlinear transport phenomena of various disordered and topological materials.
In Sec.~\ref{sec: Landauer}, we formulate the nonlinear response theory and derive the nonlinear Landauer formula.
In the obtained formula, while the linear response is given by the transmission probability through the system, the nonlinear response is given by the derivatives of the transmission probability with respect to energy.
We demonstrate that the nonlinear Landauer formula fully captures the quantum effects of the nonlinear response as a virtue of the scattering formulation.
In particular, we explore nonlinear response of the following exemplary disordered and topological materials 
and clarify the different origins that induce nonlinear electronic transport:
\begin{enumerate}
\item {\it Disordered materials} (Sec.~\ref{sec: Anderson}).---We investigate nonlinear response of disordered electron systems in one dimension.
We discover singular behavior of nonlinear conductance as a consequence of chiral or particle-hole symmetry.
We further classify the universality classes of nonlinear response in disordered electron systems according to symmetry and discuss the critical phenomena of nonlinear response due to the Anderson transitions.
\item {\it Graphene} (Sec.~\ref{sec: graphene}).---We study nonlinear quantum transport of graphene as a prototypical topological semimetal.
We show its unique nonlinear quantum transport due to topology of the Dirac point.
\item {\it Nonlinear Hall effect} (Sec.~\ref{sec: QHE}).---We discuss the nonlinear 
Hall effect in terms of the scattering theory.
We reveal a close relationship between the second-order nonlinear 
Hall conductance and shot noise.
This finding establishes a hitherto unrecognized connection between the nonlinear 
Hall response and nonequilibrium quantum fluctuations.
We further discuss the quantum effects of disorder and Anderson localization on the nonlinear 
Hall effect with particular emphasis on symmetry.
\end{enumerate}

\noindent
We conclude this work and give outlooks in Sec.~\ref{sec: discussions}.

\section{Nonlinear Landauer formula}
    \label{sec: Landauer}

\begin{figure}[t]
\centering
\includegraphics[width=86mm]{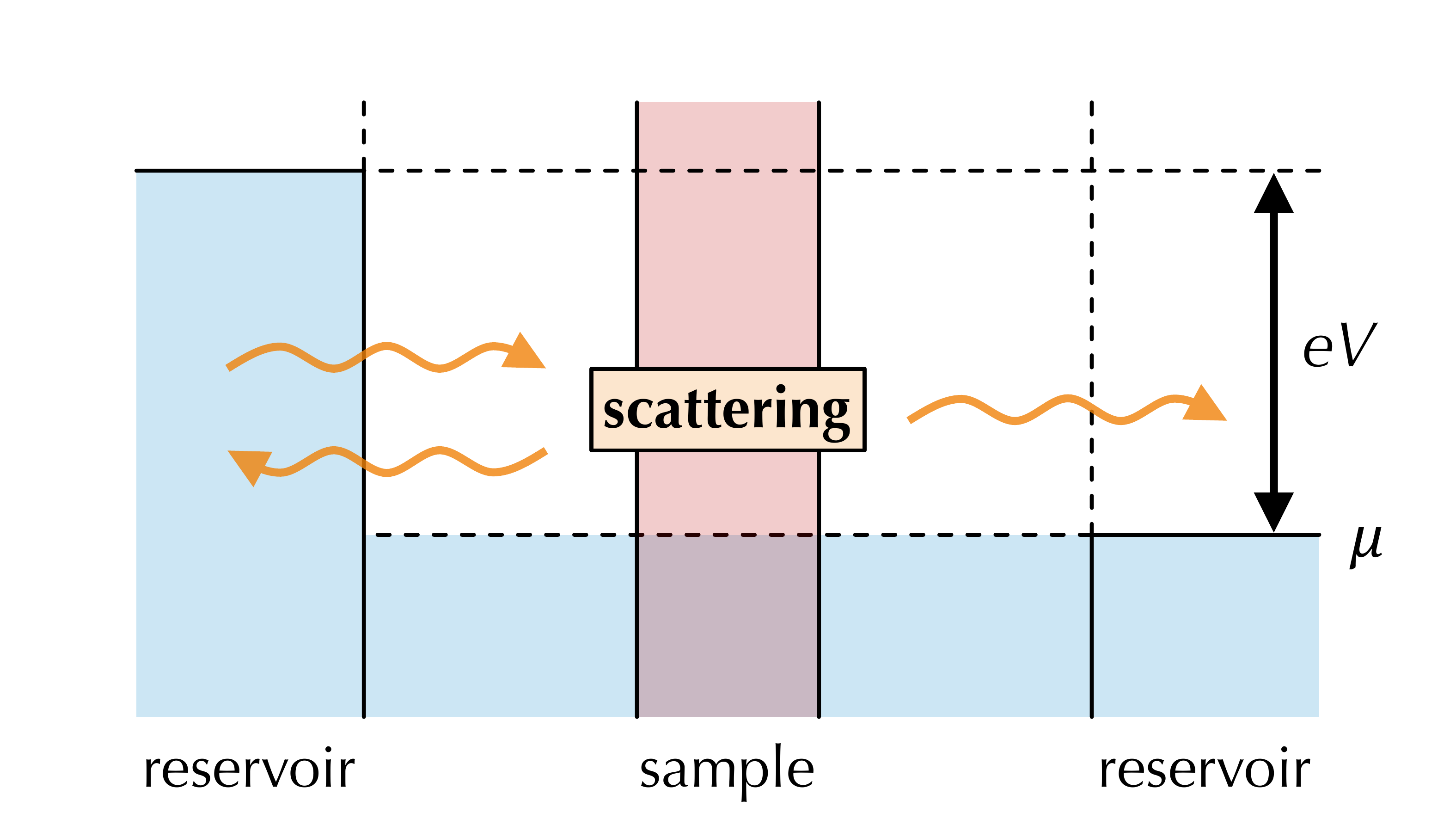} 
\caption{Scattering formulation of electrical conduction. The sample is attached to the two reservoirs at the different voltages through the ideal leads. The Landauer formula relates the electrical conduction to the transmission through the sample.}
	\label{fig: Landauer}
\end{figure}

We formulate a nonlinear response theory based on the scattering approach.
Let us consider a system of electrons that is attached to two electrodes via the ideal leads (Fig.~\ref{fig: Landauer}).
The electrodes are described by large reservoirs of electrons at thermal equilibrium, which obey the Fermi-Dirac distribution
\begin{equation}
    f_{\rm eq} \left( E \right) \coloneqq \frac{1}{e^{\beta\,(E-\mu)} + 1}
\end{equation}
with the inverse temperature $\beta$ and the chemical potential $\mu$.
We impose a voltage $V$ on the system by giving the potential difference $eV$ between the two reservoirs.
The applied voltage $V$ produces a current $I$ through the system, which generally depends nonlinearly on $V$ as
\begin{equation}
    I = \sum_{n=1}^{\infty} G_n V^n.
        \label{eq: conductance - definition}
\end{equation}
Here, $G_1$ denotes the linear conductance, and $G_n$ ($n \geq 2$) denotes the nonlinear conductance.
In the following, we show
\begin{align}
    G_n = \frac{e^{n+1}}{\left( n! \right) h} \int_{-\infty}^{\infty} \frac{d^{n-1}T}{dE^{n-1}} \left( - \frac{df_{\rm eq}}{dE} \right) dE,
        \label{eq: nonlinear conductance formula}
\end{align}
where $T = T \left( E \right)$ is the transmission probability of an electronic wave with energy $E$ through the system.
Moreover, $e > 0$ is the elementary charge, and $h$ is the Planck constant. 
At zero temperature, Eq.~(\ref{eq: nonlinear conductance formula}) reduces
to
\begin{equation}
    G_n = \frac{e^{n+1}}{\left( n! \right) h} \left. \frac{d^{n-1} T}{dE^{n-1}} \right|_{E = \mu}.
        \label{eq: nonlinear conductance formula - zero temperature}
\end{equation}
Equations~(\ref{eq: nonlinear conductance formula}) and (\ref{eq: nonlinear conductance formula - zero temperature}) constitute the nonlinear Landauer formula, which provides the nonlinear conductance in the scattering formulation.
While the linear conductance is given by the transmission probability $T = T \left( E \right)$, the nonlinear conductance is given by its derivatives.
In these formulas, the spin degrees of freedom are neglected for simplicity, which are readily recovered. 

\subsection{Derivation}
    \label{sec: Landauer - derivation}

Now, we derive the nonlinear Landauer formula in Eqs.~(\ref{eq: nonlinear conductance formula}) and (\ref{eq: nonlinear conductance formula - zero temperature}).
Let us first focus on electronic waves in the infinitesimal energy range $\left[ E, E+dE \right]$.
The current $dI_{\rm L \to R}^{\rm (in)}$ from the left reservoir to the system is 
\begin{align}
    dI_{\rm L \to R}^{\rm (in)}
    = ev\,dN_{\rm L \to R},
\end{align}
where $v$ is the velocity of the electrons, and $dN_{\rm L \to R}$ is their number.
In terms of the wave number $k$, we have $v = \hbar^{-1} dE/dk$ and $dN_{\rm L \to R} = f_{\rm eq} \left( E \left( k \right) - eV \right) dk/2\pi$.
Here, the chemical potential of the left reservoir is prepared to be $\mu + eV$.
Then, we have
\begin{align}
    dI_{\rm L \to R}^{\rm (in)}
    &= e \left( \frac{1}{\hbar} \frac{dE}{dk} \right) \left[ f_{\rm eq} \left( E\left( k \right) - eV \right) \frac{dk}{2\pi} \right] \nonumber \\
    &= \frac{e}{h} f_{\rm eq} \left( E - eV \right) dE.
\end{align}
Similarly, the current $dI_{\rm R \to L}^{\rm (in)}$ from the right reservoir to the system is 
\begin{align}
    dI_{\rm R \to L}^{\rm (in)}
    = \frac{e}{h} f_{\rm eq} \left( E \right) dE,
\end{align}
where the chemical potential of the right reservoir is $\mu$.

The incident electronic waves are scattered in the system.
Let $T_{\rm L \to R}$ ($T_{\rm R \to L}$) be the transmission probability of the system for the incident electronic wave from the left to the right (from the right to the left).
These transmission probabilities contain all information about the system.
It is also notable that they generally depend on energy $E$ of the electronic waves: $T_{\rm L \to R} = T_{\rm L \to R} \left( E \right)$ and $T_{\rm R \to L} = T_{\rm R \to L} \left( E \right)$.
Then, the total current $dI$ through the system is 
\begin{align}
    dI &= T_{\rm L\to R} \left( E \right) dI_{\rm L\to R}^{\rm (in)} - T_{\rm R\to L} \left( E \right) dI_{\rm R\to L}^{\rm (in)} \nonumber \\
    &= \frac{e}{h} \left[ T_{\rm L\to R} \left( E \right) f_{\rm eq} \left( E - eV \right) - T_{\rm R \to L} \left( E \right) f_{\rm eq} \left( E \right) \right] dE.
\end{align}

Now, considering all the electronic waves with arbitrary energy, we have
\begin{align}
    I &= \int dI = \frac{e}{h} \int_{-\infty}^{\infty} \left[ T_{\rm L\to R} \left( E \right) f_{\rm eq} \left( E - eV \right) \right. \nonumber \\
    &\qquad\qquad\qquad\qquad\qquad \left. - T_{\rm R \to L} \left( E \right) f_{\rm eq} \left( E \right) \right] dE.
\end{align}
When the system is isolated from the environment, current conservation requires the scattering matrix to be unitary (see Appendix~\ref{asec: scattering} 
for details).
As a result, the transmission probability $T_{\rm L\to R} \left( E \right)$ from the left to the right is identical to the transmission probability $T_{\rm R \to L} \left( E \right)$ from the right to the left:
\begin{align}
    T_{\rm L\to R} \left( E \right)
    = T_{\rm R \to L} \left( E \right)
    \eqqcolon T \left( E \right).
        \label{eq: transmission - unitarity}
\end{align}
This is a general consequence of unitarity of scattering matrices.
Then, the current $I$ reduces to
\begin{equation}
    I = \frac{e}{h} \int_{-\infty}^{\infty} T \left( E \right) \left[ f_{\rm eq} \left( E - eV \right) - f_{\rm eq} \left( E \right)\right] dE.
        \label{eq: Landauer - general}
\end{equation}
This is a general scattering formula of the current $I$ for the applied voltage $V$.

If we further assume that the transmission probability $T = T \left( E \right)$ is independent of energy $E$~\cite{Datta-textbook, Imry-textbook}, 
Eq.~(\ref{eq: Landauer - general}) reduces to
\begin{align}
    I &= \frac{e}{h}\,T \int_{-\infty}^{\infty} \left[ f_{\rm eq} \left( E - eV \right) - f_{\rm eq} \left( E \right)\right] dE \nonumber \\
    &= \left( \frac{e^2}{h} T \right) V,
        \label{eq: Landauer - textbook}
\end{align}
where we use 
\begin{equation}
    \int_{-\infty}^{\infty} \left[ f_{\rm eq} \left( E - eV \right) - f_{\rm eq} \left( E \right)\right] dE
    = eV.
\end{equation}
Here, the current $I$ is proportional to the voltage $V$, and the linear conductance is given as the transmission probability $T$ multiplied by the fundamental constant $e^2/h$.
Although this simplified formula is useful in obtaining the linear response, it does not explain nonlinear response. 
We note that Eq.~(\ref{eq: Landauer - textbook}) is valid as long as the transmission probability $T$ is independent of energy $E$.
Consequently, even for large $V$ and nonzero temperature $\beta < \infty$, Eq.~(\ref{eq: Landauer - textbook}) is valid, and nonlinear conductance vanishes.

Although Eq.~(\ref{eq: Landauer - textbook}) is widely used to obtain the linear conductance, the transmission probability $T$ does depend on energy $E$ in many cases.
In fact, such energy dependence of the transmission probability leads to the nonlinear response.
To obtain the nonlinear conductance, let us consider the energy dependence of the transmission probability. 
Expanding the Fermi-Dirac distribution function $f_{\rm eq}$ in terms of the applied voltage $V$, we have
\begin{align}
    f_{\rm eq} \left( E - eV \right) - f_{\rm eq} \left( E \right)
    = \sum_{n=1}^{\infty} \frac{1}{n!} \frac{d^n f_{\rm eq}}{dE^{n}} \left( -eV \right)^{n}.
\end{align}
Inserting this expansion into Eq.~(\ref{eq: Landauer - general}), we have
\begin{align}
    I = \frac{e}{h} \sum_{n=1}^{\infty} \frac{1}{n!} \left( \int_{-\infty}^{\infty} T \frac{d^n f_{\rm eq}}{dE^{n}} dE \right) \left( -eV \right)^{n}.
\end{align}
From the definition of the conductance $G_n$ in Eq.~(\ref{eq: conductance - definition}), we have
\begin{align}
    G_n = \frac{\left( -1 \right)^n e^{n+1}}{\left( n! \right) h} \int_{-\infty}^{\infty} T \frac{d^n f_{\rm eq}}{dE^{n}} dE.
\end{align}
Here, the transmission probability $T$ and its derivatives should vanish in the limit $\left| E \right| \to \infty$ for physical electron systems in solids.
Then, integration by parts leads to Eq.~(\ref{eq: nonlinear conductance formula}).
If the transmission probability $T$ is independent of energy $E$, we recover Eq.~(\ref{eq: Landauer - textbook}), i.e., $G_1 = \left( e^2/h \right) T$ and $G_n = 0$ for $n \geq 2$.
In many cases, temperature is sufficiently low in comparison with the relevant energy scale and well approximated to zero.
At zero temperature (i.e., $\beta = \infty$), we have $-df_{\rm eq}/dE = \delta \left( E - \mu \right)$,
and Eq.~(\ref{eq: nonlinear conductance formula}) reduces to Eq.~(\ref{eq: nonlinear conductance formula - zero temperature}).
Equations~(\ref{eq: nonlinear conductance formula}) and (\ref{eq: nonlinear conductance formula - zero temperature}) constitute a nonlinear extension of the Landauer formula.

According to the conventional Landauer formula in Eq.~(\ref{eq: Landauer - textbook}), the linear conductance $G_1$ is given by the transmission probability $T$.
On the other hand, according to the nonlinear Landauer formula in Eqs.~(\ref{eq: nonlinear conductance formula}) and (\ref{eq: nonlinear conductance formula - zero temperature}), the $n$th-order nonlinear conductance $G_n$ ($n\geq2$) is given by the $\left(n-1\right)$\,th derivative of the transmission probability $T$ with respect to energy $E$. 
Here, the transmission probability $T$ can be systematically calculated on the basis of the Green's function method~\cite{Datta-textbook}.
In Appendix~\ref{asec: example}, 
we investigate nonlinear transport through some exemplary potential barriers on the basis of the nonlinear Landauer formula.

Notably, the nonlinear Landauer formula in Eqs.~(\ref{eq: nonlinear conductance formula}) and (\ref{eq: nonlinear conductance formula - zero temperature}) is applicable in higher dimensions as well.
In two dimensions, for example, let us consider the diagonal conductance along the $x$ direction.
In contrast to one dimension, transport along the $x$ direction consists of many modes characterized by wave numbers $k_y$ along the $y$ direction.
In the presence of translation invariance, these modes are independent of each other.
Hence, we have a well-defined transmission probability $T_x \left( E, k_y \right)$ for electronic waves with energy $E$ and wave number $k_y$ along the $y$ direction.
The total transmission probability is given as $T_x \left( E \right) = \sum_{k_y} T_x \left( E, k_y \right)$.
In the absence of translation invariance due to disorder, the modes with different wave numbers $k_y$ interact with each other. 
Still, we can calculate the total transmission probability $T \left( E \right)$ by summing up all the interacting transmitted modes.
As a prime example of two-dimensional materials, we investigate nonlinear conductance of graphene in Sec.~\ref{sec: graphene}.
We also discuss nonlinear Hall conductance with the scattering approach in Sec.~\ref{sec: QHE}.

\subsection{Nonreciprocity}
    \label{sec: nonreciprocity}

A crucial feature of nonlinear response is nonreciprocity.
Here, the nonreciprocal response is defined by
\begin{equation}
    I \left( V \right) \neq - I \left( -V \right).
        \label{eq: nonreciprocal response}
\end{equation}
For the linear response $I \left( V \right) = G_1 V$, we always have $I \left( V \right) = - I \left( - V \right)$, and the response is reciprocal.
Thus, nonreciprocal response requires nonlinearity, especially even-order nonlinear response.
Nonreciprocal response has recently attracted growing interest, for example, in noncentrosymmetric quantum materials~\cite{Tokura-Nagaosa-review}.
It has been analyzed perturbatively in a manner similar to nonlinear optics~\cite{Shen-textbook, Boyd-textbook}.

The nonlinear Landauer formula in Eqs.~(\ref{eq: nonlinear conductance formula}) and (\ref{eq: nonlinear conductance formula - zero temperature}) provides a general understanding about nonreciprocal response in the scattering theory. 
In particular, the dominant contribution to the nonreciprocal response arises from the second-order nonlinear conductance 
\begin{equation}
    G_2 = \frac{e^2}{2h} \int_{-\infty}^{\infty} \frac{dT}{dE} \left( - \frac{df_{\rm eq}}{dE}\right) dE.
\end{equation}
Thus, in the scattering theory, the derivative of the transmission probability is crucial for nonreciprocal response.
In the following, we demonstrate that various types of materials indeed exhibit nonreciprocal response $G_2 \neq 0$. 
In particular, such nonreciprocal response can arise even without many-body interaction.
Nonreciprocal response in simple systems is discussed in Appendix~\ref{asec: example}.

\subsection{Nonlinear Boltzmann conductivity}
    \label{sec: Boltzmann}

The Landauer formula is applicable in the presence of quantum coherence, which is distinct from semiclassical approaches such as the Boltzmann transport theory~\cite{AM-textbook, Abrikosov-textbook}.
To highlight this feature, we here derive nonlinear conductivity on the basis of the Boltzmann equation and compare it with the nonlinear Landauer formula.
Let us consider a system of electrons characterized by the energy dispersion $E = E \left( \bm{k} \right)$.
In contrast to the scattering approach, the system is prepared to be in isolation from the environment including the electrodes.
Without any external field, the system is at thermal equilibrium and described by the Fermi-Dirac distribution $f_{\rm eq} = f_{\rm eq} \left( \bm{k} \right) \coloneqq 1/\left( e^{\beta \left( E \left( \bm{k} \right) - \mu\right)} + 1\right)^{-1}$ with the inverse temperature $\beta$ and the chemical potential $\mu$.
Instead of imposing the voltage by attaching the electrodes, we apply an electric field $\bm{\mathcal{E}}$ to the system.
Because of this applied electric field $\bm{\mathcal{E}}$, the distribution function $f = f \left( \bm{k} \right)$ deviates from the equilibrium distribution function $f_{\rm eq}$.
Since $\bm{\mathcal{E}}$ is uniform and static, $f$ only depends on the wave number $\bm{k}$ and is independent of space and time.
We assume that the distribution function $f$ follows the Boltzmann equation with the relaxation time approximation~\cite{AM-textbook, Abrikosov-textbook}:
\begin{equation}
    - \frac{e\bm{\mathcal{E}}}{\hbar} \cdot \frac{\partial f}{\partial \bm{k}}
    = - \frac{f-f_{\rm eq}}{\tau}.
        \label{eq: Boltzmann}
\end{equation}
The relaxation time $\tau$ depends on details of scattering processes.
While $\tau$ is generally a complicated function of wave number $\bm{k}$, we ignore such $\bm{k}$ dependence for the sake of brevity.
To make the comparison with the Landauer formula clear, we focus on one-dimensional systems in the following.
A generalization to higher-dimensional systems is straightforward.

Solving the Boltzmann equation in Eq.~(\ref{eq: Boltzmann}) perturbatively, we have
\begin{equation}
    f = f_{\rm eq} + \sum_{n=1}^{\infty} \left( \frac{e\tau \mathcal{E}}{\hbar} \right)^{n} \frac{d^{n} f_{\rm eq}}{dk^{n}}.
\end{equation}
The current density $\currentdensity$ is obtained as
\begin{align}
    \currentdensity = -e \oint v f \frac{dk}{2\pi}
\end{align}
with the velocity $v \coloneqq \hbar^{-1} dE/dk$.
Here, the integral is taken over the entire momentum space.
Now, we define the $n$th-order nonlinear conductivity $\sigma_n$ by
\begin{equation}
    \currentdensity = \sum_{n=1}^{\infty} \sigma_n \mathcal{E}^{n}
\end{equation}
in a manner similar to the $n$th-order nonlinear conductance $G_n$ in Eq.~(\ref{eq: conductance - definition}).
Generally, in $d$ dimensions, we have $\currentdensity = I/L^{d-1}$ and $\mathcal{E} = V/L$ with the length scale $L$, leading to
\begin{equation}
    \sigma_n = G_n L^{n-d+1}.
\end{equation}
In one dimension, this relationship reduces to $\sigma_n = G_n L^{n}$.
Then, the $n$th-order conductivity $\sigma_n$ is obtained as
\begin{equation}
    \sigma_n = - \frac{1}{\tau} \left( \frac{e\tau}{\hbar} \right)^{n+1} \oint  \frac{dE}{dk} \frac{d^{n} f_{\rm eq}}{dk^{n}} \frac{dk}{2\pi},
\end{equation}
where the integration is carried over the entire momentum space.
This is the response formula derived from the Boltzmann equation.
Integration by parts leads to
\begin{align}
    \sigma_n 
    &= \frac{1}{\tau} \left( -\frac{e\tau}{\hbar} \right)^{n+1} \oint \frac{d^n E}{dk^n} \left( - \frac{df_{\rm eq}}{dk}\right) \frac{dk}{2\pi} \nonumber \\
    &= \frac{1}{\tau} \left( -\frac{e\tau}{\hbar} \right)^{n+1} \oint \frac{d^{n+1} E}{dk^{n+1}} f_{\rm eq} \frac{dk}{2\pi}.
        \label{eq: nonlinear conductance - Boltzmann}
\end{align}
At zero temperature, we have $-df_{\rm eq}/dk = \left( dE/dk \right) \delta \left( E \left( k\right) - \mu \right)$, which simplifies the formula to
\begin{align}
    \sigma_n 
    &= \frac{1}{\tau} \left( -\frac{e\tau}{\hbar} \right)^{n+1} \oint \frac{d^n E}{dk^n} \frac{dE}{dk} \delta \left( E \left( k \right) - \mu \right) \frac{dk}{2\pi}. 
\end{align}
In Appendix~\ref{asec: Boltzmann}, 
we calculate linear and nonlinear conductivity of some exemplary systems on the basis of the Boltzmann equation.

As demonstrated in Eq.~(\ref{eq: nonlinear conductance - Boltzmann}), the $n$th-order Boltzmann conductivity $\sigma_n$ is given by the $n$th derivative of the energy dispersion $E = E \left( k \right)$ [i.e., the $\left( n-1 \right)$\,th derivative of the velocity $\hbar^{-1} dE/dk$].
On the other hand, as demonstrated in Eq.~(\ref{eq: nonlinear conductance formula}), the $n$th-order Landauer conductance $G_n$ is given by the $\left( n-1 \right)$\,th derivative of the transmission probability $T = T \left( E \right)$. 
These results appear to be similar to each other and show a common mathematical structure underlying the two different transport theories.
In fact, both velocity $\hbar^{-1} dE/dk$ and transmission probability $T$ describe a flow of electrons and share a similar physical interpretation, at least in the semiclassical regime.
In the fully quantum regime, however, a crucial distinction arises between $\hbar^{-1} dE/dk$ and $T$.
In fact, the transmission probability $T$ cannot be obtained solely from the energy dispersion $E$.
It also contains information about wave functions of electrons.
In the presence of strong quantum coherence, electronic transport should be described by both energy dispersion and wave functions.
Consequently, while the Landauer formula fully captures such a quantum effect, the Boltzmann equation does not. 

In fact, the Boltzmann equation is valid only in the semiclassical regime and invalid in the fully quantum regime~\cite{AM-textbook, Abrikosov-textbook}.
The Boltzmann equation assumes that electrons are particles with momenta $\hbar \bm{k}$.
In other words, packets of electronic waves are assumed to be well defined, which behave as particles.
For this assumption to be valid, the mean free path $\ell$ (i.e., the average distance between successive scattering events) needs to be much larger than the Fermi wave length $\lambda$ (i.e., $\ell \gg \lambda$).
At sufficiently high temperature, this condition is usually respected because of strong inelastic scattering, which destroys the coherence of electrons.
At low temperature, by contrast, the coherence can be strong, and the condition $\ell \gg \lambda$ can break down.
In such a fully quantum regime, electrons behave as waves and exhibit unique transport phenomena even in the linear regime, such as Anderson localization~\cite{Anderson-58}.
In this work, we use the nonlinear Landauer formula to explore nonlinear transport phenomena that have genuinely quantum nature.

\subsection{Noise}
    \label{sec: noise}

Noise provides key pieces of information about transport properties~\cite{Khlus-87, Lesovik-89, Buttiker-90, *Buttiker-92, Martin-92, Buttiker-review}.
For example, the discrete nature of electrons is captured by shot noise far from thermal equilibrium, which contrasts with the thermal (Johnson-Nyquist) noise at equilibrium.
Shot noise is also sensitive to fractional charges that accompany the fractional quantum Hall effect~\cite{Saminadayar-97, dePicciotto-97}.
As a virtue of the scattering formulation, the Landauer formula enables direct characterization of noise even far from thermal equilibrium.
Here, we derive the nonlinear contributions of shot noise in the scattering formulation.

In general, the current fluctuates with time: $I = I \left( t \right)$.
To characterize the current fluctuations, we consider the correlation function of the current $I \left( t \right)$ defined by
\begin{align}
    C \left( t \right)
    \coloneqq \overline{I \left( t \right) I \left( 0 \right)}
    - \overline{I \left( t \right)}~\overline{I \left( 0 \right)},
\end{align}
where the overline denotes the time average.
Then, the noise $S$ is defined as
\begin{align}
    S \coloneqq 2 \int_{-\infty}^{\infty} C \left( t \right) dt.
        \label{eq: noise - def}
\end{align}
Using the wave-packet approach~\cite{Martin-92}, we associate the noise $S$ with the transmission probability $T = T \left( E \right)$ (see Appendix~\ref{asec: noise} 
for details):
\begin{align}
    S &= \frac{2e^2}{h} \int_{-\infty}^{\infty} \left\{
    T^2 \left( E \right) \left[ f_{\rm eq} \left( E - eV \right) \left( 1 - f_{\rm eq} \left( E-eV \right) \right) \right. \right. \nonumber \\
    &\qquad\qquad\qquad\qquad\qquad \left. +f_{\rm eq} \left( E \right) \left( 1 - f_{\rm eq} \left( E \right) \right) \right] \nonumber \\
    &\qquad+ T \left( E \right) \left( 1 - T \left( E \right) \right) \left[ f_{\rm eq} \left( E- eV \right) \left( 1 - f_{\rm eq} \left( E \right) \right) \right.\nonumber \\
    &\qquad\qquad\quad \left.\left. + f_{\rm eq} \left( E \right) \left( 1 - f_{\rm eq} \left( E - eV \right) \right)\right] \right\} dE.
        \label{eq: noise - general}
\end{align}
The noise $S$ includes the thermal noise driven by thermal fluctuations at equilibrium.
The thermal noise is present even in the absence of an external bias voltage $V$.
The noise $S$ also includes the nonequilibrium noise that is caused by the external voltage $V$ and survives even at zero temperature.
Such nonequilibrium noise, i.e., shot noise, originates from the discrete nature of electrons.

To characterize the shot noise, let us focus on zero temperature.
Then, the distribution functions of the reservoirs reduce to the step function $f_{\rm eq} \left( E \right) = \theta \left( \mu - E \right)$, and Eq.~(\ref{eq: noise - general}) reduces to
\begin{align}
    S = \frac{2e^2}{h} \int_{\mu}^{\mu + eV} T \left( E \right) \left( 1 - T \left( E \right) \right) dE.
\end{align}
This is a general formula of the shot noise for an arbitrary external voltage $V$.
If we define the $n$th-order noise power $\sigma_n$ by
\begin{align}
    S = \sum_{n=1}^{\infty} \sigma_n V^n,
\end{align}
we have
\begin{align}
    \sigma_n = \frac{2e^{n+2}}{\left( n! \right) h} \left. \frac{d^{n-1}}{dE^{n-1}} \left[ T \left( E \right) \left( 1 - T \left( E \right) \right) \right] \right|_{E = \mu}.
        \label{eq: shot noise - sigma}
\end{align}
The first three $\sigma_n$'s are explicitly given by
\begin{align}
    \sigma_1 &= \frac{2e^3}{h} T \left( \mu \right) \left( 1 - T \left( \mu \right) \right), \\
    \sigma_2 &= \frac{e^4}{h} T' \left( \mu \right) \left( 1 - 2T \left( \mu \right) \right), \\
    \sigma_3 &= \frac{e^5}{3h} \left[ T'' \left( \mu \right) \left( 1 - 2T \left( \mu \right) \right) - 2 \left( T' \left( \mu \right) \right)^2 \right].
\end{align}
Here, $T' \left( \mu \right)$ and $T'' \left( \mu \right)$ are the first and second derivatives of the transmission probability $T \left( \mu \right)$ with respect to the chemical potential $\mu$.
Thus, the $n$th-order noise power $\sigma_n$ is given by the $\left( n -1 \right)$\,th derivative of the transmission probability $T \left( E \right)$ in a manner similar to the $n$th-order nonlinear conductance $G_n$ in Eq.~(\ref{eq: nonlinear conductance formula - zero temperature}). 

The transmission probability $T$ is often very small.
For example, when the system is subject to Anderson localization due to disorder, we have $T \ll 1$.
In such cases, we have $1 - T \simeq 1$ and hence
\begin{align}
    \sigma_n \simeq \frac{2e^{n+2}}{\left( n! \right) h} T^{\left( n-1 \right)} \left( \mu \right),
\end{align}
where $T^{\left( n-1 \right)} \left( \mu \right)$ denotes the $\left( n-1 \right)$\,th derivative of $T \left( \mu \right)$.
Since the $n$th-order nonlinear conductance $G_n$ is given by Eq.~(\ref{eq: nonlinear conductance formula - zero temperature}), we have 
\begin{align}
    \sigma_n \simeq 2eG_n.
\end{align}
In the linear regime, the relationship $\sigma_1 \simeq 2eG_1$ holds for the classical shot noise (i.e., Poisson noise)~\cite{Buttiker-review}.
Our results demonstrate that a similar relationship generally holds also for the nonlinear noise power $\sigma_n$ and the nonlinear conductance $G_n$.
By contrast, when the transmission probability $T$ is not small (i.e., $T \simeq 1$), the shot noise deviates from the Poisson noise. 
The quantum correction is evaluated by the Fano factor
\begin{equation}
    F \coloneqq \frac{S}{2eI}.
\end{equation}
The leading-order contribution of the Fano factor is known to be $1-T$.
Here, we derive the nonlinear corrections to the Fano factor as
\begin{align}
    &F 
    = 1 - T \left( \mu \right) - \frac{eT' \left( \mu \right)}{2} V \nonumber \\
    &\quad - \frac{e^2 \left[ 2 T \left( \mu \right) T'' \left( \mu \right) + \left( T' \left( \mu \right) \right)^2 \right]}{12 T \left( \mu \right)} V^2 + \mathcal{O} \left( V^3 \right).
\end{align}

Remarkably, thermal noise was experimentally measured for a quantum conductor even in the nonlinear regime~\cite{Nakamura-10, *Nakamura-11}.
In this experiment, the observed noise is mainly due to thermal fluctuations and survives even in the absence of the external voltage.
At much lower temperature, the thermal noise should be suppressed and replaced by shot noise driven by the external voltage. 
In such a nonequilibrium regime, Eq.~(\ref{eq: shot noise - sigma}) characterizes the noise of the conductor.
It also deserves further research to investigate the nonlinear contributions of shot noise for the fractional quantum Hall effect~\cite{Saminadayar-97, dePicciotto-97}.
In Sec.~\ref{sec: QHE}, we find a new connection between the shot noise and the nonlinear 
Hall conductance.

\subsection{Comparison with previous formulas}
    \label{sec: literature}

In the literature, Eq.~(\ref{eq: Landauer - general}) was used to obtain the current-voltage characteristic of specific scattering processes, such as double-barrier tunneling~\cite{Chang-Esaki-Tsu-74, Datta-textbook}.
However, it involves numerical calculations, which makes the analytical treatment and general understanding difficult.
The nonlinear Landauer formula in Eqs.~(\ref{eq: nonlinear conductance formula}) and (\ref{eq: nonlinear conductance formula - zero temperature}) elucidates universal features of nonlinear transport, as demonstrated below.

It should also be noted that the nonlinear Landauer formula in Eqs.~(\ref{eq: nonlinear conductance formula}) and (\ref{eq: nonlinear conductance formula - zero temperature}) is not directly applicable to strongly correlated electron systems.
To take into account many-body interaction, we need to consider renormalization of the current and the chemical potential difference between the electrodes.
In fact, the chemical potential difference does not necessarily coincide with the electric potential difference $eV$ in the presence of many-body interaction; the interaction-induced density redistribution also contributes to the chemical potential difference. 
The current $I$ is also subject to similar renormalization.
Such renormalization is crucial even in the linear regime~\cite{Izuyama-61, Maslov-95, Safi-95, Ponomarenko-95, Kawabata-96, *Kawabata-98, Shimizu-96}, which is needed to explain the robust quantization of the linear conductance experimentally observed in point contacts~\cite{vanWees-88, Wharam-88}, wires~\cite{Tarucha-95}, and carbon nanotubes~\cite{Frank-98}.
It is also notable that Ref.~\cite{Christen-96} presented a formula similar to Eq.~(\ref{eq: nonlinear conductance formula}), which was applied to time-reversal-symmetry-breaking transport~\cite{Sanchez-04} and thermoelectric transport~\cite{Sanchez-13}. 
However, the formula in Ref.~\cite{Christen-96} does not consider the renormalization due to many-body interaction. 
In the present work, we focus on noninteracting systems and show that new physics arises even in the absence of many-body interaction.
In particular, disorder and topology lead to unique nonlinear transport phenomena, as demonstrated in the subsequent sections.

\section{Disorder-induced nonlinear quantum transport}
    \label{sec: Anderson}

Anderson localization~\cite{Anderson-58, Thouless-74, Abrahams-79, Lee-review, Evers-review} is the disorder-induced localization of coherent waves.
In perfect crystals with translation invariance, electrons form Bloch waves that extend over the entire systems.
Disorder that breaks translation invariance of crystals leads to scattering and interference of electronic waves, resulting in the formation of localized standing waves and the suppression of transport.
Such localization, i.e., Anderson localization, originates from the quantum nature of electronic waves and serves as one of the best platforms where the Landauer formula is relevant.
Anderson localization plays an important role in transport of mesoscopic electron systems~\cite{Lee-review, Evers-review, Datta-textbook, Imry-textbook, Beenakker-review, *Beenakker-review-sc}, as well as synthetic materials of light~\cite{Schwartz-07, Lahini-08, Segev-review} and cold atoms~\cite{Billy-08, Roati-08}. 

\begin{figure}[t]
\centering
\includegraphics[width=86mm]{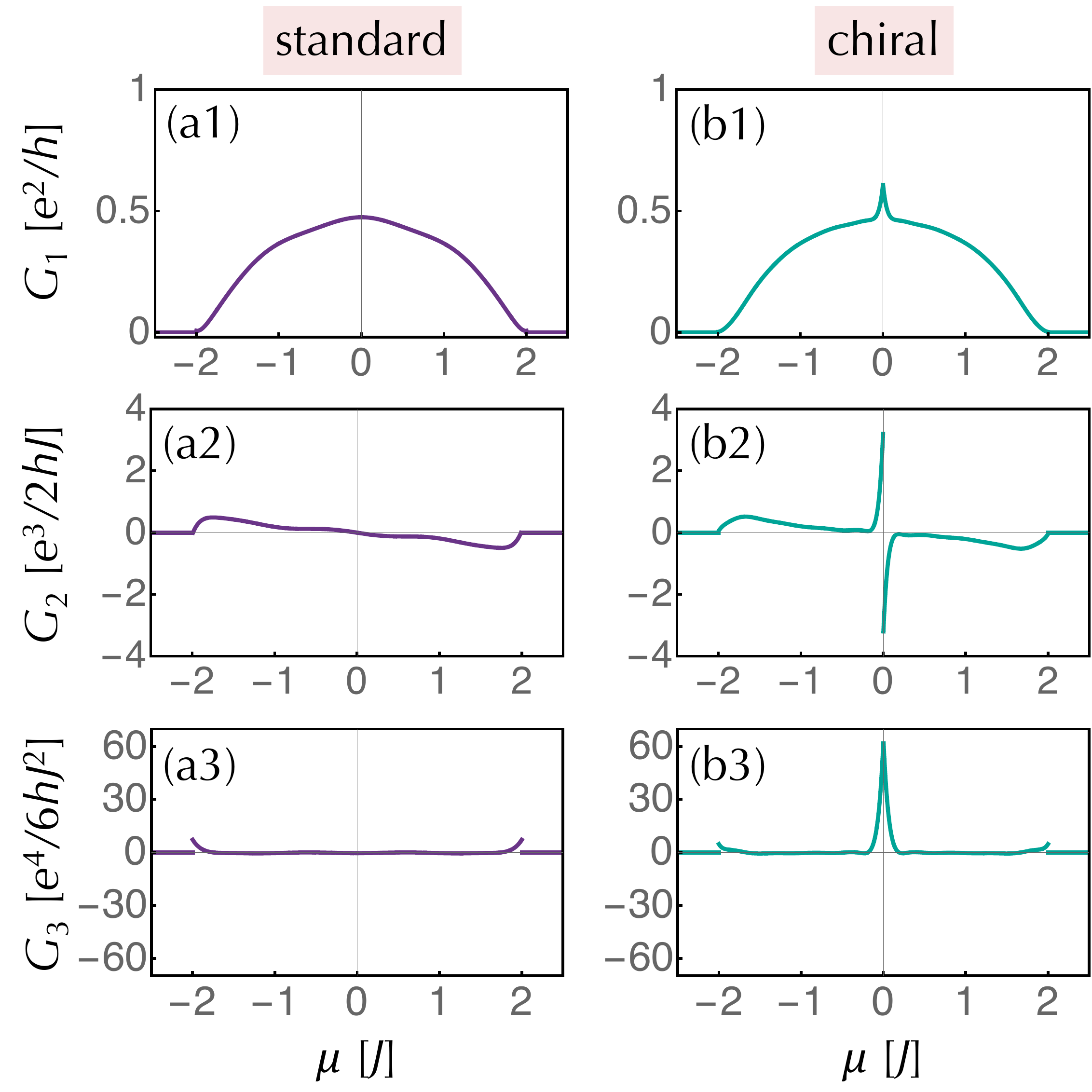} 
\caption{Linear and nonlinear conductance of the disordered chains ($L=50$, $J=1.0$, $W = 1.0$). Each datum shows the average over $100000$ samples. (a1)~The linear conductance $G_1$, (a2)~the second-order nonlinear conductance $G_2$, and (a3)~the third-order nonlinear conductance $G_3$ as functions of the chemical potential $\mu$ in the standard class. (b1)~$G_1$, (b2)~$G_2$, and (b3)~$G_3$ in the chiral class. Away from zero energy, the conductance behaves similarly in the standard and chiral classes for both linear and nonlinear regimes. At zero energy, by contrast, the conductance exhibits the singular behavior in the chiral class.}
	\label{fig: disorder}
\end{figure}

Here, we use the nonlinear Landauer formula to investigate nonlinear response of disordered electron systems that are subject to Anderson localization.
In particular, we demonstrate the singularly large nonlinear conductance in disordered chains with chiral or particle-hole symmetry (Fig.~\ref{fig: disorder}),
which provides a clear experimental signature of the disorder-induced nonlinear transport.
In addition to the numerical calculations of prototypical models, we discuss general classification of nonlinear response of disordered electron systems in one dimension based on symmetry (Table~\ref{tab: AZ}).
We further discuss the influence of the Anderson transitions on nonlinear transport in higher-dimensional disordered systems.

\subsection{Standard class}
    \label{sec: Anderson - standard}

To understand nonlinear response of disordered chains, we investigate the following lattice model in one dimension with a disordered onsite potential:
\begin{equation}
    \hat{H} = \sum_{n=1}^{L} \left[ -J \left( \hat{c}_{n+1}^{\dag} \hat{c}_{n} + \hat{c}_{n}^{\dag} \hat{c}_{n+1} \right) + V_n \hat{c}_{n}^{\dag} \hat{c}_{n} \right].
        \label{eq: Anderson model}
\end{equation}
Here, $\hat{c}_{n}$ ($\hat{c}_{n}^{\dag}$) annihilates (creates) a spinless fermion at site $n$, and $J > 0$ is the hopping amplitude.
The disordered potential $V_n \in \mathbb{R}$ is chosen uniformly from $\left[ -W/2, W/2 \right]$ with the disorder strength $W \geq 0$.
We impose the periodic boundary conditions (i.e., $\hat{c}_{L+1} = \hat{c}_{1}$, $\hat{c}_{L+1}^{\dag} = \hat{c}_{1}^{\dag}$).
This model respects time-reversal symmetry (i.e., $\hat{H}^{*} = \hat{H}$) because of the real parameters $J, V_n \in \mathbb{R}$ (see Appendix~\ref{asec: symmetry} 
for details about symmetry).

In the absence of the disordered potential (i.e., $W = 0$), the single-particle spectrum is given as
\begin{equation}
    E \left( k \right) = -2J \cos k
        \label{eq: Anderson - band}
\end{equation}
with momentum $k \in \left[ 0, 2\pi \right]$.
In such a periodic crystal, the perfect transmission (i.e., $T = 1$) is realized inside the energy band, while no transmission (i.e., $T = 0$) occurs outside the energy band:
\begin{equation}
    T \left( E \right) = \begin{cases}
    1 & \left( \left| E \right| < 2J \right); \\
    0 & \left( \left| E \right| > 2J \right).
    \end{cases}
        \label{eq: Anderson - periodic}
\end{equation}
Consequently, the linear conductance is $G_1 = e^2/h$ in the energy band $\left| E \right| < 2J$, and 
the nonlinear conductance vanishes 
except for the band edges $E = \pm 2J$.

The disordered potential deforms the band structure and changes the transport properties.
We numerically calculate the transmission probability $T = T \left( E \right)$ for the disordered chain in Eq.~(\ref{eq: Anderson model}) (see Appendix~\ref{asec: numerics - conductance} 
for details about the numerics)~\cite{Ohtsuki-review}.
Here, $T$ is a disorder-dependent statistical quantity, and we consider the average transmission probability $\braket{T}$ for many samples.  
Then, from the nonlinear Landauer formula in Eq.~(\ref{eq: nonlinear conductance formula - zero temperature}), we obtain the average linear and nonlinear conductance, as summarized in Fig.~\ref{fig: disorder}\,(a).
In this model, all the eigenstates are localized even for small disorder.
This behavior is consistent with the scaling theory of Anderson localization~\cite{Abrahams-79}, which predicts the absence of delocalization in one-dimensional disordered systems.
As shown in Fig.~\ref{fig: disorder}\,(a1), the transmission probability clearly deviates from the behavior of the periodic crystal in Eq.~(\ref{eq: Anderson - periodic}).
The linear conductance $G_1$ reaches the maximum at the band center $\mu = 0$, gradually decreases away from the band center, and vanishes outside the energy band.
This behavior is consistent with the density of states and the localization behavior.
In fact, the density of states near the band center is larger than that near the band edges; the eigenstates near the band center are more delocalized than those near the band edges.

As a consequence of this behavior, the second-order nonlinear conductance $G_2$, which is given by the derivative of $G_1 = G_1 \left( \mu \right)$, vanishes at the band center $\mu = 0$ and grows around the band edges.
Since the nonlinear response vanishes (i.e., $G_2 = 0$) in the perfect crystal, the nonlinear response $G_2 \neq 0$ is induced by disorder.
It also accompanies nonreciprocal response, as discussed in Sec.~\ref{sec: nonreciprocity}.
In a manner similar to $G_2$, the third-order nonlinear response $G_3$ grows near the band edges.
This linear and nonlinear response is general features of one-dimensional electron systems in the standard class (i.e., symmetry classes that only involve time-reversal symmetry; namely, classes A, AI, and AII in Table~\ref{tab: AZ}).

For sufficiently weak disorder and a sufficiently large system length, the distribution of the transmission probability $T$ is analytically obtained by the random-matrix approach (see Appendix~\ref{asec: DMPK - standard} 
for details)~\cite{Anderson-80, Dorokhov-82, Mello-88, Beenakker-review, *Beenakker-review-sc}.
Using the obtained probability distribution, we have the average transmission probability
\begin{align}
    \braket{T}
    \simeq \sqrt{\frac{8\xi}{\pi L}}~e^{-L/2\xi}
        \label{eq: Anderson - T - standard}
\end{align}
with the energy-dependent localization length
\begin{align}
    \xi = \frac{2 \left( 4J^2 - E^2 \right)}{\braket{V_n^2}}.
\end{align}
When the disordered potential $V_n$ is distributed uniformly in $\left[ -W/2, W/2 \right]$, we have
\begin{equation}
    \braket{V_n^2}
    = \int_{-W/2}^{W/2} V^2 \frac{dV}{W}
    = \frac{W^2}{12}.
\end{equation}
The transmission probability $\braket{T}$ decreases exponentially as a function of the system length $L$, which is a hallmark of Anderson localization.
The localization length $\xi$ and the transmission probability $\braket{T}$ become maximal at the band center $E = 0$ and decrease away from it.
Moreover, the large nonlinear response arises near the band edges (see Fig.~\ref{fig: DMPK} in Appendix~\ref{asec: DMPK - standard}). 
These analytical results are consistent with the numerical results in Fig.~\ref{fig: disorder}\,(a).

\subsection{Chiral class}

While the transport properties discussed in the preceding subsection~\ref{sec: Anderson - standard} are universal for one-dimensional electron systems in the standard class, symmetry changes the universality class of Anderson localization.
In one dimension, relevant symmetry is chiral (sublattice) symmetry, which enables delocalized zero-energy modes even in the presence of disorder~\cite{Dyson-53, Stone-81A, *Stone-81B, Gade-91, *Gade-93, AZ-97, Brouwer-98, *Brouwer-00}. 
To understand the role of chiral symmetry for nonlinear quantum transport, we investigate the following lattice model in one dimension with random hopping:
\begin{equation}
    \hat{H} = - \sum_{n=1}^{L} \left( J + \disorderedhopping \right) \left( \hat{c}_{n+1}^{\dag} \hat{c}_{n} + \hat{c}_{n}^{\dag} \hat{c}_{n+1} \right).
        \label{eq: Dyson model}
\end{equation}
Similarly to the model in Eq.~(\ref{eq: Anderson model}),
this model reduces to Eqs.~(\ref{eq: Anderson - band}) and (\ref{eq: Anderson - periodic}) in the absence of the random hopping.
The random hopping amplitude $\disorderedhopping \in \mathbb{R}$ is chosen uniformly from $\left[ -W/2, W/2 \right]$ with the disorder strength $W \geq 0$.

A crucial difference between the two models is chiral symmetry (see Appendix~\ref{asec: symmetry} 
for details about symmetry).
In fact, the model in Eq.~(\ref{eq: Dyson model}) respects chiral symmetry
\begin{equation}
    \hat{\cal S} \hat{H} \hat{\cal S}^{-1} 
    = \hat{H},
        \label{eq: CS}
\end{equation}
where the antiunitary operator $\hat{\cal S}$ is defined by
\begin{equation}
    \hat{\cal S} \hat{c}_{n} \hat{\cal S}^{-1} 
    = \left( -1 \right)^{n} \hat{c}_{n}^{\dag}
\end{equation}
and
\begin{equation}
    \forall\,z \in \mathbb{C}
    \quad
    \hat{\cal S} z \hat{\cal S}^{-1}
    = z^{*}.
\end{equation}
Chiral symmetry imposes a special constraint on the eigenstates with zero energy.
On the other hand, the onsite potential $V_n \hat{c}_{n}^{\dag} \hat{c}_{n}$ in Eq.~(\ref{eq: Anderson model}) breaks chiral symmetry in Eq.~(\ref{eq: CS}).

Chiral symmetry changes the transport properties of disordered electron systems. 
We numerically calculate the linear and nonlinear conductance of the chiral-symmetric model in Eq.~(\ref{eq: Dyson model}), as summarized in Fig.~\ref{fig: disorder}\,(b).
Away from zero energy, the conductance behaves similarly to the standard class in Fig.~\ref{fig: disorder}\,(a) for both linear and nonlinear regimes.
However, the situation changes around zero energy.
As shown in Fig.~\ref{fig: disorder}\,(b1), the linear conductance $G_1$ exhibits a singular peak at zero energy.
The change of $G_1$ is rapid and indifferentiable as a function of energy.
Consequently, the nonlinear conductance, which is given by the derivatives of the linear conductance, exhibits singularly large values at zero energy [Fig.~\ref{fig: disorder}\,(b2, b3)].
This singularly large nonlinear conductance serves as a clear experimental signature of the disorder-induced nonlinear transport that can be distinguished from other contributions.

The singularity of zero modes is a unique feature of disordered electron systems with chiral symmetry, which dates back to the work by Dyson~\cite{Dyson-53}. 
In fact, the transmission probability $\braket{T}$ decreases only algebraically at zero energy (see Appendix~\ref{asec: DMPK - chiral} 
for details):
\begin{equation}
    \braket{T} \left( E = 0 \right) \simeq \sqrt{\frac{2\ell}{\pi L}}
        \label{eq: Anderson - T - chiral}
\end{equation}
with the mean free path
\begin{equation}
    \ell \coloneqq \frac{J^2}{\braket{\left( \Delta J_{n, n+1} \right)^2}}.
\end{equation}
The distinction between Eqs.~(\ref{eq: Anderson - T - standard}) and (\ref{eq: Anderson - T - chiral}) reflects from the different distributions of the transmission probability $T$.
As a consequence of the power-law decay in Eq.~(\ref{eq: Anderson - T - chiral}), the transmission of zero modes decays slowly in comparison with other modes with nonzero energy.
In fact, zero modes never exhibit Anderson localization even in the presence of disorder.
This anomalous delocalization of zero modes is protected by chiral symmetry.
The singularity of zero modes leads to the singularly large nonlinear response.

\subsection{Classification in one dimension}

The behavior of the nonlinear conductance discussed in the preceding subsections is not specific to the models in Eqs.~(\ref{eq: Anderson model}) and (\ref{eq: Dyson model}) but constitutes general features of disordered electron systems.
Anderson localization of disordered electron systems is generally understood by the tenfold internal-symmetry classification based on time-reversal symmetry, particle-hole (charge-conjugation) symmetry, and chiral (sublattice) symmetry (Table~\ref{tab: AZ}; see also Appendix~\ref{asec: symmetry} 
for details)~\cite{AZ-97, Evers-review, Beenakker-review-sc, Ryu-review}.
Using this symmetry classification, we also classify the universal behavior of nonlinear response of disordered electron systems in one dimension.

\begin{table}[t]
	\centering
	\caption{Symmetry classification of disordered electron systems in one dimension. The tenfold internal-symmetry classification is based on time-reversal symmetry (TRS), particle-hole symmetry (PHS), and chiral symmetry (CS). For the entries of TRS and PHS, the signs $\pm 1$ denote the signs of the symmetry operators. For the entries of CS, $0$  and $1$ describe the absence and presence of CS, respectively. We show the five symmetry classes for which the singularity of the linear and nonlinear conductance arises at zero energy (i.e., classes BDI, AIII, CII, DIII, and D). In the three of these five symmetry classes, the singularity arises only for an odd number of channels (i.e., classes BDI, AIII, and CII).}
		\label{tab: AZ}
     \begin{tabular}{cccccc} \hline \hline
    ~Class~ & ~TRS~ & ~PHS~ & ~CS~ & ~Singularity~ & ~Parity~ \\ \hline
    AI & $+1$ & $0$ & $0$ & & \\ 
    A & $0$ & $0$ & $0$ & & \\ 
    AII & $-1$ & $0$ & $0$ &  & \\ \hline
    BDI & $+1$ & $+1$ & $1$ & $\checkmark$ & $\checkmark$ \\ 
    AIII & $0$ & $0$ & $1$ & $\checkmark$ & $\checkmark$ \\ 
    CII & $-1$ & $-1$ & $1$ & $\checkmark$ & $\checkmark$ \\ \hline
    CI & $+1$ & $-1$ & $1$ & & \\ 
    C & $0$ & $-1$ & $0$ & & \\ 
    DIII & $-1$ & $+1$ & $1$ & $\checkmark$ & \\ 
    D & $0$ & $+1$ & $0$ & $\checkmark$ & \\ \hline \hline
    \end{tabular}
\end{table}

In the standard class (i.e., classes A, AI, and AII), nonlinear conductance generally behaves in a manner similar to the model in Eq.~(\ref{eq: Anderson model}).
In fact, the model in Eq.~(\ref{eq: Anderson model}) only respects time-reversal symmetry and belongs to class AI.
Hence, no singular behavior appears in the nonlinear response in the standard class.
In the chiral class (i.e., classes AIII, BDI, and CII), on the other hand, nonlinear conductance generally behaves in a manner similar to the model in Eq.~(\ref{eq: Dyson model}).
Consistently, the model in Eq.~(\ref{eq: Dyson model}) belongs to class BDI.
Thus, electron systems in the chiral class generally exhibit the singularity of nonlinear response at zero energy.
Notably, such singular behavior arises only for an odd number of channels~\cite{Brouwer-98, *Brouwer-00}.

The classification in Table~\ref{tab: AZ} allows us to predict possible other symmetry classes in which strong nonlinear response arises.
In addition to the chiral class, the singular behavior of nonlinear conductance arises in classes DIII and D, which describe spinful and spinless superconducting wires, respectively.
Remarkably, zero modes of superconductors in classes DIII and D obey the statistics of Majorana fermions~\cite{Kitaev-01, Alicea-review, Sato-review}.
Thus, the Majorana zero modes in disordered topological superconductors should exhibit the singular behavior of the nonlinear response.
It is worthwhile to further study this nonlinear transport phenomenon with specific models.

\subsection{Anderson transitions}

As described above, all eigenstates are subject to Anderson localization even for infinitesimal disorder in one dimension, except for zero modes protected by chiral or particle-hole symmetry.
In higher-dimensional systems, by contrast, eigenstates can stay delocalized for small disorder and exhibit the Anderson transitions between delocalization and localization for critical disorder.
Correspondingly, the systems can have mobility edges $E = E_{\rm c}$ in the spectra, across which delocalized eigenstates turn localized.
The linear conductance $G_1$ exhibits the critical behavior around the mobility edge:
\begin{equation}
    G_1 \propto \left| E - E_{\rm c} \right|^{s}.
\end{equation}
The critical exponent $s$ depends solely on symmetry and dimension, characterizing the universality class of the Anderson transitions.
In a number of noninteracting electron systems, the critical behavior is determined solely by one relevant parameter~\cite{Abrahams-79}.
Under the one-parameter scaling, the critical exponent $s$ of the linear conductance $G_1$ is associated with the critical exponent $\nu$ of the localization length $\xi$ by $s = \left( d-2 \right) \nu$~\cite{Wegner-76}.
In three dimensions, for example, the critical exponents are numerically obtained as $\nu \simeq 1.57$ in the presence of time-reversal symmetry (i.e., 3D class AI) and $\nu \simeq 1.44$ in the absence of time-reversal symmetry (i.e., 3D class A)~\cite{Slevin-97, *Slevin-99}.

In contrast to the linear regime, the effect of the Anderson transitions is not well understood in the nonlinear regime. 
From the nonlinear Landauer formula in Eq.~(\ref{eq: nonlinear conductance formula - zero temperature}), we understand the behavior of nonlinear conductance around the Anderson transitions.
In fact, the $n$th-order nonlinear conductance $G_n$ should exhibit the critical behavior
\begin{align}
    G_{n} \propto \left| E - E_{\rm c} \right|^{s-n}
\end{align}
near the mobility edge $E = E_{\rm c}$.
For example, in time-reversal-invariant systems in three dimensions, we have from $s\simeq 1.57$~\cite{Slevin-97, *Slevin-99}
\begin{align}
    G_2 &\propto \frac{dG_1}{dE} \propto \left| E - E_{\rm c} \right|^{0.57}, \\
    G_3 &\propto \frac{d^2G_1}{dE^2} \propto \left| E - E_{\rm c} \right|^{-0.43},
\end{align}
and so on.
Remarkably, while $G_1$ and $G_2$ vanish at the mobility edge $E = E_{\rm c}$, the higher-order nonlinear conductance $G_n$ ($n \geq 3$) diverges at $E = E_{\rm c}$.
This singular behavior is similar to the Dyson singularity of zero modes in one-dimensional disordered systems, as discussed in the preceding subsections.
Meanwhile, away from the mobility edges, the nonlinear response does not grow since the transmission probability changes only gradually.

\section{Nonlinear transport and quantum anomaly in graphene}
    \label{sec: graphene}

\begin{figure*}[t]
\centering
\includegraphics[width=144mm]{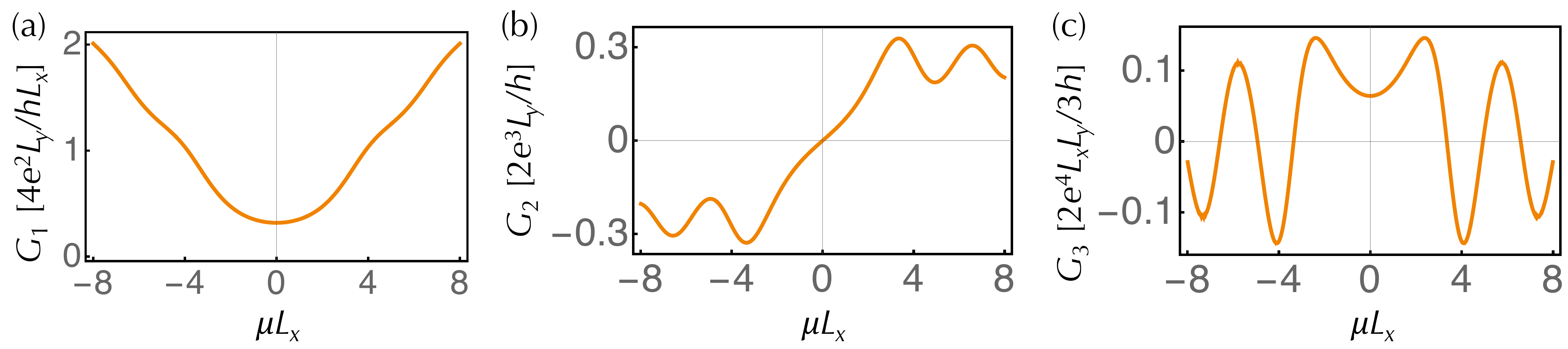} 
\caption{Linear and nonlinear conductance of graphene ($\hbar v = 1$). (a)~The linear conductance $G_1$, (b)~the second-order nonlinear conductance $G_2$, and (c)~the third-order nonlinear conductance $G_3$ as functions of the chemical potential $\mu$. At the Dirac point $\mu = 0$, the conductance exhibits the universal behavior such as $G_1 = ( 4e^2/h)\,( L_y/L_x )\,( 1/\pi )$, $G_2 = 0$, and $G_3 = ( 2e^4/3h )\,( L_x L_y )\,( 0.20 \cdots /\pi)$.}
	\label{fig: graphene}
\end{figure*}

Topological materials are quantum materials that exhibit nontrivial topology in wave functions~\cite{Kane-review, Zhang-review, Ryu-review, Vishwanath-review}.
Among them, graphene, a single layer of carbon atoms on a two-dimensional honeycomb lattice, is a prototypical topological semimetal~\cite{graphene-NG-review, graphene-DasSarma-review}.
It is also a building block of van der Waals heterostructures~\cite{vdW-review}.
The low-energy electronic band structure of graphene is generally described by the continuum Dirac Hamiltonian
\begin{equation}
H \left( \bm{k} \right)
= \hbar v \left( k_x \sigma_x + k_y \sigma_y \right),
    \label{eq: Hamiltonian - graphene}
\end{equation} 
where $v > 0$ is the Fermi velocity, and $\sigma_i$'s ($i = x, y, z$) are Pauli matrices.
The two bands touch at the Dirac point $E=0$ with the relativistic dispersion $E \left (\bm{k} \right) = \pm \hbar v \left| \bm{k}\right|$.
The two-dimensional Dirac Hamiltonian in Eq.~(\ref{eq: Hamiltonian - graphene}) describes a single valley degree of freedom, which appears twice in momentum space with opposite chirality.
This description is valid as long as the intervalley scattering is irrelevant.
Notably, it also describes a two-dimensional surface mode of three-dimensional topological insulators~\cite{Fu-Kane-Mele-07, Moore-Balents-07, Roy-09}, and our discussions below are applicable also to such topological surface modes.

A unique characteristic of graphene appears in transport phenomena. 
In particular, the linear transport at the Dirac point originates from nontrivial topology, or equivalently, quantum anomaly~\cite{Ludwig-94, Fendley-01, Ostrovsky-07, Ryu-07-Z2, Fu-12}, and is robust against disorder~\cite{Ando-98a, *Ando-98b, Bardarson-07, Nomura-Koshino-Ryu-07}.
Since quantum coherence can be maintained because of the high tunability, graphene is among the best platforms in which the Landauer formula plays a major role.
Here, we investigate nonlinear conductance of graphene on the basis of the nonlinear Landauer formula and demonstrate a unique quantum effect in the nonlinear response (Fig.~\ref{fig: graphene}).
These discussions also provide representative calculations of electrical conduction in two-dimensional quantum materials based on the scattering formulation.

\subsection{Linear conductance}

We consider graphene described by Eq.~(\ref{eq: Hamiltonian - graphene}) of lengths $L_x$ and $L_y$ along the $x$ and $y$ directions, respectively.
In the $x$ direction, we attach the two electrodes to the graphene.
In the $y$ direction, on the other hand, no electrodes are attached, and transmitted modes along the $x$ direction are specified by the transverse momentum $k_y$.
The scattering problem of graphene is solvable, as shown in Appendix~\ref{asec: graphene - derivation}~\cite{Katsnelson-06, Tworzydlo-06}.
The transmission probability $T \left( E, k_y \right)$ for the given energy $E$ and transverse momentum $k_y$ is obtained as
\begin{equation}
    T \left( E, k_y \right)
    =  \frac{1}{\cos^2 k_x L_x + \left( E/\hbar v k_x \right)^2 \sin^2 k_x L_x}
\end{equation}
with 
\begin{equation}
    k_x \coloneqq \sqrt{\left( \frac{E}{\hbar v} \right)^2 - k_y^2}.
        \label{eq: graphene - kx}
\end{equation}
Here, $k_y$ is always real-valued, but $k_x$ is imaginary-valued for $\left| E \right| < \hbar v \left|k_y\right|$.
In the limit $L_y \to \infty$, the total transmission probability is given by
\begin{align}
    T \left( E \right)
    = \frac{L_y}{\pi} \int_0^{\infty} T \left( E, k_y \right) dk_y.
\end{align}

According to the Landauer formula, the linear and nonlinear conductance is obtained by the transmission probability $T = T \left( E \right)$.
The linear conductance $G_1$ at zero temperature is given from Eq.~(\ref{eq: nonlinear conductance formula - zero temperature}) by
\begin{equation}
    G_1 \left( \mu \right)
    = \frac{4e^2}{h} T \left( \mu \right),
\end{equation}
where we include the multiplication by four to take into account the valley and spin degrees of freedom.

From these formulas, the linear conductance $G_1$ is obtained, as shown in Fig.~\ref{fig: graphene}\,(a).
Away from the Dirac point $\mu = 0$, $G_1$ grows linearly with the chemical potential $\mu$ besides the small oscillations.
Near the Dirac point $\mu = 0$, by contrast, $G_1$ is no longer linear in $\mu$; it takes a nonzero minimum value at the Dirac point.
This behavior qualitatively agrees with the experimental results~\cite{Novoselov-04, *Novoselov-05, Zhang-05}.
While it seems unfeasible to explicitly obtain $G_1$ for the arbitrary chemical potential $\mu$, we analytically obtain the asymptotic behavior (see Appendix~\ref{asec: graphene - asymptotics} 
for detailed derivations).
At the Dirac point $\mu = 0$, we have
\begin{align}
    T \left( E = 0 \right)
    = \frac{L_y}{\pi} \int_0^{\infty} \frac{dk_y}{\cosh^2 k_y L_x}
    = \frac{L_y}{\pi L_x},
\end{align}
which gives~\cite{Katsnelson-06, Tworzydlo-06}
\begin{equation}
    G_1 \left( \mu = 0 \right)
    = \frac{4e^2}{h} \frac{L_y}{L_x} \frac{1}{\pi}.
        \label{eq: graphene - G1 - zero}
\end{equation}
Away from the Dirac point (i.e., $\left| \mu \right| \to \infty$), the linear conductance $G_1$ behaves as
\begin{equation}
    G_1 \left( \mu \right)
    \simeq \frac{4e^2}{h} \frac{L_y}{L_x} \frac{a}{\pi} \frac{\left| \mu \right| L_x}{\hbar v}
    \quad \left( \left| \mu  \right| \to \infty \right),
        \label{eq: graphene - G1 - largemu}
\end{equation}
where the numerical coefficient is
\begin{equation}
    a \coloneqq 2-\sqrt{2}\,\mathrm{arccoth}\,\sqrt{2}
    = 0.753550 \cdots.
        \label{eq: graphene - a}
\end{equation}
These asymptotic results are compatible with the numerical results in Fig.~\ref{fig: graphene}\,(a).

If Eq.~(\ref{eq: graphene - G1 - largemu}) were valid even at the Dirac point $\mu = 0$, the linear conductance $G_1$ would vanish.
This behavior is consistent with the density of states, which gets larger for the larger chemical potential and vanishes at zero chemical potential.
It is also similar to the semiclassical conductance derived with the Boltzmann equation~\cite{Nomura-06, *Nomura-07, Ando-06}.
Hence, the vanishing linear conductance $G_1$ at the Dirac point, including Eq.~(\ref{eq: graphene - G1 - largemu}), is a semiclassical result of electronic transport; 
nonvanishing $G_1$ even at the Dirac point, including Eq.~(\ref{eq: graphene - G1 - zero}), signals a genuinely quantum effect.
In fact, at the Dirac point $\mu = 0$, the wave number $k_x = \ii k_y$ along the conducting direction is pure imaginary, which means that $G_1$ in Eq.~(\ref{eq: graphene - G1 - zero}) originates solely from quantum tunneling (see also Appendix~\ref{asec: graphene - derivation} 
for details).
Furthermore, the nonzero linear conductance in Eq.~(\ref{eq: graphene - G1 - zero}) is a direct consequence of nontrivial topology of wave functions~\cite{Ludwig-94, Fendley-01, Ostrovsky-07, Ryu-07-Z2, Fu-12}.
From the field-theoretical perspective, this is equivalent to the quantum anomaly of the underlying quantum field theory.
Thus, the linear transport at the Dirac point is a direct experimental signature of the nontrivial quantum effect in graphene.
The Landauer formula enables its theoretical characterization.

\subsection{Second-order nonlinear conductance}

The second-order nonlinear conductance $G_2$ is obtained by the derivative of the transmission probability $T = T \left( E \right)$ as
\begin{equation}
    G_2 \left( \mu \right)
    = \frac{2e^3}{h} \left. \frac{dT}{dE} \right|_{E = \mu}.
\end{equation}
Figure~\ref{fig: graphene}\,(b) shows $G_2$ as a function of the chemical potential $\mu$.
Away from the Dirac point, the second-order nonlinear conductance $G_2$ is nearly constant besides the small oscillations.
As shown in Appendix~\ref{asec: graphene - asymptotics}, 
this behavior is asymptotically obtained as
\begin{align}
    G_2 \left( \mu \right)
    \simeq \frac{2e^3}{h} L_y \frac{a}{\pi} \frac{\mathrm{sgn} \left( \mu \right)}{\hbar v}
    \quad \left( \left| \mu \right| \to \infty \right)
\end{align}
with the numerical constant $a$ in Eq.~(\ref{eq: graphene - a}).
This is the semiclassical result that can be obtained also from the Boltzmann equation.
Near the Dirac point, by contrast, such a semiclassical result is no longer valid.
In fact, $G_2$ decreases and behaves as
\begin{equation}
    G_2 \left( \mu \right)
    \simeq \frac{2e^3}{h} L_y \frac{c}{\pi} \frac{\mu L_x}{\left( \hbar v \right)^2}
    \quad \left( \mu \simeq 0 \right)
\end{equation}
with the numerical constant
\begin{align}
    c 
    &\coloneqq - \int_0^{\infty} \frac{1 + 2\tanh^2 x - \cosh^2 x}{x^2 \cosh^2 x} dx \nonumber \\
    &= 0.201876 \cdots.
        \label{eq: graphene - c}
\end{align}
The vanishing behavior of the second-order nonlinear conductance $G_2$ at the Dirac point is a direct signature of nontrivial topology, which is experimentally observable.
Because of the nonperturbative nature of topology, this behavior is expected to be immune to disorder.
It is also notable that nonzero $G_2 \neq 0$ away from the Dirac point implies the nonreciprocal response, as discussed in Sec.~\ref{sec: nonreciprocity}.
The reciprocal response at the Dirac point, accompanied by $G_2$ = 0, is a consequence of the nontrivial topology.

\subsection{Third-order nonlinear conductance}

The third-order nonlinear conductance $G_3$ is obtained as
\begin{equation}
    G_3 \left( \mu \right)
    = \frac{2e^3}{3h} \left. \frac{d^2T}{dE^2} \right|_{E = \mu}.
\end{equation}
In the semiclassical regime away from the Dirac point, $G_3$ oscillates around zero as a function of the chemical potential $\mu$, as shown in Fig.~\ref{fig: graphene}\,(c).
The oscillation of $G_3$ is more pronounced than $G_1$ and $G_2$.
Around the Dirac point $\mu = 0$, the third-order nonlinear conductance $G_3$ exhibits different behavior again because of nontrivial topology.
It takes a nonzero value,
\begin{equation}
    G_3 \left( \mu = 0 \right)
    = \frac{2e^4}{3h} L_x L_y \frac{c}{\pi} \left( \hbar v \right)^{-2},
\end{equation}
where the numerical constant $c$ is given as Eq.~(\ref{eq: graphene - c}) (see Appendix~\ref{asec: graphene - asymptotics} 
for derivations).
This nonzero value is also a direct experimental signature of the nontrivial topology of graphene.

In summary, the linear and nonlinear electronic transport of graphene is obtained as
\begin{align}
    I \simeq \frac{4e^2}{h} \frac{L_y}{L_x} \frac{1}{\pi} \left[ V + \frac{c}{6} \left( \frac{e L_x}{\hbar v} \right)^2 V^3 \right]~\left( \mu = 0 \right)
\end{align}
at the Dirac point and
\begin{align}
    I \simeq \frac{4e^2}{h} \frac{L_y}{L_x} \frac{a}{\pi} \left[ \left| \mu \right| V + \frac{\mathrm{sgn} \left( \mu \right)}{2} eL_x V^2 \right]~\left( \left| \mu \right| \to \infty \right)
\end{align}
away from the Dirac point.
It is worthwhile to take trigonal warping terms~\cite{Ando-98a, *Ando-98b, Koshino-09} into consideration.
It also merits further study to revisit our results in terms of the Kubo formula and its nonlinear extensions.

\section{Nonlinear 
Hall effect}
    \label{sec: QHE}

The quantum Hall effect is a prototypical topological phenomenon~\cite{Klitzing-80, Laughlin-81, Halperin-82, TKNN-82, Buttiker-88-QHE, Haldane-88, QHE-review}.
While it was extensively studied in the linear regime, the nonlinear 
Hall effect has attracted growing interest in recent years~\cite{Sodemann-15, Ma-19, Kang-19, He-19, Du-19PRL, *Du-19, *Du-21, Nandy-19, Xiao-19, Singh-20, Zhang-20, Resta-21}.
For an applied voltage $V$ and the concomitant Hall current $I^{\rm H}$, the linear and nonlinear Hall conductance is defined by
\begin{align}
I^{\rm H} = G^{\rm H}_1 V + G^{\rm H}_2 V^2 + \cdots,
\end{align}
where $G^{\rm H}_1$ is the linear Hall conductance, and $G^{\rm H}_2$ is the second-order nonlinear Hall conductance.
The second-order nonlinear Hall conductance exhibits unique transport properties that have no analogs in the linear Hall conductance.
For example, in the presence of time-reversal symmetry, $G^{\rm H}_1$ vanishes, but $G^{\rm H}_2$ survives.
Moreover, $G^{\rm H}_2$ contains new information about electron systems, such as the Berry curvature dipole.
The second-order nonlinear 
Hall effect was experimentally observed in WTe$_{\text 2}$~\cite{Ma-19, Kang-19}.

The nonlinear 
Hall effect was studied on the basis of the Boltzmann equation with the relaxation time approximation (see Appendix~\ref{asec: Boltzmann - QHE} 
for details)~\cite{Sodemann-15}.
Although this approach is likely to be valid at high temperature, it cannot capture the genuinely quantum nature of the nonlinear Hall effect. 
Here, we develop a scattering theory of the nonlinear 
Hall effect and fully capture its quantum nature. 
In particular, we demonstrate that the second-order nonlinear 
Hall conductance is related to shot noise (see also Sec.~\ref{sec: noise} for details about shot noise).
We also discuss the effect of disorder and Anderson localization on the nonlinear 
Hall effect, including the significance of symmetry.

\subsection{Scattering formulation of the nonlinear 
Hall effect}
    \label{sec: QHE - bulk}

We consider a generic electron system in two dimensions of lengths $L_x$ and $L_y$ along the $x$ and $y$ directions, respectively.
The system is characterized by the energy dispersion $E = E \left( \bm{k} \right)$ and the Berry curvature $\Omega = \Omega \left( \bm{k} \right)$.
We attach the two reservoirs to the system and apply a bias voltage $V$ along the $x$ direction.
The two reservoirs at the left and right are described by the Fermi-Dirac distributions $f_{\rm eq} \left( E - eV \right)$ and $f_{\rm eq} \left( E \right)$, respectively.
On the other hand, the system is subject to the periodic boundary conditions along the $y$ direction.
Our discussions can also be straightforwardly generalized to three-dimensional systems.

The applied bias voltage $V$ leads to the diagonal current $I$ along the $x$ direction and the Hall current $I^{\rm H}$ along the $y$ direction.
These currents arise from the velocity
\begin{equation}
    \bm{v} \left( \bm{k} \right)
    = \frac{1}{\hbar} \frac{\partial E \left( \bm{k} \right)}{\partial \bm{k}}
    - \frac{e}{\hbar} \left( \bm{\mathcal{E}} \times \bm{n} \right) \Omega \left( \bm{k} \right).
        \label{eq: anomalous velocity}
\end{equation}
Here, $\bm{\mathcal{E}}$ is an electric field applied to the system.
In two dimensions, $\bm{n}$ is the unit vector along the $z$ direction, and we have $\bm{\mathcal{E}} \times \bm{n} = \left( \mathcal{E}_y, - \mathcal{E}_x \right)$.
As discussed in Sec.~\ref{sec: Landauer - derivation}, the diagonal current $I$ arises from the first term and is given as the nonlinear Landauer formula in Eqs.~(\ref{eq: nonlinear conductance formula}) and (\ref{eq: nonlinear conductance formula - zero temperature}).
By contrast, the Hall current $I^{\rm H}$ arises from the second term, the anomalous velocity~\cite{Nagaosa-review, Xiao-review, Vanderbilt-textbook}.
The anomalous velocity originates from the nontrivial topology of wave functions [i.e., Berry curvature $\Omega \left( \bm{k} \right)$] and plays a key role in topological transport phenomena.
While the first term in Eq.~(\ref{eq: anomalous velocity}) can also contribute to electrical conduction in the $y$ direction, it should vanish in the presence of time-reversal symmetry or space-inversion symmetry.
Hence, we focus on the anomalous velocity in the following.

To obtain the anomalous velocity in Eq.~(\ref{eq: anomalous velocity}), we need to evaluate the electric field $\mathcal{E}$ inside the system.
Naively, we may consider 
\begin{equation}
    \mathcal{E} = \frac{V}{L_x}.
        \label{eq: E-V}
\end{equation}
However, Eq.~(\ref{eq: E-V}) is valid only in the semiclassical regime, in which inelastic scattering destroys quantum coherence of electronic transport and leads to Ohm's law.
In the presence of quantum coherence, by contrast, the semiclassical result in Eq.~(\ref{eq: E-V}) is no longer valid.

In general, the total voltage drop $V$ is due to the voltage drop through the system and the voltage drop in the electrodes (or equivalently, the voltage drop at the contacts between the system and the reservoirs).
In the aforementioned semiclassical regime, the voltage drop through the system dominates the total voltage drop.
In the quantum regime, the voltage drop in the electrodes is comparable with that across the system as a consequence of quantum coherence.
In particular, when the perfect transmission $T=1$ is realized in a clean system, no voltage drop should arise in the system, and the voltage drop should arise only in the electrodes.
This is a characteristic of coherent transport of electrons in the quantum regime.
For a generic case, the dominant contribution of the voltage drop through the system is given as $\left( 1-T \right) V$ with the transmission probability $T$~\cite{Datta-textbook, Imry-textbook}.
In actual experiments, the voltage drop $\left( 1 - T \right) V$ through the system is measured in the four-terminal setup, while the total voltage drop $V$ is measured in the two-terminal setup.
Consequently, the electric field $\mathcal{E}$ in the system is given as
\begin{equation}
    \mathcal{E} = \left( 1 - T \right) \frac{V}{L_x} + \mathcal{O} \left( V^2 \right).
        \label{eq: E-V - v2}
\end{equation}
For the perfect transmission $T=1$, no electric field appears in the system (i.e., $\mathcal{E} = 0$), and the voltage drop arises only in the electrodes, which is consistent with the above discussions.
On the other hand, for the low transmission probability $T \ll 1$, the electric field $\mathcal{E}$ approaches the semiclassical result in Eq.~(\ref{eq: E-V}).

Now, we take the anomalous velocity into consideration and calculate the nonlinear Hall conductance in the scattering theory.
Let us first focus on the transmitted modes with the wave numbers in $\left[ k_x, k_x + dk_x \right] \times \left[ k_y, k_y + dk_y \right]$.
Then, similarly to Sec.~\ref{sec: Landauer - derivation}, the numbers $dN_{\rm L\to R}$ and $dN_{\rm R\to L}$ of the electronic waves from the left to the right and from the right to the left are respectively given as  
\begin{align}
    dN_{\rm L\to R} &= T \left( \bm{k} \right) f_{\rm eq} \left( E \left( \bm{k} \right) - eV \right) \frac{L_y d^2k}{\left( 2\pi\right)^2}, \\
    dN_{\rm R\to L} &= T \left( \bm{k} \right) f_{\rm eq} \left( E \left( \bm{k} \right) \right) \frac{L_y d^2k}{\left( 2\pi\right)^2}.
\end{align}
Here, $T = T \left( \bm{k} \right)$ is the transmission probability as a function of the wave numbers $\bm{k}$ and calculated from the given Hamiltonian $H = H \left( \bm{k} \right)$ in a manner similar to graphene (see Sec.~\ref{sec: graphene} for details).
Because of the anomalous velocity in Eq.~(\ref{eq: anomalous velocity}), these transmitted modes contribute to the Hall current $dI^{\rm H}$ along the $y$ direction
\begin{align}
    dI^{\rm H}
    &= e \left( \frac{e}{\hbar} \mathcal{E} \Omega \left( \bm{k} \right) \right) \left( dN_{\rm L\to R} - dN_{\rm R \to L} \right) \nonumber \\
    &\simeq \frac{e^2 V}{\hbar} \frac{L_y}{L_x} T \left( \bm{k} \right) \left( 1 - T \left( \bm{k} \right) \right) \Omega \left( \bm{k} \right) \nonumber \\
    &\quad \times \left( f_{\rm eq} \left( E \left( \bm{k} \right) -eV \right) - f_{\rm eq} \left( E \left( \bm{k} \right) \right) \right) \frac{d^2k}{\left( 2\pi\right)^2},
\end{align}
where Eq.(\ref{eq: E-V - v2}) is used. Using the expansion
\begin{align}
    &f_{\rm eq} \left( E \left( \bm{k} \right) -eV \right) - f_{\rm eq} \left( E \left( \bm{k} \right) \right) \nonumber \\
    &\qquad= - eV \frac{\partial f_{\rm eq} \left( E \left( \bm{k} \right) \right)}{\partial E\left( \bm{k} \right)}
    + \mathcal{O} \left( V^2 \right),
\end{align}
we have
\begin{align}
    I^{\rm H}
    &= \int dI^{\rm H}
    \simeq \frac{e^3 V^2}{\hbar} \frac{L_y}{L_x} \oint_{\rm BZ} T \left( \bm{k} \right) \left( 1 - T \left( \bm{k} \right) \right) \Omega \left( \bm{k} \right) \nonumber \\
    &\qquad\qquad\qquad \times \left( - \frac{\partial f_{\rm eq} \left( E \left( \bm{k} \right) \right)}{\partial E\left( \bm{k} \right)} \right) \frac{d^2k}{\left( 2\pi\right)^2}.
\end{align}
The momentum integral is taken over the whole Brillouin zone.
From this equation, we see that the linear Hall conductance vanishes.
The leading-order contribution is quadratic for the applied voltage $V$, which gives the scattering formula of the second-order nonlinear Hall conductance
\begin{align}
    G_{2}^{\rm H}
    &= \frac{e^3}{\hbar} \frac{L_y}{L_x} \oint_{\rm BZ} T \left( \bm{k} \right) \left( 1 - T \left( \bm{k} \right) \right) \Omega \left( \bm{k} \right) \nonumber \\
    &\qquad\qquad\qquad \times \left( - \frac{\partial f_{\rm eq} \left( E \left( \bm{k} \right) \right)}{\partial E\left( \bm{k} \right)} \right) \frac{d^2k}{\left( 2\pi\right)^2}.
        \label{eq: GH2}
\end{align}
At zero temperature, we have $-\partial f_{\rm eq}/\partial E = \delta \left( E - \mu \right)$, and the formula further reduces to
\begin{align}
    G_{2}^{\rm H}
    &= \frac{e^3}{\hbar} \frac{L_y}{L_x} \oint_{\rm BZ} T \left( \bm{k} \right) \left( 1 - T \left( \bm{k} \right) \right) \Omega \left( \bm{k} \right) \nonumber \\
    &\qquad\qquad\qquad\qquad\quad \times \delta \left( E \left( \bm{k} \right) - \mu \right) \frac{d^2k}{\left( 2\pi\right)^2}.
        \label{eq: GH2 - zero temperature}
\end{align}
The momentum integral is taken only at the Fermi surface $E \left( \bm{k} \right) = \mu$, which implies the Fermi-liquid nature of the second-order nonlinear Hall conductance~\cite{Haldane-04}.
Equations~(\ref{eq: GH2}) and (\ref{eq: GH2 - zero temperature}) are the general formulas of the second-order nonlinear 
Hall conductance in the scattering theory.

As shown in Eqs.~(\ref{eq: GH2}) and (\ref{eq: GH2 - zero temperature}), the second-order nonlinear Hall conductance $G_2^{\rm H}$ is obtained as the integral of $T \left( 1 - T \right)$ multiplied by the Berry curvature $\Omega$ on the Fermi surface $E \left( \bm{k} \right) = \mu$.
As discussed above, $T$ in the formula is due to the transmission into the system along the $x$ direction, and $1-T$ is due to the anomalous velocity along the $y$ direction.
This formula is relevant to coherent electronic transport in the quantum regime.
In such a quantum regime, the Boltzmann equation, which does not take quantum coherence into consideration, is no longer valid, and the semiclassical theory is not applicable.
From Eqs.~(\ref{eq: GH2}) and (\ref{eq: GH2 - zero temperature}), $G_2^{\rm H}$ gets small for the low transmission probability $T \ll 1$.
It gets small also for large $T$ and vanishes for the perfect transmission $T = 1$.
These results are unique to quantum electronic transport.

\subsection{Nonlinear 
Hall response and nonequilibrium quantum fluctuations}

The product $T \left( 1 - T \right)$ reminds us of shot noise, which is discussed in Sec.~\ref{sec: noise}.
In fact, in two dimensions, the dominant contribution of shot noise $S$ at zero temperature reads $S \simeq \sigma_0 V$ with
\begin{align}
    \sigma_0 = \frac{2e^3}{h} L_y \int_{0}^{2\pi} T \left( \mu, k_y \right) \left( 1 - T \left( \mu, k_y \right) \right) \frac{dk_y}{2\pi},
        \label{eq: QHE - shot noise}
\end{align}
where $T \left( \mu, k_y \right)$ is the transmission probability as a function of the chemical potential $\mu$ and the wave number $k_y$ along the $y$ direction.
Since transport properties are determined solely by the transmission probability $T$, the similarity between Eqs.~(\ref{eq: GH2 - zero temperature}) and (\ref{eq: QHE - shot noise}) implies an intimate relationship between the second-order nonlinear Hall conductance $G_2^{\rm H}$ and the shot noise power $\sigma_0$.

Suppose that the transmission probability depends solely on the Fermi energy: $T = T \left( \mu \right)$.
This simplification is similar to the approximation that neglects the energy dependence of $T$ for the conventional Landauer formula in Eq.~(\ref{eq: Landauer - textbook})~\cite{Datta-textbook, Imry-textbook}, which are likely to capture the dominant contribution of transport properties.
Then, the second-order nonlinear conductance in Eq.~(\ref{eq: GH2}) reduces to 
\begin{align}
    G_{2}^{\rm H}
    &\simeq \frac{e^3}{\hbar} \frac{L_y}{L_x} T \left( 1 - T \right) C,
        \label{eq: GH2 - 86}
\end{align}
where the topological term $C$ is the integral of the Berry curvature on the Fermi surface (or equivalently, the integral of the Berry curvature dipole below the Fermi energy):
\begin{align}
    C \coloneqq \oint_{\rm BZ} \Omega \left( \bm{k} \right) \delta \left( E \left( \bm{k} \right) - \mu \right) \frac{d^2k}{\left( 2\pi\right)^2}.
\end{align}
Meanwhile, the shot noise power in Eq.~(\ref{eq: QHE - shot noise}) reduces to
\begin{align}
    \sigma_0 \simeq \frac{2e^3}{h} L_y T \left( 1- T \right).
        \label{eq: sigma0 - 88}
\end{align}
Comparing Eqs.~(\ref{eq: GH2 - 86}) and (\ref{eq: sigma0 - 88}), we find
\begin{align}
    G_{2}^{\rm H} = \frac{\pi C}{L_x} \sigma_0.
\end{align}
The proportional coefficient depends solely on the topological property $C$ of the system, not on the transport property $T$.
This fact suggests a universal underlying mechanism between $G_2^{\rm H}$ and $\sigma_0$.

The thermal noise at equilibrium is related to the linear diagonal conductance via the fluctuation-dissipation theorem~\cite{Kubo-textbook}.
Our results establish a new connection between nonequilibrium quantum fluctuations and nonlinear transport.
This relationship is crucially different from the conventional fluctuation-dissipation theorem.
In fact, the nonlinear Hall conductance does not accompany energy dissipation in contrast to the linear diagonal conductance.
Moreover, as also discussed in Sec.~\ref{sec: noise}, shot noise originates from the quantum nature far from thermal equilibrium in contrast to the thermal noise at equilibrium.
Such nonequilibrium quantum noise cannot be discussed in the linear response theory, which can only describe electrical conduction near thermal equilibrium.
Our new fundamental relationship between the nonequilibrium quantum fluctuations and the nonlinear quantum  Hall response is due to the scattering formulation.

\subsection{Edge transport}
    \label{sec: chiral edge}
    
As discussed in the previous subsection~\ref{sec: QHE - bulk}, the bulk of two-dimensional materials exhibits the nonlinear Hall response at the leading order, and the linear Hall response vanishes.
At first sight, this fact seems to be incompatible with the quantized linear Hall conductance in the insulating phase~\cite{Klitzing-80}.
This apparent inconsistency is resolved if we consider the electric conduction at the boundary of the system~\cite{Halperin-82, Buttiker-88-QHE}.
If the wave function of the gapped bulk exhibits nontrivial topology (i.e., Chern number), the chiral edge modes appear at the boundary (i.e., bulk-boundary correspondence)~\cite{Kane-review, Zhang-review, Ryu-review}.
The number of these chiral edge modes coincides with the Chern number $C_1 \in \mathbb{Z}$ of the bulk wave function.
When the system is attached to the reservoirs along the $x$ direction and subject to the periodic boundary conditions along the $y$ direction in a manner similar to Sec.~\ref{sec: QHE - bulk}, pairs of the chiral edge modes are localized at the boundary between the system and the reservoirs and move unidirectionally along the $y$ direction, giving rise to the Hall response.
Their energy dispersion reads
\begin{equation}
    E \left( k \right) = \pm \hbar v k
        \label{eq: QHE - edge - dispersion}
\end{equation}
with the Fermi velocity $v > 0$.
Importantly, backscattering is forbidden for the chiral edge modes.
The absence of backscattering is a unique feature due to chiral anomaly~\cite{Adler-69, Bell-69, Peskin-textbook}, which contrasts with normal electron systems on lattices such as Eqs.~(\ref{eq: Anderson model}) and (\ref{eq: Dyson model}).
As a result, the chiral edge modes realize the perfect transmission
\begin{equation}
    T \left( E \right) = 1
\end{equation}
for arbitrary energy $E$.
Since $T$ does not depend on $E$, the nonlinear conductance vanishes even for an arbitrary external voltage $V$. 
In fact, from the nonlinear Landauer formula in Eq.~(\ref{eq: nonlinear conductance formula - zero temperature}), the current $I$ due to the chiral edge mode is given by
\begin{align}
    I = C_1 \frac{e^2}{h} V,
        \label{eq: QHE - quantization}
\end{align}
which shows the quantized linear conductance $G_1 = C_1 e^2/h$.
In Appendix~\ref{aeq: Boltzmann - chiral}, 
we also obtain the semiclassical conductivity of the chiral edge modes on the basis of the Boltzmann equation.
Also in such a semiclassical regime, the chiral edge modes only exhibit the linear conductivity.

The vanishing nonlinear Hall response of the chiral edge modes underlies the experimental observation of the robust quantized Hall conductance.
If the chiral edge modes gave rise to the nonlinear Hall response, the quantization of the Hall response in Eq.~(\ref{eq: QHE - quantization}) would be fragile against the nonlinear contribution.
Furthermore, it is immune to disorder and lattice deformation.
If the lattice spacing $a$ is introduced, the energy dispersion of the chiral edge modes reads
\begin{align}
    E \left( k \right) = \pm \frac{\hbar v}{a} \sin ka
        \label{eq: QHE - edge - dispersion (lattice)}
\end{align}
instead of Eq.~(\ref{eq: QHE - edge - dispersion}).
For $a \to 0$, Eq.~(\ref{eq: QHE - edge - dispersion (lattice)}) reduces to Eq.~(\ref{eq: QHE - edge - dispersion}).
Even in such a case, backscattering is forbidden, and the Hall conductance is quantized as in Eq.~(\ref{eq: QHE - quantization}).
The robust quantization of the linear Hall conductance originates from chiral anomaly~\cite{Adler-69, Bell-69, Peskin-textbook}.

Therefore, the quantized linear Hall response is due to transport at the boundary.
In such an insulating phase, the nonlinear 
Hall response discussed in the preceding subsection~\ref{sec: QHE - bulk} is exponentially suppressed because of the very low transmission probability $T \ll 1$ due to the bulk energy gap.
The nonlinear 
Hall response in the metallic phase arises only at the bulk, which contrasts with the linear quantum Hall response at the boundary.
These theoretical results can be confirmed experimentally by local measurement of the linear and nonlinear response at the bulk and boundary.
It is yet another advantage of the Landauer formula to distinguish the edge response from the bulk response and to address actual experimental situations.
The above results show the rich interplay of the linear and nonlinear response with topology.

\subsection{Effect of disorder}
    \label{sec: QHE - disorder}

In Refs.~\cite{Du-19, Nandy-19, Xiao-19}, the second-order nonlinear Hall conductance of disordered electron systems was calculated on the basis of the Boltzmann equation.
There, the nonlinear Hall conductance never exhibits the exponential suppression even for arbitrary disorder.
However, quantum coherence should lead to Anderson localization and suppress electronic transport for sufficiently low temperature and sufficiently strong disorder.

The nonlinear Landauer formula enables us to consider the effect of Anderson localization on the nonlinear 
Hall effect.
We understand its behavior using the scattering formula in Eqs.~(\ref{eq: GH2}) and (\ref{eq: GH2 - zero temperature}).
In particular, when the system is subject to Anderson localization because of sufficiently strong disorder, the second-order nonlinear Hall conductance $G_{2}^{\rm H}$ in Eqs.~(\ref{eq: GH2}) and (\ref{eq: GH2 - zero temperature}) is approximated by $T \left( 1-T \right) \simeq T$.
Thus, $G_{2}^{\rm H}$ should behave similarly to the linear diagonal conductance $G_1 \propto T$.
From this fact, we understand the qualitative behavior of the nonlinear 
Hall effect in disordered electron systems.

In generic disordered electron systems in two dimensions, even infinitesimal disorder drives the systems into Anderson localization and leads to the exponential suppression of the linear diagonal conductance $G_1$ ~\cite{Abrahams-79, Gorkov-79, Altshuler-80}.
In such a generic situation, the second-order nonlinear Hall conductance $G_{2}^{\rm H}$ should decay exponentially with respect to the system size even for infinitesimal disorder.
This behavior contrasts with the robust quantization of the linear Hall conductance in the insulating phase (see the previous subsection~\ref{sec: chiral edge} for details).

Even in two-dimensional electron systems, additional symmetry can enable delocalization and Anderson transitions.
In particular, in time-reversal-invariant materials with spin-orbit coupling, the systems respect time-reversal symmetry $T^{-1} H^{*} T = H$ with a unitary matrix $T$ satisfying $T^{*}T = -1$ and belong to the symplectic class (class AII; see Appendix~\ref{asec: symmetry} 
for details about the symmetry classification).
In the symplectic class, systems exhibit antilocalization and Anderson transitions even in two dimensions~\cite{Hikami-80}, which is relevant to the $\mathbb{Z}_2$ topological insulators featuring the quantum spin Hall effect~\cite{Kane-Mele-05-QSH, *Kane-Mele-05-Z2, BHZ-06, Konig-07}.
This also implies possible unique behavior of the nonlinear quantum spin Hall effect, which deserves further research.
Moreover, chiral symmetry gives rise to singular transport phenomena in disordered electron systems~\cite{Gade-91, *Gade-93} (see also Sec.~\ref{sec: Anderson} for details about the one-dimensional case).
Consequently, the second-order nonlinear Hall conductance should also exhibit singular behavior in the presence of chiral symmetry.

As well as symmetry, topology changes the universality classes of Anderson localization.
For example, as discussed in Sec.~\ref{sec: graphene}, nontrivial topology at the Dirac point leads to the robust delocalization of graphene for arbitrary disorder~\cite{Ando-98a, *Ando-98b, Bardarson-07, Nomura-Koshino-Ryu-07}.
Similarly, topology at the Fermi surface can give rise to unconventional transport phenomena for the nonlinear 
Hall effect.
The Berry curvature that appears in the momentum integral of Eqs.~(\ref{eq: GH2}) and (\ref{eq: GH2 - zero temperature}) should play a significant role in such topological phenomena.

It should be noted that the quantum effect of disorder may be captured also by the Kubo formula and its nonlinear extensions.
In the linear regime, the weak-localization correction was quantified by perturbative calculations with diagrammatic combinatorics~\cite{Gorkov-79, Altshuler-80, Hikami-80}.
However, such perturbative calculations work only for weak disorder and cannot characterize Anderson transitions correctly.
No research has ever succeeded in systematically characterizing nonlinear response of disordered electron systems exhibiting Anderson localization in the perturbative approach.
While a recent work~\cite{Du-21} discussed the nonlinear 
Hall effect perturbatively, it only investigated diagrams at the lowest order and hence failed to capture Anderson localization.
To correctly capture Anderson localization, we need to consider an infinite number of relevant diagrams that yield nontrivial contributions to self energy in a manner similar to the linear regime~\cite{Gorkov-79, Altshuler-80, Hikami-80}.
The Landauer formula is free from these conceptual and technical difficulties and enables simple characterization of linear and nonlinear response because of the scattering formulation.

\section{Discussions}
    \label{sec: discussions}

Electronic transport lies at the heart of condensed matter physics.
However, it has been studied mainly in the linear regime, and a unified understanding about nonlinear electronic transport has yet to be established.
We have developed a general nonlinear response theory in the scattering approach.
We have derived the nonlinear Landauer formula, which enables us to obtain the nonlinear conductance by the transmission probability.
The aim of this work is to understand nonlinear response of disordered and topological materials on the basis of the nonlinear Landauer formula.
For disordered electron systems, we have found universal behavior of the nonlinear conductance and generally classified the universality classes on the basis of symmetry.
As a prime example of topological materials, we have studied the nonlinear conductance of graphene and shown their unique behavior due to the quantum anomaly.
Furthermore, we have developed a scattering theory of the nonlinear 
Hall effect and discovered a universal relationship between the nonlinear 
Hall response and the nonequilibrium quantum fluctuations.

Before closing, we discuss several outlooks.
We have classified the nonlinear response of disordered electron systems in one dimension, as shown in Sec.~\ref{sec: Anderson}.
It is worthwhile to further investigate nonlinear conductance of higher-dimensional disordered electron systems that exhibit the Anderson transitions.
Moreover, although the one-parameter scaling~\cite{Abrahams-79} is usually valid in noninteracting electron systems, it can be violated in the presence of additional topology~\cite{Khmel'nitskii-83, Levine-83, *Pruisken-84, *Pruisken-88, Altland-14, *Altland-15}.
Thus, nonlinear response of disordered topological materials is also worth further research. 
In particular, while we have studied nonlinear transport of clean graphene in Sec.~\ref{sec: graphene}, it is of interest to investigate the effect of disorder on the nonlinear transport of graphene.
Away from the Dirac point, disorder induces Anderson localization and strongly suppresses electrical conduction.
At the Dirac point, by contrast, the linear conductance is immune to disorder~\cite{Ando-98a, *Ando-98b, Bardarson-07, Nomura-Koshino-Ryu-07} as long as it does not mix the valleys.
This anomalous delocalization is a consequence of the nonperturbative nature of topology or quantum anomaly~\cite{Ludwig-94, Fendley-01, Ostrovsky-07, Ryu-07-Z2, Fu-12}.
Thus, the nonlinear response is expected to exhibit singular behavior at the Dirac point of disordered graphene.
It is also significant to study the influence of symmetry and disorder on the nonlinear 
Hall effect, as discussed in Sec.~\ref{sec: QHE}.

Nonreciprocity is a hallmark of the nonlinear response, as discussed in Sec.~\ref{sec: nonreciprocity}.
Notably, nonreciprocity can arise even if the transmission probabilities are reciprocal [i.e., Eq.~(\ref{eq: transmission - unitarity})].
Equation~(\ref{eq: transmission - unitarity}) is ensured solely by unitarity of scattering matrices due to current conservation (see Appendix~\ref{asec: scattering} 
for details).
While time-reversal symmetry is relevant to the quantum phases of the transmission amplitudes, it is not necessary for Eq.~(\ref{eq: transmission - unitarity}).
In the presence of dissipation, by contrast, the scattering processes are no longer unitary.
The coupling to the external environment results in, for example, the violation of charge conservation or the destruction of quantum coherence.
In such a case, Eq.~(\ref{eq: transmission - unitarity}) is violated, and nonreciprocal response can arise because of the nonreciprocal transmission probabilities: $T_{\rm L\to R} \neq T_{\rm R\to L}$.
Dissipative nonreciprocity brings about unique physical phenomena, such as coherent perfect absorption~\cite{Chong-10, *Chong-11, Longhi-10}, unidirectional invisibility~\cite{Mostafazadeh-09, Lin-11}, and destruction of Anderson localization~\cite{Kawabata-21}.
It is also relevant to a quantum point contact for ultracold atoms~\cite{Dogra-19, *Corman-19}.
Nonreciprocity in open systems is fundamentally different from nonreciprocity discussed in this work, the latter of which arises from the energy dependence of the transmission probability in closed systems with coherent and conservative scattering processes.
It merits further study to develop a unified understanding about nonreciprocal response in closed and open systems.

While we have focused on disordered and topological materials in this work, it is worthwhile to consider the effect of many-body interaction~\cite{Kane-Fisher-92L, *Kane-Fisher-92B, *Kane-Fisher-94, Furusaki-93, Giamarchi-textbook}.
There, the renormalization due to the many-body interaction~\cite{Izuyama-61, Maslov-95, Safi-95, Ponomarenko-95, Kawabata-96, *Kawabata-98, Shimizu-96} should be significant, as discussed in Sec.~\ref{sec: literature}.
It is also of interest to study nonperturbative transport phenomena in the scattering theory, including the Landau-Zener transition~\cite{LL-textbook, Zener-32}.
While the nonlinear Landauer formula in this work is concerned with the two-terminal case, it can be straightforwardly generalized to the multi-terminal case in a manner similar to the linear regime~\cite{Buttiker-86}.
In this respect, a recent work~\cite{Kane-21} has discovered the quantization of the $\left( d+1 \right)$-terminal nonlinear conductance in a $d$-dimensional ballistic metal. 
It is significant to develop a scattering theory of the quantized nonlinear conductance.

In statistical physics, it is significant to develop a general understanding about nonlinear response theories.
In the linear regime, the Landauer formula was shown to be equivalent to the Kubo formula by taking into account the electrodes~\cite{Stone-88}.
This correspondence is related to the conformal field theory~\cite{Ryu-07}.
It merits further research to explore a similar correspondence for nonlinear response.
We also note that Ref.~\cite{Watanabe-20} introduced the nonlinear Drude weights and showed that the $n$th-order nonlinear Drude weight is given by the $\left( n+1 \right)$\,th derivative of many-body eigenenergy with respect to a vector potential.
This result is similar to the nonlinear Boltzmann conductivity discussed in Sec.~\ref{sec: Boltzmann}, although Ref.~\cite{Watanabe-20} does not depend on the Boltzmann equation but the limiting time evolutions such as quench and adiabatic processes~\cite{Oshikawa-03}.
It is worthwhile to further study the nonlinear Drude weights in the scattering theory.

\section*{Acknowledgements}
We thank Shinsei Ryu and David S\'anchez for helpful discussions.
K.K. is supported by the Japan Society for the Promotion of Science (JSPS) through KAKENHI Grant No.~JP19J21927 and the Overseas Research Fellowship, and by the Gordon and Betty Moore Foundation through Grant GBMF8685 toward the Princeton theory program.
M.U. is supported by KAKENHI Grant Nos.~JP18H01145 and JP22H01152 from the JSPS.

\bigskip
{\it Note added.}---After the completion of this work, we became aware of Ref.~\cite{Wei-21}, which numerically investigates the nonlinear 
Hall effect in the four-terminal setup. 
While the nonlinear Hall effect in Ref.~\cite{Wei-21} originates from the Coulomb potential, the nonlinear Hall effect in this work originates from the Berry curvature and the concomitant anomalous velocity.


\appendix



 \section{Scattering and transfer matrices}
    \label{asec: scattering}

We summarize basic properties of the scattering theory.
We consider a system connected to two ideal leads.
Waves incident on the system from the left and right are respectively
\begin{align}
a_{\rm in}^{+} &\coloneqq \left( a_{1}^{+}~a_{2}^{+}~\cdots~a_{N}^{+} \right)^{T}, \\
b_{\rm in}^{-} &\coloneqq \left( b_{1}^{-}~b_{2}^{-}~\cdots~b_{N}^{-} \right)^{T}.
\end{align}
Here, $N$ is the number of channels. 
Similarly, the reflected and transmitted waves scattered to the right and left are respectively
\begin{align}
b_{\rm out}^{+} &\coloneqq \left( b_{1}^{+}~b_{2}^{+}~\cdots~b_{N}^{+} \right)^{T},\\
a_{\rm out}^{-} &\coloneqq \left( a_{1}^{-}~a_{2}^{-}~\cdots~a_{N}^{-} \right)^{T}.
\end{align}
For these incident and scattered waves, the scattering matrix $S$ is defined by
\begin{equation}
\begin{pmatrix} 
	a_{\rm out}^{-} \\ b_{\rm out}^{+} \\
\end{pmatrix}
= S \begin{pmatrix}
	a_{\rm in}^{+} \\ b_{\rm in}^{-} \\
\end{pmatrix},\quad
S \coloneqq \begin{pmatrix} 
	r_{\rm L} & t_{\rm L} \\
	t_{\rm R} & r_{\rm R} \\ 
	\end{pmatrix},
\end{equation}
where $r_{\rm L}$ ($r_{\rm R}$) is an $N \times N$ invertible matrix that describes the reflection from the left to the left (from the right to the right), and $t_{\rm R}$ ($t_{\rm L}$) is an $N \times N$ invertible matrix that describes the transmission from the left to the right (from the right to the left). 
Similarly, the transfer matrix $M$ is defined by
\begin{equation}
\begin{pmatrix}
	b_{\rm out}^{+} \\ b_{\rm in}^{-} \\
\end{pmatrix}
= M \begin{pmatrix}
	a_{\rm in}^{+} \\ a_{\rm out}^{-} \\
	\end{pmatrix}.
\end{equation}
The scattering matrix $S$ and the transfer matrix $M$ contain the same information on the scattering process.
In fact, from the definitions of $S$ and $M$, we have
\begin{align}
a_{\rm out}^{-} &= r_{\rm L} a_{\rm in}^{+} + t_{\rm L} b_{\rm in}^{-},\\
b_{\rm out}^{+} &= t_{\rm R} a_{\rm in}^{+} + r_{\rm R} b_{\rm in}^{-}, \\
b_{\rm out}^{+} &= M_{11} a_{\rm in}^{+} + M_{12} a_{\rm out}^{-},\\
b_{\rm in}^{-} &= M_{21} a_{\rm in}^{+} + M_{22} a_{\rm out}^{-},
\end{align}
leading to
\begin{equation}
M = \begin{pmatrix} 
	t_{\rm R} - r_{\rm R} t_{\rm L}^{-1} r_{\rm L} & r_{\rm R} t_{\rm L}^{-1} \\
	- t_{\rm L}^{-1} r_{\rm L} & t_{\rm L}^{-1} \\
	\end{pmatrix},
	    \label{aeq: transfer matrix}
\end{equation}
and 
\begin{align}
r_{\rm L} &= - M_{22}^{-1} M_{21},\\
r_{\rm R} &= M_{12} M_{22}^{-1},\\
t_{\rm L} &= M_{22}^{-1},\\
t_{\rm R} &= M_{11} - M_{12} M_{22}^{-1} M_{21}.
\end{align}

Suppose that the system is closed and isolated from the environment.
Then, the norms of the waves are conserved under the scattering (i.e., current conservation):
\begin{equation}
\left| a_{\rm in}^{+} \right|^{2} + \left| b_{\rm in}^{-} \right|^{2}
= \left| a_{\rm out}^{-} \right|^{2} + \left| b_{\rm out}^{+} \right|^{2}.
\end{equation}
As a result, the scattering matrix $S$ is unitary:
\begin{equation}
S^{\dag} S = S S^{\dag} = 1.
\end{equation}
With the transmission amplitudes $t_{\rm R}$, $t_{\rm L}$ and the reflection amplitudes $r_{\rm L}$, $r_{\rm R}$, unitarity of $S$ is equivalent to
\begin{align}
    1 &= S^{\dag} S
    = \begin{pmatrix}
r_{\rm L}^{\dag} r_{\rm L} + t_{\rm R}^{\dag} t_{\rm R} 
& r_{\rm L}^{\dag} t_{\rm L} + t_{\rm R}^{\dag} r_{\rm R} \\
t_{\rm L}^{\dag} r_{\rm L} + r_{\rm R}^{\dag} t_{\rm R} 
& t_{\rm L}^{\dag} t_{\rm L} + r_{\rm R}^{\dag} r_{\rm R}
\end{pmatrix},\\
    1 &= S S^{\dag}
    = \begin{pmatrix}
r_{\rm L} r_{\rm L}^{\dag} + t_{\rm L} t_{\rm L}^{\dag} 
& r_{\rm L} t_{\rm R}^{\dag} + t_{\rm L} r_{\rm R}^{\dag} \\
t_{\rm R} r_{\rm L}^{\dag} + r_{\rm R} t_{\rm L}^{\dag}
& t_{\rm R} t_{\rm R}^{\dag} + r_{\rm R} r_{\rm R}^{\dag}
\end{pmatrix},
\end{align}
which further leads to
\begin{align}
    t_{\rm R} t_{\rm R}^{\dag} = t_{\rm L} t_{\rm L}^{\dag},\quad
    t_{\rm R} t_{\rm R}^{\dag} + r_{\rm L} r_{\rm L}^{\dag} 
    = t_{\rm L} t_{\rm L}^{\dag} + r_{\rm R} r_{\rm R}^{\dag}
    = 1.
\end{align}
Thus, in the presence of unitarity, the transmission probability from the left to the right (i.e., sum of the eigenvalues of $t_{\rm R} t_{\rm R}^{\dag}$) is the same as the transmission probability from the right to the left (i.e., sum of the eigenvalues of $t_{\rm L} t_{\rm L}^{\dag}$).
In terms of the transfer matrix $M$, we have
\begin{align}
\left| a_{\rm in}^{+} \right|^{2} - \left| a_{\rm out}^{-} \right|^{2}
&= \left| b_{\rm out}^{+} \right|^{2} - \left| b_{\rm in}^{-} \right|^{2} \nonumber \\
&= \left( \begin{array}{@{\,}c@{\,}} 
	b_{\rm out}^{+} \\ b_{\rm in}^{-} \\
	\end{array} \right)^{\dag} \sigma_{z} \left( \begin{array}{@{\,}c@{\,}} 
	b_{\rm out}^{+} \\ b_{\rm in}^{-} \\
	\end{array} \right) \nonumber \\
&= \left( \begin{array}{@{\,}c@{\,}} 
	a_{\rm in}^{+} \\ a_{\rm out}^{-} \\
	\end{array} \right)^{\dag} M^{\dag} \sigma_{z} M \left( \begin{array}{@{\,}c@{\,}} 
	a_{\rm in}^{+} \\ a_{\rm out}^{-} \\
	\end{array} \right)
\end{align}
with a Pauli matrix $\sigma_{z}$, and the transfer matrix $M$ is pseudo-unitary:
\begin{equation}
\sigma_{z} M^{\dag} \sigma_{z}^{-1} = M^{-1}.
\end{equation}
On the other hand, if the system exchanges energy or particles with the environment, the scattering matrix $S$ is no longer unitary, and the transfer matrix is no longer pseudo-unitary~\cite{Kawabata-21}.

\section{Illustrative examples}
    \label{asec: example}

\begin{figure*}[t]
\centering
\includegraphics[width=172mm]{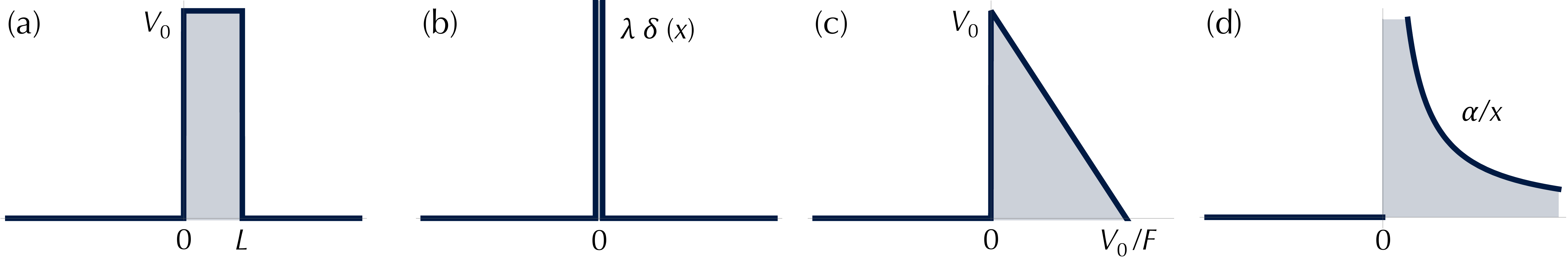} 
\caption{Potential barriers. (a)~Square potential $V \left( x \right) = V_0$ ($0 < x < L$). (b)~Delta potential $V \left( x \right) = \lambda \delta \left( x \right)$. (c)~Linear potential $V \left( x \right) = V_0 - Fx$ ($x > 0$). (d)~Coulomb potential $V \left( x \right) = \alpha/x$ ($x > 0$).}
	\label{fig: potential}
\end{figure*}

\begin{figure*}[t]
\centering
\includegraphics[width=172mm]{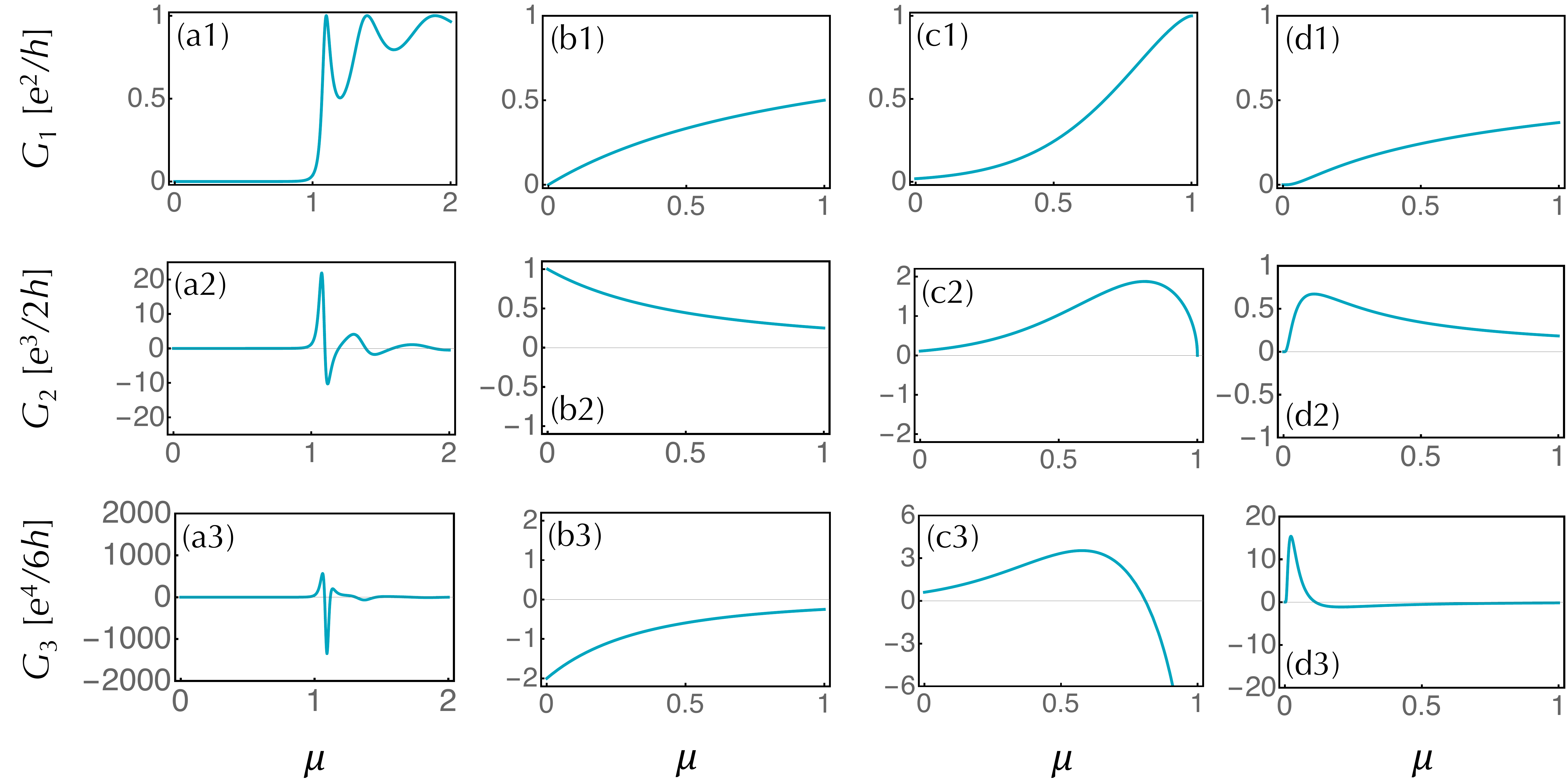} 
\caption{Linear and nonlinear conductance through potential barriers. (a1)~The linear conductance $G_1$, (a2)~the second-order nonlinear conductance $G_2$, and (a3)~the third-order nonlinear conductance $G_3$ for the square potential as functions of the chemical potential $\mu$ ($V_0 = 1$, $\sqrt{2mV_0}L/\hbar = 10$). (b1)~$G_1$, (b2)~$G_2$, and (b3)~$G_3$ for the delta potential ($m\lambda^2/2\hbar^2 = 1$). (c1)~$G_1$, (c2)~$G_2$, and (c3)~$G_3$ for the linear potential ($V_0 = 1$, $4\sqrt{2mV_0^3}/3\hbar F = 4$). (d1)~$G_1$, (d2)~$G_2$, and (d3)~$G_3$ for the Coulomb potential ($2\pi^2 m\alpha^2/\hbar^2 = 1$).}
	\label{fig: scattering}
\end{figure*}

The Landauer formula provides a useful way to obtain the linear and nonlinear conductance of quantum materials.
Here, we illustrate this fact with simple examples.
We investigate scattering of a quantum particle with mass $m$ and charge $-e$ through a potential barrier $V \left( x \right)$ in one-dimensional continuum space~\cite{LL-textbook}.
The wave function $\psi \left( x \right)$ is described by the Schr\"odinger equation
\begin{equation}
    - \frac{\hbar^2}{2m} \frac{d^2}{dx^2} \psi \left( x \right) + V \left( x \right) \psi \left( x \right) = E \psi \left( x \right)
\end{equation}
for given energy $E \geq 0$.
We solve this scattering problem for some exemplary potentials (Fig.~\ref{fig: potential}) and obtain the linear and nonlinear conductance (Fig.~\ref{fig: scattering}).
We assume zero temperature and use the Landauer formula in Eq.~(\ref{eq: nonlinear conductance formula - zero temperature}).
The conductance behaves differently depending on the details of the potentials.
As discussed in Sec.~\ref{sec: nonreciprocity}, nonreciprocity is one of the hallmarks of nonlinear response.
We demonstrate that nonreciprocity, which is characterized by the even-ordered nonlinear conductance including $G_2$, indeed arises even in such simple systems.
    
\subsection{Square potential}
    \label{asec: square potential}

We begin with scattering through the square potential
\begin{equation}
    V \left( x \right) = \left\{ \begin{array}{lc}
        0 & ~~\left( x \leq 0 \right); \\
        V_0 & ~~\left( 0 < x < L \right); \\
        0 & ~~\left( x \geq L \right),
    \end{array} \right.
        \label{eq: square potential}
\end{equation}
with $V_0 \geq 0$ [Fig.~\ref{fig: potential}\,(a)].
Let the wave function be 
\begin{equation}
    \psi \left( x \right) = \left\{ \begin{array}{lc}
        e^{\ii k_0 x} + re^{-\ii k_0 x} & ~~\left( x < 0 \right); \\
        a e^{\ii kx} + b e^{-\ii kx} & ~~\left( 0 < x < L \right); \\
        t e^{\ii k_0\,(x-L)} & ~~\left( x > L \right),
    \end{array} \right.
\end{equation}
with the amplitudes $a, b, r, t \in \mathbb{C}$ and the wave numbers
\begin{equation}
    k_0 \coloneqq \frac{\sqrt{2mE}}{\hbar},\quad
    k \coloneqq \frac{\sqrt{2m \left( E - V \right)}}{\hbar}.
\end{equation}
The wave number $k$ in the presence of the square potential is real for $E \geq V_0$ and imaginary for $0 \leq E < V_0$.
The boundary conditions at $x = 0$ 
\begin{equation}
    \psi \left( -0 \right) = \psi \left( +0 \right),\quad
    \left. \frac{d\psi}{dx} \right|_{x=-0} = \left. \frac{d\psi}{dx} \right|_{x=+0}
\end{equation}
reduce to
\begin{equation}
    1 + r = a + b,\quad
    k_0 \left( 1 - r \right) = k \left( a - b \right),
        \label{aeq: square1}
\end{equation}
and the boundary conditions at $x = L$ 
\begin{equation}
    \psi \left( L-0 \right) = \psi \left( L+0 \right),\quad
    \left. \frac{d\psi}{dx} \right|_{x=L-0} = \left. \frac{d\psi}{dx} \right|_{x=L+0}
\end{equation}
reduce to
\begin{equation}
    a e^{\ii kL} + b e^{-\ii kL} = t,\quad
    k \left( a e^{\ii kL} - b e^{-\ii kL} \right) = k_0 t.
        \label{aeq: square2}
\end{equation}
Combining Eqs.~(\ref{aeq: square1}) and (\ref{aeq: square2}), we have
\begin{align}
    a &= - \frac{2 \left( 1 + k/k_0 \right) e^{-\ii kL}}{\left( 1 - k/k_0 \right)^2 e^{\ii kL} - \left( 1 + k/k_0 \right)^2 e^{-\ii kL}},\\
    b &= \frac{2 \left( 1 - k/k_0 \right) e^{\ii kL}}{\left( 1 - k/k_0 \right)^2 e^{\ii kL} - \left( 1 + k/k_0 \right)^2 e^{-\ii kL}}, \\
    r &= \frac{\left( 1- k^2/k_0^2 \right) \left( e^{\ii kL} - e^{-\ii kL} \right)}{\left( 1 - k/k_0 \right)^2 e^{\ii kL} - \left( 1 + k/k_0 \right)^2 e^{-\ii kL}}, \\
    t &= - \frac{4k/k_0}{\left( 1 - k/k_0 \right)^2 e^{\ii kL} - \left( 1 + k/k_0 \right)^2 e^{-\ii kL}}.
\end{align}
From these amplitudes, the transmission probability $T = \left| t \right|^2$ is obtained as
\begin{align}
    T = \left[ 1 + \frac{V_0^2}{4E \left( V_0 - E \right)} \sinh^2 \left( \frac{\sqrt{2m \left( V_0 - E \right)}\,L}{\hbar}\right) \right]^{-1}
    \label{aeq: scattering - square tunneling}
\end{align}
for $0 \leq E < V_0$ and
\begin{align}
    T = \left[ 1 + \frac{V_0^2}{4E \left( E - V_0 \right)} \sin^2 \left( \frac{\sqrt{2m \left( E - V_0 \right)}\,L}{\hbar}\right) \right]^{-1}
        \label{aeq: square - transmission}
\end{align}
for $E > V_0$.

Figure~\ref{fig: scattering}\,(a) shows the linear conductance $G_1$ and the nonlinear conductance $G_2, G_3$ for the square potential.
When the incident wave has smaller energy than the potential barrier (i.e., $\mu < V_0$), the transmission is due to quantum tunneling and suppressed for $L \gg L_0$.
As a result, the linear and nonlinear conductance is small for $\mu < V_0$.
On the other hand, when the incident wave has larger energy than the potential barrier (i.e., $\mu \geq V_0$), it can transmit through the potential barrier.
Because of the wave nature of the transport, the perfect transmission $T \left( E \right) = 1$ occurs only for the sequence of the resonant energy.
Consequently, the nonlinear conductance oscillates as a function of the chemical potential $\mu$.
For large $L$, however, the oscillation gets faster, and the derivatives of Eq.~(\ref{aeq: square - transmission}) get smaller on average.
Such average behavior is obtained by
\begin{align}
T &\sim \int_{0}^{2\pi} \left[ 1 + \frac{V_0^2}{4E \left( E - V_0 \right)} \sin^2 \theta \right]^{-1} \frac{d\theta}{2\pi} \nonumber \\
&= \left[ \left( 1 + \frac{V_0^2}{8E \left( E - V_0 \right)} \right)^2 - \left( \frac{V_0^2}{8E \left( E - V_0 \right)} \right)^2 \right]^{-1/2} \nonumber \\
&\simeq 1 - \frac{V_0^2}{8E \left( E - V_0 \right)}
\end{align}
for $L \gg \hbar/\sqrt{2mV_0}$ and $E \gg V_0$.
Here, we use the formula
\begin{align}
    \int_{0}^{2\pi} \frac{1}{a+b\cos\theta} \frac{d\theta}{2\pi}
    = \frac{1}{\sqrt{a^2-b^2}}
\end{align}
with $a > b > 0$.

\subsection{Delta potential}
    \label{asec: delta potential}

We next investigate scattering through the delta potential
\begin{equation}
    V \left( x \right) = \lambda \delta \left( x \right),
        \label{eq: delta potential}
\end{equation}
where $\lambda \geq 0$ is the strength of the potential barrier [Fig.~\ref{fig: potential}\,(b)].
Clearly, the transmission is enhanced for small $\lambda$ and suppressed for large $\lambda$. 
The square potential in Eq.~(\ref{eq: square potential}) reduces to the delta potential in Eq.~(\ref{eq: delta potential}) for $\lambda \coloneqq V_0 L$ and $L \to 0$.
Let the wave function be
\begin{equation}
    \psi \left( x \right) = \left\{ \begin{array}{lc}
        e^{\ii kx} + re^{-\ii kx} & ~~\left( x < 0 \right); \\
        t e^{\ii kx} & ~~\left( x > 0 \right),
    \end{array} \right.
\end{equation}
with the wave number $k \coloneqq \sqrt{2mE}/\hbar$, as well as the transmission amplitude $t \in \mathbb{C}$ and the reflection amplitude $r \in \mathbb{C}$.
The boundary conditions at $x = 0$ read
\begin{align}
    \psi \left( +0 \right) = \psi \left( -0 \right),
\end{align}
and
\begin{align}
    - \frac{\hbar^2}{2m} \left[ \frac{d}{dx} \psi \left( x \right) \right]_{x=-0}^{x=+0} + \lambda \psi \left( 0 \right) = 0.
\end{align}
Combining these equations, we have
\begin{equation}
    r = \frac{1}{\ii \hbar^2 k/m\lambda - 1},\quad
    t = \frac{1}{1 + \ii m\lambda/\hbar^2 k}.
\end{equation}
Thus, the transmission probability $T$ is obtained as
\begin{align}
    T = \left| t \right|^2 
    &= \frac{1}{1 + \left( m\lambda/\hbar^2 k\right)^2} \nonumber \\
    &= \frac{1}{1 + m\lambda^2/2\hbar^2 E} \nonumber \\
    &= \frac{1}{1 + E_0/E},
\end{align}
where we define the renormalized potential strength $E_0$ as
\begin{align}
    E_0 \coloneqq \frac{m\lambda^2}{2\hbar^2}.
\end{align}
This result is equivalent to the transmission probability in Eq.~(\ref{aeq: scattering - square tunneling}) with $\lambda = V_0 L$ and $L \to 0$.

The transmission probability $T$ monotonically grows with increasing $E \geq 0$ and reaches one for $E \to \infty$ [Fig.~\ref{fig: scattering}\,(b)].
At zero energy $E=0$, we have $T = 0$, i.e., no transmission.
The transmission probability $T$ changes most rapidly near zero energy $E=0$. 
Around $E = 0$, we have the expansion
\begin{align}
    T = - \sum_{n=1}^{\infty} \left( - \frac{E}{E_0} \right)^{n},
\end{align}
which means $\left. T \right|_{E=0} = 0$ and
\begin{equation}
    \left. \frac{d^n T}{dE^{n}} \right|_{E = 0}
    = \frac{\left( -1 \right)^{n+1} n!}{E_0^{n}}
\end{equation}
for $n \geq 1$.
This $n$th derivative yields the nonlinear conductance $G_{n}$ at zero energy:
\begin{align}
    G_n \left( \mu = 0 \right) = \frac{e^{n+1}}{\left( n! \right) h} \left. \frac{d^{n-1} T}{dE^{n-1}} \right|_{E = 0}
    = \frac{\left( -1 \right)^n}{n} \frac{e^{n+1}}{h} \frac{1}{E_0^{n-1}}.
\end{align}
The current $I$ is given as
\begin{align}
    I &= \frac{e^3}{2hE_0} V^2 - \frac{e^4}{3hE_0^2} V^3 + \cdots, \nonumber \\
    &= - \frac{eE_0}{h} \log \left( 1+\frac{eV}{E_0} \right) + \frac{e^2}{h} V,
\end{align}
which shows strong nonlinearity and nonreciprocity.
The integral of the transmission probability $T$ is also obtainable as
\begin{align}
    \int_{E_{-}}^{E_{+}} T \left( E \right) dE
    &= \int_{E_{-}}^{E_{+}} \left( 1 - \frac{E_{0}}{E + E_{0}}\right) dE \nonumber \\
    &= E_{+} - E_{-} - E_{0} \log \frac{E_{+} + E_{0}}{E_{-} + E_{0}}
\end{align}
for $E_{\pm} \geq 0$.

\subsection{Linear potential}
    \label{asec: linear potential}
    
As a more nontrivial example, we investigate scattering through the linear potential [Fig.~\ref{fig: potential}\,(c)]
\begin{align}
    V \left( x \right) = \left\{ \begin{array}{lc}
        0 & ~~\left( x < 0 \right); \\
        V_0 - Fx & ~~\left( x > 0 \right). 
    \end{array}
    \right.
\end{align}
Here, $V_0 \geq 0$ and $F \geq 0$ are the height and gradient of the potential barrier, respectively.
The transmission is enhanced for small $V_0$ or large $F$ and suppressed for large $V_0$ or small $F$.

While this scattering problem is exactly solvable with the Airy functions, we here obtain the transmission probability $T \left( E \right)$ on the basis of the Wentzel-Kramers-Brillouin (WKB) approximation~\cite{LL-textbook}.
The WKB approximation is well justified for $0 \leq E \leq V_0$ and $T \left( E \right) \ll 1$.
Because of the generality of the WKB approximation, the linear and nonlinear conductance for more complicated potentials can be calculated in a similar manner.
On the basis of the WKB approximation, the transmission probability $T$ through a barrier $V \left( x \right) > E$ is generally obtained as
\begin{align}
    T \simeq \exp \left[ - \frac{2\sqrt{2m}}{\hbar} \int_{x_{-}}^{x_{+}} \sqrt{V \left( x \right) - E}~dx \right],
    \label{aeq: WKB - general}
\end{align}
where $x_{+}$ and $x_{-}$ ($x_{+} \geq x_{-}$) are the two turning points defined by $V \left( x_{+} \right) = V \left( x_{-} \right) = E$. 
For the linear potential, we have
\begin{equation}
    x_{-} = 0,\quad
    x_{+} = \frac{V_0 - E}{F},
\end{equation}
and then
\begin{align}
    T &\simeq \exp \left[ - \frac{2\sqrt{2m}}{\hbar} \int_{0}^{\left( V_0 - E \right)/F} \sqrt{V_0 - E - Fx}~dx  \right] \nonumber \\
    &= \exp \left[ - v \left( 1 - \frac{E}{V_0} \right)^{3/2} \right],
        \label{aeq: linear - transmission}
\end{align}
where we define the renormalized potential strength $v$ as
\begin{equation}
    v \coloneqq \frac{4\sqrt{2mV_0^3}}{3\hbar F}.
\end{equation}

The obtained transmission probability $T$ monotonically grows with increasing $0 \leq E \leq V_0$.
Using the Landauer formula, we calculate the linear and nonlinear conductance with $T = T \left( E \right)$.
Figure~\ref{fig: scattering}\,(c) shows the linear conductance $G_1$ and the nonlinear conductance $G_2, G_3$ for the linear potential.
In contrast to the delta potential, the nonlinear conductance vanishes for small chemical potential $\mu$.
Around $E = 0$, for example, we have
\begin{align}
    \left. T \right|_{E=0} &= e^{-v}, \\
    \left. \frac{dT}{dE} \right|_{E=0} &= \frac{3v e^{-v}}{2V_0}, \\
    \left. \frac{d^2T}{dE^2} \right|_{E=0} &= \frac{3v \left( 3v-1 \right) e^{-v}}{4V_0^2}.
\end{align}
Around $E = V_0$, on the other hand, we have
\begin{align}
    \left. T \right|_{E=V_0} &= 1, \\
    \left. \frac{dT}{dE} \right|_{E=V_0} &= 0, \\
    \left. \frac{d^2T}{dE^2} \right|_{E=V_0} &= - \infty.
\end{align}
It should be noted that the WKB approximation and the consequent formula in Eq.~(\ref{aeq: linear - transmission}) may not be justified around $E = V_0$.

Since we have
\begin{align}
    \frac{d^2T}{dE^2} \propto 3v \sqrt{\frac{\left( V_0 - E \right)^3}{V_0}} - V_0,
\end{align}
the first derivative $dT/dE$ gets largest for
\begin{equation}
    \frac{E}{V_0} = 1 - \left( 3v \right)^{-2/3},
\end{equation}
at which we have
\begin{align}
    \mathrm{max} \left( \frac{dT}{dE} \right) 
    &= \frac{\left( 3v\right)^{2/3}}{2e^{1/3}V_0} \nonumber \\ 
    &= \left( 0.745223 \cdots \right) \frac{v^{2/3}}{V_0}.
\end{align}
This gives the maximum of the second-order nonlinear conductance $G_2$:
\begin{align}
    \mathrm{max} \left( G_2 \right)
    &= \frac{e^3}{2h}\,\mathrm{max} \left( \frac{dT}{dE} \right) \nonumber \\
    &= \left( 0.745223 \cdots \right) \frac{e^3}{2h} \frac{v^{2/3}}{V_0}.
\end{align}
Similarly, since we have
\begin{align}
    \frac{d^3T}{dE^3} \propto 9v^2 \left( 1 - \frac{E}{V_0} \right)^3 - 9v \left( 1 - \frac{E}{V_0} \right)^{3/2} - 1,
\end{align}
the second derivative $d^2T/dE^2$ gets largest for
\begin{align}
    \frac{E}{V_0} &= 1 - \left( \frac{\sqrt{117}-9}{2} v \right)^{-2/3} \nonumber \\
    &= 1 - \left( 1.0662 \cdots \right) v^{-2/3},
\end{align}
at which we have
\begin{equation}
    \mathrm{max} \left( \frac{d^2T}{dE^2} \right) = \left( 0.556247 \cdots \right) \frac{v^{4/3}}{V_0^2}.
\end{equation}
This gives the maximum of the third-order nonlinear conductance $G_3$:
\begin{align}
    \mathrm{max} \left( G_3 \right)
    &= \frac{e^4}{6h}\,\mathrm{max} \left( \frac{d^2T}{dE^2} \right) \nonumber \\
    &= \left( 0.556247 \cdots \right) \frac{e^4}{6h} \frac{v^{4/3}}{V_0^2}.
\end{align}
These analytic results are compatible with the numerical results in Fig.~\ref{fig: scattering}\,(c).
The maximum nonlinear conductance gets larger with increasing $v$, which means large nonlinearity and nonreciprocity for the strong potential barrier.

\subsection{Coulomb potential}
    \label{asec: Coulomb potential}

We finally investigate scattering through the Coulomb potential
\begin{align}
    V \left( x \right) = \left\{ \begin{array}{lc}
        0 & ~~\left( x < 0 \right); \\
        \alpha/x & ~~\left( x > 0 \right),
    \end{array}
    \right.
\end{align}
where $\alpha$ is the strength of the potential [Fig.~\ref{fig: potential}\,(d)].
Historically, this scattering problem was relevant to the physics of the $\alpha$ decay~\cite{Gamow-28}.
Similarly to the linear potential, we use the WKB approximation and obtain the transmission probability $T$ for $E \geq 0$.
The turning points for the Coulomb potential are 
\begin{align}
    x_{-} = 0,\quad
    x_{+} = \frac{\alpha}{E},
\end{align}
and then the transmission probability $T$ in Eq.~(\ref{aeq: WKB - general}) is obtained as
\begin{align}
    T &\simeq \exp \left[ - \frac{2\sqrt{2m}}{\hbar}  \int_0^{\alpha/E} \sqrt{\frac{\alpha}{x} - E}\,dx\right] \nonumber \\
    &= \exp \left[ - \frac{2\sqrt{2m}}{\hbar}  \frac{\alpha}{\sqrt{E}} \int_0^{1} \sqrt{\frac{1}{x} - 1}\,dx\right].
\end{align}
The integral is analytically calculated by introducing $\theta$ by $x \eqqcolon \cos^2 \theta$:
\begin{align}
    \int_0^{1} \sqrt{\frac{1}{x} - 1}\,dx
    = 2 \int^{\pi/2}_{0} \sin^2 \theta~d\theta
    = \frac{\pi}{2}.
\end{align}
Then, we have
\begin{align}
    T &\simeq \exp \left( - \frac{\pi \alpha \sqrt{2m}}{\hbar} \frac{1}{\sqrt{E}} \right) \nonumber \\
    &\eqqcolon \exp \left( - \sqrt{\frac{E_0}{E}} \right),
\end{align}
where we define the renormalized potential strength as
\begin{equation}
    E_0 \coloneqq \frac{2\pi^2 m \alpha^2}{\hbar^2}.
\end{equation}

The transmission probability $T$ monotonically grows with increasing $E$.
Figure~\ref{fig: scattering}\,(d) shows the linear conductance $G_1$ and the nonlinear conductance $G_2, G_3$ for the Coulomb potential.
Because of the singular behavior at zero chemical potential $\mu = 0$, the linear conductance $G_1$ is suppressed for small $\mu$.
We have
\begin{align}
    \frac{dT}{dE} &= \frac{1}{2} \sqrt{\frac{E_0}{E^3}} e^{-\sqrt{E_0/E}}, \\
    \frac{d^2T}{dE^2} &= \frac{E_0}{4E^3} \left( 1  - 3 \sqrt{\frac{E}{E_0}} \right) e^{-\sqrt{E_0/E}}, \\
    \frac{d^3T}{dE^3} &= \frac{1}{8} \sqrt{\frac{E_0^3}{E^9}} \left( 1 - 9 \sqrt{\frac{E}{E_0}} + \frac{15E}{E_0} \right) e^{-\sqrt{E_0/E}}.
\end{align}
The first derivative $dT/dE$, which yields the second-order nonlinear conductance $G_2$, gets largest at $E/E_0 = 1/9$ and takes
\begin{align}
    \mathrm{max} \left( \frac{dT}{dE} \right)
    = \frac{27}{2e^3E_0}
    = \frac{0.672125\cdots}{E_0}.
\end{align}
The second derivative $d^2T/dE^2$, which yields the third-order nonlinear conductance $G_3$, gets largest and smallest at
\begin{align}
\frac{E}{E_0} &= \frac{\left( 17 - 3\sqrt{21} \right)}{150} = 0.0216818 \cdots,\\
\frac{E}{E_0} &= \frac{\left( 17 + 3\sqrt{21} \right)}{150} = 0.204985 \cdots,
\end{align}
respectively, and takes
\begin{align}
    \mathrm{max} \left(  \frac{d^2T}{dE^2} \right)
    &= \frac{15.384 \cdots}{E_0^2}, \\
    \mathrm{min} \left(  \frac{d^2T}{dE^2} \right)
    &= - \frac{1.14219 \cdots}{E_0^2}.
\end{align}
These large nonlinear conductance also means strong nonlinearity and nonreciprocity for the Coulomb potential around zero chemical potential.

\section{Boltzmann equation}
    \label{asec: Boltzmann}

We investigate the linear and nonlinear response of exemplary systems on the basis of the Boltzmann equation.
We define the linear conductivity $\sigma_1$ and nonlinear conductivity $\sigma_n$ ($n \geq 2$) by
\begin{align}
    \currentdensity = \sum_{n=1}^{\infty} \sigma_n \mathcal{E}^{n}
\end{align}
with the current density $\currentdensity$ and the electric field $\mathcal{E}$.
In one dimension, the current density $\currentdensity$ is equivalent to the current $I$, and the conductance $G_n$ defined by Eq.~(\ref{eq: conductance - definition}) is given as $G_n = \sigma_n/L^n$.
As discussed in Sec.~\ref{sec: Boltzmann}, on the basis of the Boltzmann equation with the relaxation time approximation
\begin{equation}
    - \frac{e\mathcal{E}}{\hbar} \frac{\partial f}{\partial k} = - \frac{f - f_{\rm eq}}{\tau},
\end{equation}
the $n$th-order conductivity $\sigma_n$ is given as
\begin{align}
    \sigma_n &= - \frac{1}{\tau} \left( \frac{e\tau}{\hbar} \right)^{n+1} \oint \frac{d E}{d k}  
    \frac{d^n f_{\rm eq}}{d k^n}
    \frac{dk}{2\pi}
        \label{aeq: conductivity - Boltzmann - 1}
\end{align}
for a given energy dispersion $E = E \left( k \right)$.
Here, the integral $\oint dk/2\pi$ denotes $\int_{-\infty}^{\infty} dk/2\pi$ for free space and $\int_0^{2\pi} dk/2\pi$ for a lattice.
If the boundary terms are negligible, $\sigma_n$ further reduces to
\begin{align}
    \sigma_n 
    &= \frac{1}{\tau} \left( - \frac{e\tau}{\hbar} \right)^{n+1} \oint \frac{d^{n+1} E}{d k^{n+1}} f_{\rm eq} \frac{dk}{2\pi}.
        \label{aeq: conductivity - Boltzmann - 2}
\end{align}

\subsection{Free fermion}

We begin with a free fermion
\begin{equation}
    E \left( k \right) = \frac{\hbar^2 k^2}{2m}.
        \label{aeq: free - Boltzmann}
\end{equation}
Since $f_{\rm eq}$ decays rapidly for $k \to \pm \infty$, we can use Eq.~(\ref{aeq: conductivity - Boltzmann - 2}).
For $n=1$, Eq.~(\ref{aeq: conductivity - Boltzmann - 2}) reduces to 
\begin{align}
    \sigma_1 = \frac{e^2\tau}{\hbar^2} \int_{-\infty}^{\infty} \left( \frac{\hbar^2}{m} \right) f_{\rm eq}\,\frac{dk}{2\pi}
    = \frac{ne^2\tau}{m}
        \label{aeq: Drude}
\end{align}
with the number density 
\begin{equation}
    n \coloneqq \int_{-\infty}^{\infty} f_{\rm eq}\,\frac{dk}{2\pi}.
\end{equation}
This is the Drude conductivity~\cite{AM-textbook, Abrikosov-textbook}.
For even $n$, the nonlinear conductivity $\sigma_n$ vanishes because of time-reversal symmetry $E \left( k \right) = E \left( -k \right)$.
Even if $n$ is odd, the nonlinear conductivity $\sigma_n$ vanishes because of $d^n E/d k^n = 0$ for $n \geq 3$.

While Eq.~(\ref{aeq: free - Boltzmann}) describes a free fermion in free space, a free fermion on a lattice is described by
\begin{equation}
    E \left( k \right) = \frac{\hbar^2}{ma^2} \left( 1 - \cos ka \right)
\end{equation}
with the lattice spacing $a$.
This energy dispersion reduces to Eq.~(\ref{aeq: free - Boltzmann}) for $ka \to 0$.
Similarly to the previous case, the nonlinear conductivity $\sigma_n$ for even $n$ vanishes because of time-reversal symmetry.
Since the momentum integral is taken on the Brillouin zone $\left[ 0, 2\pi \right]$, we can use Eq.~(\ref{aeq: conductivity - Boltzmann - 2}) again.
Since we have
\begin{align}
    \frac{d^{n+1} E}{d k^{n+1}}
    &= \frac{\hbar^2}{ma^2} \left( -1\right)^{(n-1)/2} a^{n+1} \cos ka \nonumber \\
    &= \frac{\left( -1\right)^{(n-1)/2} \hbar^2 a^{n-1}}{m} \left( 1 - \frac{ma^2}{\hbar^2} E \right)
\end{align}
for odd $n$, the nonlinear conductivity $\sigma_n$ in Eq.~(\ref{aeq: conductivity - Boltzmann - 2}) is obtained as
\begin{align}
    \sigma_n &= \frac{\left( -1\right)^{(n-1)/2} \hbar^2 a^{n-1}}{m\tau} \left( \frac{e\tau}{\hbar} \right)^{n+1} \left( n - \frac{ma^2}{\hbar^2} \bar{E} \right)
\end{align}
with the energy density at equilibrium:
\begin{equation}
\bar{E} \coloneqq \int_0^{2\pi} E f_{\rm eq}\,\frac{dk}{2\pi}.
\end{equation}
In the continuum limit $a \to 0$, only $\sigma_1$ survives and gives $\sigma_1 = ne^2\tau/m$, which is the Drude conductivity in Eq.~(\ref{aeq: Drude}).
For $a \neq 0$, on the other hand, we have
\begin{align}
    \sigma_1 &= \frac{e^2\tau}{m} \left( n - \frac{ma^2}{\hbar^2} \bar{E} \right), \\
    \sigma_3 &= - \frac{a^2 e^4 \tau^3}{\hbar^2 m} \left( n - \frac{ma^2}{\hbar^2} \bar{E} \right),
\end{align}
and so on. 
Thus, the nonlinear conductivity arises because of the lattice effect.
Formally, we have
\begin{align}
\currentdensity &= \sum_{n=1}^{\infty} \sigma_n \mathcal{E}^n \nonumber \\
&= \sigma_1 \mathcal{E} \left[ \sum_{n=0}^{\infty} \left( - \frac{a^2e^2\tau^2}{\hbar^2} \mathcal{E}^{2} \right)^n \right] \nonumber \\
&=\sigma_1 \frac{\mathcal{E}}{1 + \left( ae\tau \mathcal{E}/\hbar\right)^2}.
\end{align}
It should be noted that this expansion is valid only for $\left| ae\tau \mathcal{E}/\hbar \right| \ll 1$.

\subsection{Chiral fermion}
    \label{aeq: Boltzmann - chiral}

We next investigate a chiral fermion
\begin{equation}
    E \left( k \right) = \hbar v k.
\end{equation}
In this case, $f_{\rm eq}$ does not vanish for $k \to - \mathrm{sgn} \left( v \right) \infty$, and Eq.~(\ref{aeq: conductivity - Boltzmann - 2}) is no longer valid. 
This is a consequence of the quantum anomaly of the chiral fermion.
Then, we instead use Eq.~(\ref{aeq: conductivity - Boltzmann - 1}), which yields
\begin{align}
    \sigma_1 &= - \frac{e^2\tau v}{\hbar} \int_{-\infty}^{\infty} \frac{d f_{\rm eq}}{d k} \frac{dk}{2\pi} 
    = \frac{e^2\tau \left|v\right|}{h}.
\end{align}
For $n \geq 2$, the nonlinear conductivity vanishes because of $d^n f_{\rm eq}/d k^n \to 0$ for $k \to \pm \infty$.

In the Landauer formula, the linear conductance $G_1 = \sigma_1/L$ for the chiral fermion is quantized to be $e^2/h$, as discussed in Sec.~\ref{sec: chiral edge}.
From the Boltzmann equation, this quantization of the conductance is obtained for 
\begin{equation}
    \tau \left| v \right| = L,
\end{equation} 
which means that the mean free path $\tau \left| v \right|$ is equal to the system length $L$.
However, this condition is never realized since the Boltzmann equation assumes the large mean free path in comparison with the wave length of electronic waves, as discussed in Sec.~\ref{sec: Boltzmann}.
Still, the vanishing nonlinear conductivity is consistent with the results based on the Landauer formula.

\subsection{Nonlinear Hall effect}
    \label{asec: Boltzmann - QHE}

In Sec.~\ref{sec: QHE}, we discuss the linear and nonlinear Hall effect in the scattering formulation.
For comparison, we here derive the linear and nonlinear Hall effect on the basis of the Boltzmann equation~\cite{Sodemann-15}. 

We focus on a two-dimensional system with periodic boundaries in all the directions.
The system is characterized by the energy dispersion $E = E \left( \bm{k} \right)$ and the Berry curvature $\Omega = \Omega \left( \bm{k} \right)$.
Let us impose a constant electric field $\bm{\mathcal{E}}$ on the system.
The distribution function $f = f \left( \bm{k} \right)$ is assumed to be described by the Boltzmann equation
\begin{equation}
    - \frac{e\bm{\mathcal{E}}}{\hbar} \cdot \frac{\partial f}{\partial \bm{k}} = - \frac{f - f_{\rm eq}}{\tau}.
\end{equation}
Expanding $f$ in terms of $\bm{\mathcal{E}}$, we have
\begin{align}
    f = f_{\rm eq} + \frac{e\tau \bm{\mathcal{E}}}{\hbar} \cdot \frac{\partial f_{\rm eq}}{\partial \bm{k}} + \mathcal{O} \left( \bm{\mathcal{E}}^2 \right).
        \label{aeq: Hall - Boltzmann}
\end{align}
The current density $\bm{\currentdensity}$ is obtained as
\begin{align}
    \bm{\currentdensity} = - e \oint_{\rm BZ} \bm{v} f\,\frac{d^2 k}{\left( 2\pi \right)^2},
\end{align}
where the momentum integral is taken over the entire Brillouin zone.
The velocity $\bm{v} \left( \bm{k} \right)$ is defined as
\begin{equation}
    \bm{v} = \frac{1}{\hbar} \frac{\partial E}{\partial \bm{k}} - \frac{e}{\hbar} \left( \bm{\mathcal{E}} \times \bm{n} \right) \Omega
\end{equation}
with the unit vector $\bm{n}$ perpendicular to the two-dimensional system.
The second contribution is the anomalous velocity due to the Berry curvature $\Omega$~\cite{Nagaosa-review, Xiao-review, Vanderbilt-textbook}.

In the following, we assume that the electric field $\bm{\mathcal{E}}$ is applied along the $x$ direction ($\mathcal{E}_x \coloneqq \left| \bm{\mathcal{E}} \right|$).
Then, the current along the $y$ direction, i.e., the Hall current, is give as
\begin{align}
    \currentdensity_y = - \frac{e^2 \mathcal{E}_x}{\hbar} \oint_{\rm BZ} \Omega f\,\frac{d^2k}{\left( 2\pi \right)^2}.
\end{align}
We define the linear and nonlinear Hall conductivity $\sigma_{n}^{\rm H}$ by
\begin{equation}
    \currentdensity_y = \sigma_{1}^{\rm H} \mathcal{E}_x + \sigma_{2}^{\rm H} \mathcal{E}_{x}^2 + \mathcal{O} \left( \mathcal{E}_x^3 \right).
\end{equation}
The linear Hall conductivity $\sigma_{1}^{\rm H}$ reads
\begin{align}
    \sigma_{1}^{\rm H} = - \frac{e^2}{\hbar} \oint_{\rm BZ} \Omega f_{\rm eq}\,\frac{d^2k}{\left( 2\pi \right)^2}.
\end{align}
At zero temperature, the momentum integral is taken below the Fermi energy.
In particular, when a band gap is open and the Fermi energy is inside it, the integral reduces to the Chern number $C_1$~\cite{Kane-review, Zhang-review, Ryu-review, Vishwanath-review}:
\begin{align}
    C_1 \coloneqq - \oint_{E \left( \bm{k} \right) < \mu} \Omega\,\frac{d^2k}{2\pi} \in \mathbb{Z}.
\end{align}
Then, the linear Hall conductivity is quantized to be
\begin{align}
    \sigma_{1}^{\rm H} = C_1 \frac{e^2}{2\pi \hbar} = C_1 \frac{e^2}{h}.
\end{align}

From Eq.~(\ref{aeq: Hall - Boltzmann}), the second-order nonlinear Hall conductivity $\sigma_{2}^{\rm H}$ reads
\begin{align}
    \sigma_2^{\rm H} = - \frac{e^{3} \tau}{\hbar^2} \oint_{\rm BZ} \Omega
    \frac{\partial f_{\rm eq}}{\partial k_x}
    \frac{d^2k}{\left( 2\pi \right)^2}.  
        \label{aeq: Boltzmann - Hall - second}
\end{align}
Since $\partial f_{\rm eq}/\partial \bm{k}$ takes a sharp peak on the Fermi energy, $\sigma_{2}^{\rm H}$ is roughly evaluated as the Berry curvature $\Omega \left( \bm{k} \right)$ on the Fermi surface.
Meanwhile, Eq.~(\ref{aeq: Boltzmann - Hall - second}) reduces to
\begin{equation}
    \sigma_2^{\rm H} = \frac{e^{3} \tau}{\hbar^2} \oint_{\rm BZ} \frac{\partial \Omega}{\partial k_x} f_{\rm eq}\,\frac{d^2k}{\left( 2\pi \right)^2}.
\end{equation}
Thus, $\sigma_2^{\rm H}$ is given as the dipole moment $\partial \Omega/\partial \bm{k}$ of the Berry curvature at equilibrium.

In general, the $n$th-order nonlinear Hall conductivity $\sigma_n^{\rm H}$ is given by the $\left( n-1 \right)$\,th derivative of the Berry curvature $\Omega \left( \bm{k} \right)$ at equilibrium.
In fact, we have
\begin{align}
    \sigma_n^{\rm H}
    &= - \frac{e^{n+1}\tau^{n-1}}{\hbar^n} \oint_{\rm BZ}
    \Omega 
    \frac{\partial^{n-1} f_{\rm eq}}{\partial k^{n-1}_{x}}
    \frac{d^2 k}{\left( 2\pi \right)^2} \nonumber \\
    &=\frac{e}{\tau} \left( - \frac{e\tau}{\hbar} \right)^{n} \oint_{\rm BZ} 
    \frac{\partial^{n-1}\Omega}{\partial k_x^{n-1}} 
    f_{\rm eq}\,\frac{d^2 k}{\left( 2\pi \right)^2}.
\end{align}

\section{Wave-packet approach of noise}
    \label{asec: noise}

We derive the general formula of noise in Eq.~(\ref{eq: noise - general}), using the wave-packet approach~\cite{Martin-92}.    
To characterize the fluctuations of the time-dependent current $I = I \left( t \right)$, we focus on the electronic waves in the infinitesimal energy range $\left[ E, E+dE \right]$ and the long time interval $\left[ -T/2, T/2 \right]$. 
We assume that the current consists of a collection of  $N\gg 1$ pulse waves:
\begin{equation}
    dI \left( t \right) = \sum_{n=1}^{N} g_n i \left( t- n\tau \right).
        \label{aeq: wave-packet approach}
\end{equation}
Here, $i$ describes the current induced by a pulse wave of a single electron, satisfying
\begin{equation}
    \int_{-T/2}^{T/2} i \left( t \right) dt = e.
\end{equation}
Moreover, $g_n$ denotes the integer that describes the transmission of the $n$th pulse wave: we have $g_n = 1$ for the transmitted pulse waves and $g_n = -1$ for the reflected pulse waves.
The current noise is represented by the statistical uncertainty of $g_n$.
We choose the time interval between pulse waves as
\begin{equation}
    \tau \coloneqq \frac{h}{dE} \ll T
\end{equation}
such that the number of the pulse waves is
\begin{equation}
    N = \frac{T}{\tau} = \frac{TdE}{h} \gg 1.
\end{equation}
Because of this choice of the pulse waves, each pulse wave is independent of each other.
As a result, we have
\begin{equation}
    \braket{g_m g_n} = \braket{g_m} \braket{g_n} +\,( \braket{g_n^2} - \braket{g_n}^2 )\,\delta_{mn},
\end{equation}
where the angle brackets denote the ensemble average.

It follows from Eq.~(\ref{eq: noise - def}) that the noise $dS$ due to the electronic waves in $\left[ E, E+dE \right]$ reads
\begin{align}
    dS &\coloneqq \lim_{T\to\infty} \frac{2}{T} \int_{-T/2}^{T/2} \int_{-T/2}^{T/2} dC \left( t, t' \right) dtdt'\nonumber \\
    &= \lim_{T\to\infty} \frac{2}{T} \left( \braket{dQ^2} - \braket{dQ}^2 \right),
\end{align}
where $dC$ is the correlation function of the current $dI$,
\begin{equation}
    dC \left( t, t' \right) \coloneqq \braket{dI \left( t \right) dI \left( t' \right)} - \braket{dI \left( t \right)} \braket{dI \left( t' \right)},
\end{equation}
and $dQ$ is the total electric charge,
\begin{equation}
    dQ \coloneqq \int_{-T/2}^{T/2} dI \left( t \right) dt.
\end{equation}
Now, using the representation in Eq.~(\ref{aeq: wave-packet approach}), we have
\begin{align}
    \braket{dQ} &= \int_{-T/2}^{T/2} \braket{dI \left( t \right)} dt \nonumber \\
    &= \sum_{n=1}^{N} \braket{g_n} \int_{-T/2}^{T/2} i \left( t- n\tau \right) dt \nonumber \\
    &= e N \braket{g_{n}} 
\end{align}
and
\begin{align}
    \braket{dQ^2} &= \int_{-T/2}^{T/2} \int_{-T/2}^{T/2} \braket{dI \left( t \right) dI \left( t' \right)} dt dt' \nonumber \\
    &= e^2 \sum_{m, n=1}^{N} \braket{g_m g_n} \nonumber \\
    &= e^2 \left[ \left( N \braket{g_n} \right)^2 + N \left( \braket{g_n^2} - \braket{g_n}^2 \right) \right],
\end{align}
resulting in
\begin{equation}
    dS = \frac{2e^2}{h} \left( \braket{g_n^2} - \braket{g_n}^2 \right) dE.
\end{equation}
Thus, the current noise $dS$ reduces to the variance of $g_n$. 

Similarly to the Landauer formula, the variance of $g_n$ is evaluated by the transmission probability $T = T \left( E \right)$, as well as the distribution functions $f_{\rm L}$ and $f_{\rm R}$ of the respective reservoirs at the left and right.
The probability of $g_n = 1$ is $f_{\rm L} \left( 1 - f_{\rm R} \right) T$, while the probability of $g_n = -1$ is $f_{\rm R} \left( 1 - f_{\rm L} \right) T$.
Using these facts, we have
\begin{align}
    \braket{g_n}
    &= f_{\rm L} \left( 1-f_{\rm R} \right) T - f_{\rm R} \left( 1 - f_{\rm L} \right) T  \nonumber \\
    &= \left( f_{\rm L} - f_{\rm R} \right) T, \\
    \braket{g_n^2} 
    &= f_{\rm L} \left( 1-f_{\rm R} \right) T + f_{\rm R} \left( 1 - f_{\rm L} \right) T,
\end{align}
which further leads to
\begin{align}
    &\braket{g_n^2} - \braket{g_n}^2 \nonumber \\
    &\quad= \left[ f_{\rm L} \left( 1-f_{\rm R} \right) T + f_{\rm R} \left( 1 - f_{\rm L} \right) T\right] -\left( f_{\rm L} - f_{\rm R} \right)^2 T^2 \nonumber \\
    &\quad=\left[ f_{\rm L} \left( 1- f_{\rm L} \right) + f_{\rm R} \left( 1-f_{\rm R} \right) \right] T^2 \nonumber \\
    &\qquad\quad + \left[ f_{\rm L} \left( 1- f_{\rm R} \right) + f_{\rm R} \left( 1-f_{\rm L} \right) \right] T \left( 1-T \right).
\end{align}
The sum of $dS$ for all electronic waves with arbitrary energy yields Eq.~(\ref{eq: noise - general}), which is the general formula of the current noise $S$.
We note in passing that the average of the infinitesimal current $dI$ reads
\begin{align}
    \overline{\braket{dI}}
    &= \sum_{n=1}^{N} \braket{g_n} \overline{i \left( t-n \tau \right)} \nonumber \\
    &= \frac{e}{h} T \left( f_{\rm L} - f_{\rm R} \right) dE,
\end{align}
which reproduces the Landauer formula in Eqs.~(\ref{eq: nonlinear conductance formula}) and (\ref{eq: nonlinear conductance formula - zero temperature}).

\section{Numerics of the conductance for lattice models}
    \label{asec: numerics - conductance}

We calculate the conductance for lattice models.
Let $\psi_n$ be the single-particle wave function at site $n$. 
In general, the Schr\"odinger equation reads
\begin{align}
    - J_{n-1, n} \psi_{n-1} 
    + V_n \psi_{n}
    - J_{n, n+1} \psi_{n+1} 
    = E \psi_{n}.
        \label{aeq: random Schrodinger}
\end{align}
Here, $V_n$ is the potential energy at site $n$, and $J_{n, n+1}$ is the hopping amplitude between sites $n$ and $n+1$.
The Anderson model in Eq.~(\ref{eq: Anderson model}) reduces to Eq.~(\ref{aeq: random Schrodinger}) for $J_{n, n+1} = J$, and the Dyson model in Eq.~(\ref{eq: Dyson model}) reduces to Eq.~(\ref{aeq: random Schrodinger}) for $V_n = 0$ and $J_{n, n+1} = J + \disorderedhopping$.
Equation~(\ref{aeq: random Schrodinger}) is equivalent to
\begin{equation}
    \begin{pmatrix}
    \psi_{n+1} \\ \psi_{n}
    \end{pmatrix} = M_n \begin{pmatrix}
    \psi_{n} \\ \psi_{n-1}
    \end{pmatrix}
\end{equation}
with
\begin{equation}
    M_n \coloneqq \begin{pmatrix}
    \left( V_n - E \right)/J_{n, n+1} 
    & - J_{n-1,n}/J_{n,n+1} \\
    1 & 0
    \end{pmatrix}.
\end{equation}
In this representation, the Schr\"odinger equation is viewed as the spatial evolution of the wave function $\left( \psi_n~\psi_{n-1} \right)^{T}$ through the system described by the transfer matrix $M_n$.

Now, we decompose the wave function $\psi_n$ as the superposition of a forward-traveling wave $\propto e^{\ii kn}$ and a backward-traveling wave $\propto e^{-\ii kn}$:
\begin{equation}
    \psi_{n} = c_{+} e^{\ii kn} + c_{-} e^{-\ii kn}
        \label{aeq: plane wave}
\end{equation}
with coefficients $c_{+}, c_{-} \in \mathbb{C}$. 
The wave number $k$ is defined by the energy dispersion of the clean system:
\begin{equation}
    E \eqqcolon - 2J \cos k.
\end{equation}
For $E$ inside the energy band (i.e., $\left| E \right| \leq 2 \left|J \right|$), the wave number $k$ is real-valued, and the two waves $c_+ e^{\ii kn}$ and $c_{-} e^{-\ii kn}$ propagate in the opposite directions; 
for $E$ outside the energy band (i.e., $\left| E \right| > 2 \left| J \right|$), $k$ is complex-valued, and the two waves are localized and cannot transfer energy or particles.
Equation~(\ref{aeq: plane wave}) is rewritten as
\begin{align}
    \begin{pmatrix}
    \psi_{n+1} \\ \psi_{n}
    \end{pmatrix} &= \begin{pmatrix}
    1 & 1 \\
    e^{-\ii k} & e^{\ii k}
    \end{pmatrix} \begin{pmatrix}
    c_{+} e^{\ii k \left( n+1 \right)} \\
    c_{-} e^{-\ii k \left( n+1 \right)}
    \end{pmatrix} \nonumber \\
    &\eqqcolon Q \begin{pmatrix}
    c_{+} e^{\ii k \left( n+1 \right)} \\
    c_{-} e^{-\ii k \left( n+1 \right)}
    \end{pmatrix},
\end{align}
which leads to
\begin{align}
    \begin{pmatrix}
    c_{+} e^{\ii k \left( L+1 \right)} \\
    c_{-} e^{-\ii k \left( L+1 \right)}
    \end{pmatrix} 
    &= Q^{-1} \begin{pmatrix}
    \psi_{L+1} \\ \psi_{L}
    \end{pmatrix} \nonumber \\
    &= Q^{-1} M_{L} \begin{pmatrix}
    \psi_{L} \\ \psi_{L-1}
    \end{pmatrix} \nonumber \\
    &= Q^{-1} M_{L} M_{L-1} \cdots M_{1} \begin{pmatrix}
    \psi_{1} \\ \psi_{0}
    \end{pmatrix} \nonumber \\
    &= Q^{-1} M_{L} M_{L-1} \cdots M_{1} Q \begin{pmatrix}
    c_{+} e^{\ii k} \\
    c_{-} e^{-\ii k}
    \end{pmatrix}.
\end{align}
Consequently, the transfer matrix $\mathcal{M}_{L}$ of the system with length $L$ is given as
\begin{equation}
    \mathcal{M}_L = Q^{-1} M_L M_{L-1} \cdots M_{1} Q.
\end{equation}
The transfer matrix $\mathcal{M}_{L}$ is related to the transmission amplitudes by Eq.~(\ref{aeq: transfer matrix}).
Thus, we obtain the transmission probability $T_{L}$ as
\begin{equation}
     T_{L} = \frac{1}{\left| \left[ \mathcal{M}_{L} \right]_{22} \right|^2}.
\end{equation}
In this manner, we calculate the transmission probability $T_{L}$ for a given lattice model and each energy $E$.
The direct applicability even in the presence of disorder is one of the advantages of the scattering formulation.
In addition, the conductance can be straightforwardly calculated even for finite systems, which is another advantage of the scattering formulation. 
These advantages contrast with other transport theories including the Kubo formula.
For the numerical results in Fig.~\ref{fig: disorder}, we take the ensemble average $\braket{T_{L}}$ for many samples.

\section{Random-matrix theory of quantum transport in disordered chains}
    \label{asec: DMPK}

We develop a random-matrix theory of quantum transport for disordered chains~\cite{Anderson-80, Dorokhov-82, Mello-88, Beenakker-review, *Beenakker-review-sc}.
We obtain the distribution of the transmission probability for disordered chains by the functional renormalization group equations, which further yields the linear and nonlinear conductance according to the Landauer formula.
This approach is valid for sufficiently weak and sufficiently complicated disorder.
The obtained probability distribution does not depend on specific details of systems but universally depends on symmetry.
In particular, chiral symmetry changes the universality classes, as demonstrated below.

\subsection{Standard class}
    \label{asec: DMPK - standard}

We begin with the Anderson model in Eq.~(\ref{eq: Anderson model}).
As described in Appendix~\ref{asec: numerics - conductance}, the transmission probability $T_L$ of the system with length $L$ is given as
\begin{equation}
     T_{L} = \frac{1}{\left| \left[ \mathcal{M}_{L} \right]_{22} \right|^2},
\end{equation}
where the transfer matrix $\mathcal{M}_{L}$ is defined as
\begin{equation}
    \mathcal{M}_L \coloneqq Q^{-1} M_L M_{L-1} \cdots M_{1} Q
        \label{aeq: DMPK - transfer}
\end{equation}
with 
\begin{align}
    Q &\coloneqq \begin{pmatrix}
    1 & 1 \\
    e^{-\ii k} & e^{\ii k}
    \end{pmatrix}, \label{aeq: DMPK - Q} \\
    M_n &\coloneqq \begin{pmatrix}
    \left( V_n - E \right)/J 
    & - 1 \\
    1 & 0
    \end{pmatrix}. \label{aeq: DMPK - Mn}
\end{align}
Here, the wave number $k$ is related to the energy $E$ by the dispersion relation
\begin{equation}
    E = - 2J \cos k.
\end{equation}
When the energy $E$ is outside the energy band (i.e., $\left| E \right| > 2 \left| J \right|$), the wave number $k$ is pure imaginary, and the transmission is exponentially suppressed.
In the following, we assume $\left| E \right| \leq 2 \left| J \right|$ and hence the real wave number $k \in \mathbb{R}$.

As described in Appendix~\ref{asec: scattering}, current conservation imposes a constraint on the transfer matrix $\mathcal{M}_{L}$.
In fact, current conservation leads to unitarity of the scattering matrix $S$, or equivalently, pseudo-unitarity of the transfer matrix:
\begin{align}
    \sigma_z \mathcal{M}_{L}^{\dag} \sigma_z^{-1} = \mathcal{M}_{L}^{-1}
\end{align}
with a Pauli matrix $\sigma_z$.
While this is a general constraint on the transfer matrix, we can also confirm it explicitly by the matrix representation in Eqs.~(\ref{aeq: DMPK - transfer}), (\ref{aeq: DMPK - Q}), and (\ref{aeq: DMPK - Mn}).
Because of pseudo-unitarity, we can perform the following polar decomposition of the transfer matrix $\mathcal{M}_{L}$~\cite{Horn-textbook}:
\begin{align}
    \mathcal{M}_{L} &= \begin{pmatrix}
    u_{L} & 0 \\
    0 & u'_{L}
    \end{pmatrix} \begin{pmatrix}
    \cosh x_{L} & \sinh x_{L} \\
    \sinh x_{L} & \cosh x_{L}
    \end{pmatrix} \begin{pmatrix}
    v_{L} & 0 \\
    0 & v'_{L}
    \end{pmatrix} \nonumber \\
    &
    = \begin{pmatrix}
    u_L v_L \cosh x_{L} & u_L v'_L \sinh x_{L} \\
    u'_L v_L \sinh x_{L} & u'_L v'_L \cosh x_{L}
    \end{pmatrix},
        \label{aeq: polar decomposition}
\end{align}
where $u_{L}, u'_{L}, v_{L}, v'_{L} \in \mathbb{C}$ are independent complex numbers satisfying
\begin{equation}
    \left| u_{L} \right|^2 = 
    \left| u'_{L} \right|^2 =
    \left| v_{L} \right|^2 =
    \left| v'_{L} \right|^2 = 1.
\end{equation}
These complex numbers describe the quantum phases of the transmission and reflection amplitudes.
On the other hand, the nonnegative number $x_{L} \geq 0 $ is related to the transmission probability by
\begin{align}
    T_{L} = \frac{1}{\cosh^2 x_{L}}.
\end{align}
This parametrization is useful for obtaining the functional renormalization group equations.

Additional symmetry can further impose a constraint on the transfer matrix $\mathcal{M}_{L}$~\cite{Beenakker-review}.
In particular, the Anderson model in Eq.~(\ref{eq: Anderson model}) respects time-reversal symmetry (see Appendix~\ref{asec: symmetry} for details about symmetry).
In the presence of time-reversal symmetry, the scattering matrix $S$ is required to satisfy
\begin{equation}
    S^T = S,
\end{equation}
and the transfer matrix $\mathcal{M}_{L}$ is required to satisfy
\begin{equation}
    \sigma_x \mathcal{M}_{L}^{*} \sigma_x^{-1} = \mathcal{M}_{L}.
\end{equation}
This constraint is equivalent to
\begin{equation}
    u'_{L} = u^{*}_{L},\quad
    v'_{L} = v^{*}_{L}.
        \label{aeq: DMPK - TRS}
\end{equation}
In terms of the scattering matrix $S$, it is also equivalent to $t_{\rm L} = t_{\rm R}$.
Notably, the constraints due to unitarity and time-reversal symmetry are applicable to arbitrary energy $E$, which contrast with chiral or particle-hole symmetry.

Now, we consider the incremental changes of the transmission probability $T_L$ and derive the probability distribution of the statistical variable $T_L$.
The transfer matrix $\mathcal{M}_{L+1}$ is related to $\mathcal{M}_{L}$ by
\begin{equation}
    \mathcal{M}_{L+1} = Q^{-1} M_{L+1} Q \mathcal{M}_{L}
\end{equation}
with 
\begin{align}
    &Q^{-1} M_{L+1} Q 
    = \begin{pmatrix}
    e^{\ii k} & 0 \\
    0 & e^{-\ii k}
    \end{pmatrix} + \frac{V_{L+1}}{2\ii J \sin k} \begin{pmatrix}
    e^{\ii k} & e^{\ii k} \\
    - e^{-\ii k} & - e^{-\ii k}
    \end{pmatrix}, 
\end{align}
which leads to
\begin{align}
    &\left[ \mathcal{M}_{L+1} \right]_{22}
    = v'_{L} \biggl[ u'_L \cosh x_L  \nonumber \\
    &\qquad \left.- \frac{V_{L+1}}{2\ii J\sin k} \left( u_L \sinh x_{L} + u'_{L} \cosh x_{L} \right) \right] e^{-\ii k}.
\end{align}
In the absence of disorder, we have $\left[ \mathcal{M}_{L} \right]_{22} = e^{-\ii kL}$ and hence the perfect transmission $T_{L} = 1/\left| \left[ \mathcal{M}_{L} \right]_{22} \right|^2 = 1$.
The disordered potential leads to scattering between the plane waves, which further results in Anderson localization.
For sufficiently weak disorder, we expand $x_{L+1}$ in terms of $V_{L+1}/J$:
\begin{align}
    x_{L+1} = x_{L} 
    + c_1 \left( \frac{V_{L+1}}{J}\right)
    + \frac{c_{2}}{2} \left( \frac{V_{L+1}}{J} \right)^2
    + \mathcal{O} \left( V_{L+1}^3 \right).
\end{align}
Using
\begin{align}
    &\cosh x_{L+1}
    = \cosh x_{L}
    + c_1 \sinh x_{L} \left( \frac{V_{L+1}}{J} \right) \nonumber \\
    &\quad+ \frac{c_1^2 \cosh x_{L} + c_2 \sinh x_{L}}{2} \left( \frac{V_{L+1}}{J} \right)^2
    + \mathcal{O} \left( V_{L+1}^3 \right),
\end{align}
we have
\begin{align}
    c_1 &= -\frac{\mathrm{Im} \left( u'_L/u_L \right)}{2 \sin k}, \\
    c_2 &= \frac{1}{2\sin^2 k} \left[ \frac{\left( \mathrm{Re} \left( u'_L/u_L \right) \right)^2}{\tanh \left( 2x_{L} \right)} + \mathrm{Re} \left( u'_L/u_L \right) \right].
\end{align}

We assume that the disordered potential is sufficiently complicated such that the quantum phases of the scattered waves are distributed in a completely random manner.
Under this assumption, the complex numbers $u_L$ and $u'_L$, which describe the quantum phases of the scattered waves, are distributed uniformly on the unit circle in the complex plane.
This fact yields
\begin{equation}
    \braket{\left( \mathrm{Re} \left( u'_L/u_L \right) \right)^2} 
    = \braket{\left( \mathrm{Im} \left( u'_L/u_L \right) \right)^2}
    = 1/2,
        \label{aeq: DMPK - average - standard}
\end{equation}
where the brackets denote the ensemble average.
Using this formula, we obtain the moments of the evolution $\Delta x \coloneqq x_{L+1} - x_{L}$ as
\begin{align}
    \braket{\Delta x} &= \frac{\braket{c_2} \braket{V_{n}^2}}{2J^2}
    = \frac{\braket{V_{n}^2}}{8J^2 \sin^2 k} \frac{1}{\tanh \left( 2x \right)}, \label{aeq: DMPK - standard - dx} \\
    \braket{\left( \Delta x \right)^2} &= \frac{\braket{c_1^2} \braket{V_{n}^2}}{J^2}
    = \frac{\braket{V_{n}^2}}{8J^2 \sin^2 k}, \label{aeq: DMPK - standard - dx2} \\
    \braket{\left( \Delta x \right)^n} &= 0 \qquad \left( n \geq 3 \right). \label{aeq: DMPK - standard - dx3}
\end{align}
These relations determine the probability distribution $P = P \left( x, L \right)$ of the statistical variable $x$.
In fact, from Eq.~(\ref{aeq: DMPK - standard - dx3}), $P \left( x, L \right)$ should obey the Fokker-Planck equation
\begin{equation}
    \frac{\partial}{\partial L} P
    = - \frac{\partial}{\partial x} \braket{\Delta x} P+ \frac{1}{2} \frac{\partial^2}{\partial x^2} \braket{\left( \Delta x \right)^2} P,
\end{equation}
which further reduces to
\begin{align}
    \frac{8 J^2 \sin^2 k}{\braket{V_n^{2}}} \frac{\partial}{\partial L} P
    = \frac{1}{2} \frac{\partial}{\partial x} \sinh \left( 2x \right) \frac{\partial}{\partial x} \frac{P}{\sinh \left( 2x \right)}
        \label{aeq: DMPK - standard - FP}
\end{align}
from Eqs.~(\ref{aeq: DMPK - standard - dx}) and (\ref{aeq: DMPK - standard - dx2}).
This Fokker-Planck equation depends solely on the single length scale defined as
\begin{equation}
    \xi \coloneqq \frac{8J^2 \sin^2 k}{\braket{V_n^2}}
    = \frac{2 \left( 4J^2 - E^2 \right)}{\braket{V_n^2}}.
\end{equation}
This is a manifestation of the one-parameter scaling~\cite{Abrahams-79}.
The probability distribution $P \left( x, L \right)$ contains all information about the transport properties of the disordered system.
The Fokker-Planck equation~(\ref{aeq: DMPK - standard - FP}) controls the behavior of $P \left( x, L \right)$.
Although we begin with the specific model in Eq.~(\ref{eq: Anderson model}), the Fokker-Planck equation~(\ref{aeq: DMPK - standard - FP}) is universal for one-dimensional disordered electron systems in the standard class.

While the Fokker-Planck equation~(\ref{aeq: DMPK - standard - FP}) is exactly solvable for arbitrary $L$~\cite{Abrikosov-81}, it is sufficient for our purposes to focus on the asymptotic behavior for $L \to \infty$.
In such a limit, we have $T \ll 1$ and hence $x \gg 1$.
As a result, Eq.~(\ref{aeq: DMPK - standard - FP}) is simplified to
\begin{align}
    \xi \frac{\partial P}{\partial L}
    \simeq - \frac{\partial P}{\partial x} + \frac{1}{2} \frac{\partial^2 P}{\partial x^2}.
\end{align}
This is the standard diffusion equation with a drift term, which is straightforwardly solved as
\begin{align}
    P \left( x, L \right)
    \simeq \frac{1}{\sqrt{2\pi L/\xi}} \exp \left[ - \frac{\left( x - L/\xi \right)^2}{2L/\xi} \right].
        \label{aeq: P - standard}
\end{align}
This solution satisfies the normalization condition $\int_{0}^{\infty} P \left( x, L \right) dx = 1$ in the limit $L \to \infty$.
Thus, the statistical variable $x$ obeys the normal distribution with the mean $L/\xi$ and the variance $L/\xi$.
Because of $T\simeq 4e^{-2x}$ in the limit $L \to \infty$, the transmission probability $T$ obeys the log-normal distribution.
Consequently, the typical value of the transmission probability $T$ is obtained as
\begin{equation}
    T_{\rm typ}
    \coloneqq e^{\braket{\log T}}
    \simeq 4 e^{-2\braket{x}}   
    = 4 e^{-2L/\xi}.
        \label{aeq: T-typ}
\end{equation}
Moreover, the average value of $T$ is obtained as
\begin{align}
    \braket{T}
    &\simeq 4\int_0^{\infty} e^{-2x} P \left( x, L \right) dx 
    = \sqrt{\frac{8\xi}{\pi L}} e^{-L/2\xi},
        \label{aeq: T-av}
\end{align}
leading to Eq.~(\ref{eq: Anderson - T - standard}).
The difference between the average and typical values is due to rare realizations of atypically large transmission probabilities.
In other words, it is a consequence of the broad distribution of the log-normal distribution.

\begin{figure*}[t]
\centering
\includegraphics[width=144mm]{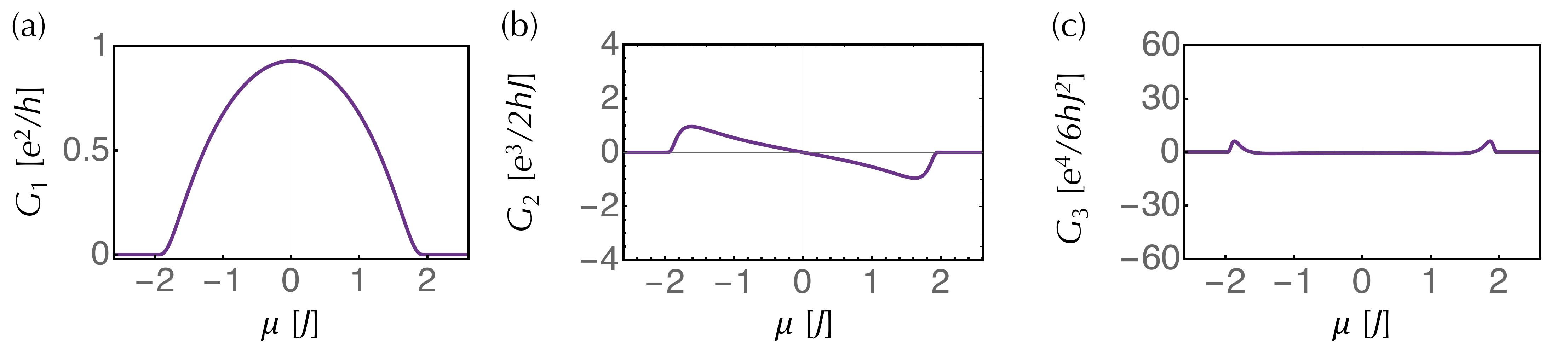}
\caption{Linear and nonlinear conductance of the disordered chain in the standard class ($L =50$, $J=1.0$, $W = 1.0$). The formula in Eq.~(\ref{eq: Anderson - T - standard}) based on the random-matrix approach is used. (a)~The linear conductance $G_1$, (b)~the second-order nonlinear conductance $G_2$, and (c)~the third-order nonlinear conductance $G_3$ as functions of the chemical potential $\mu$.}
	\label{fig: DMPK}
\end{figure*}

On the basis of the Landauer formula in Eq.~(\ref{eq: nonlinear conductance formula - zero temperature}), we calculate the linear and nonlinear conductance from the average conductance $\braket{T} = \braket{T} \left( E \right)$, as shown in Fig.~\ref{fig: DMPK}.
The transmission probability gets largest at the band center $E = 0$ and decreases away from the band center.
Consequently, the nonlinear conductance grows near the band edges.
This behavior is qualitatively consistent with the numerical results in Fig.~\ref{fig: disorder}.
The quantitative difference is due to a finite-size effect.
Although Eq.~(\ref{eq: Anderson - T - standard}) assumes $L \gg \xi$, this assumption is not applicable for the parameters in Figs.~\ref{fig: disorder} and \ref{fig: DMPK}.

We note in passing that the above results imply that the mean free path $\ell$ is comparable with the localization length $\xi$ in one dimension.
Physically, this means that the localization occurs after a couple of scattering events.
As discussed in Sec.~\ref{sec: Boltzmann}, the semiclassical Boltzmann equation assumes that the mean free path $\ell$ is much larger than the Fermi wave length $\lambda$.
The above results show that this assumption and the concomitant Boltzmann equation actually break down in one dimension.

\subsection{Chiral class}
    \label{asec: DMPK - chiral}

We next consider the Dyson model in Eq.~(\ref{eq: Dyson model}).
In contrast with the Anderson model in Eq.~(\ref{eq: Anderson model}), the Dyson model respects chiral symmetry (see also Appendix~\ref{asec: symmetry} for details about symmetry)~\cite{Beenakker-review-sc}.
In the presence of chiral symmetry, the scattering matrix $S$ is required to satisfy
\begin{equation}
    S^{\dag} \left( E \right) = S \left( - E \right),
\end{equation}
and the transfer matrix $\mathcal{M}_{L}$ is required to satisfy
\begin{equation}
    \sigma_x \mathcal{M}_{L} \left( E \right) \sigma_x^{-1} = \mathcal{M}_{L} \left( - E \right).
\end{equation}
This constraint is equivalent to
\begin{align}
    u'_{L} \left( E \right) = u_{L} \left( - E \right),\quad
    v'_{L} \left( E \right) = v_{L} \left( - E \right).
\end{align}
While time-reversal symmetry imposes a constraint on each eigenmode, chiral symmetry imposes a constraint on each pair of eigenmodes with opposite eigenenergy.
Exceptionally, eigenmodes with zero energy $E=0$ are subject to the special constraint due to chiral symmetry.
In fact, for $E=0$, chiral symmetry leads to $u'_L = u_L$ and $v'_L = v_L$, both of which are real in the presence of additional time-reversal symmetry [see Eq.~(\ref{aeq: DMPK - TRS})].
This constraint changes the universality class of Anderson localization and enables delocalization even in one dimension.
For example, chiral symmetry invalidates the formula in Eq.~(\ref{aeq: DMPK - average - standard}), which is crucial for the derivation of the Fokker-Planck equation~(\ref{aeq: DMPK - standard - FP}) in the standard class.
To consider the new universality class due to chiral symmetry, we focus on zero modes (i.e., $E=0$) in the following.

Because of chiral symmetry, the polar decomposition in Eq.~(\ref{aeq: polar decomposition}) is simplified to
\begin{align}
    \mathcal{M}_{L} 
    &= u_{L} v_{L} \begin{pmatrix}
    \cosh x_{L} & \sinh x_{L} \\
    \sinh x_{L} & \cosh x_{L} 
    \end{pmatrix} \nonumber \\
    &= u_{L} v_{L} e^{x_{L} \sigma_x}.
\end{align}
In contrast with the standard class, we allow the statistical variable $x \in \mathbb{R}$ to be negative.
Furthermore, for the zero modes of the Dyson model in Eq.~(\ref{eq: Dyson model}), the transfer matrix $M_n$ reads
\begin{align}
    M_n &= \begin{pmatrix}
    0 & - \left( J + \Delta J_{n-1, n} \right)/\left( J + \Delta J_{n, n+1} \right) \\
    1 & 0
    \end{pmatrix}
\end{align}
with the disordered hopping amplitude $\Delta J_{n, n+1}$. 
In a manner similar to the standard class, we assume that the disordered hopping is sufficiently weak and sufficiently complicated.
Then, the moments of the evolution $\Delta x \coloneqq x_{L+1} - x_{L}$ of the statistical variable $x_{L}$ are obtained as
\begin{align}
    \braket{\Delta x} &= 0, \\
    \braket{\left( \Delta x \right)^2} &= \frac{\braket{\left( \Delta J_{n, n+1} \right)^2}}{J^2}, \\
    \braket{\left( \Delta x \right)^n} &= 0 \qquad \left( n \geq 3 \right).
\end{align}
Notably, the first moment vanishes, which contrasts with the standard class [see Eq.~(\ref{aeq: DMPK - standard - dx}) for comparison].
From these moments, the Fokker-Planck equation that describes the probability distribution $P = P \left( x, L \right)$ of the statistical variable $x$ reads
\begin{equation}
    \frac{\partial P}{\partial L}
    = \frac{1}{2\ell} \frac{\partial^2 P}{\partial x^2}
        \label{aeq: DMPK - chiral - FP}
\end{equation}
with the relevant length scale (i.e., mean-free path)
\begin{align}
    \ell \coloneqq \frac{J^2}{\braket{\left( \Delta J_{n, n+1} \right)^2}}.
\end{align}
Equation~(\ref{aeq: DMPK - chiral - FP}) is clearly different from the Fokker-Planck equation~(\ref{aeq: DMPK - standard - FP}) for the standard class.
Still, Eq.~(\ref{aeq: DMPK - chiral - FP}) depends solely on the single parameter $L/\ell$, manifesting the one-parameter scaling~\cite{Abrahams-79} also in the chiral class.

The diffusion equation~(\ref{aeq: DMPK - chiral - FP}) is solved as
\begin{align}
    P \left( x, L \right)
    = \frac{1}{\sqrt{2\pi L/\ell}}
    \exp \left( - \frac{x^2}{2L/\ell} \right)
\end{align}
under the initial condition $P \left( x, L=0 \right) = \delta \left( x \right)$ and the normalization condition $\int_{-\infty}^{\infty} P \left( x, L \right) dx = 1$.
In contrast to the probability distribution for the standard class [Eq.~(\ref{aeq: P - standard})], the statistical variable $x$ is most probable for $x=0$.
From the obtained probability distribution, the average conductance is 
\begin{align}
    \braket{T}
    = \int_{-\infty}^{\infty} \frac{P \left( x, L \right)}{\cosh^2 x} dx
    \sim \sqrt{\frac{2}{\pi L/\ell}}
\end{align}
in the limit $L \to \infty$.
On the other hand, the typical conductance is 
\begin{align}
    T_{\rm typ}
    \coloneqq e^{\braket{\log T}}
    \sim e^{- \sqrt{8L/\pi \ell}}
\end{align}
in the limit $L \to \infty$.

This universal behavior is different from the behavior for the standard class in Eqs.~(\ref{aeq: T-typ}) and (\ref{aeq: T-av}) and originates from chiral symmetry.
Chiral-symmetry-breaking perturbations change the universality class and replace Eq.~(\ref{aeq: DMPK - chiral - FP}) with the Fokker-Planck equation~(\ref{aeq: DMPK - standard - FP}) for the standard class.
Similarly, away from zero energy, chiral symmetry is no longer relevant.
Then, the universality class reduces to the standard class, and the Fokker-Planck equation~(\ref{aeq: DMPK - standard - FP}) follows.
Consequently, the nonlinear conductance, which is obtained by the derivatives of the transmission probability, gets singularly large at zero energy, as shown in Fig.~\ref{fig: disorder}.
In general, the crossover between the standard class and the chiral class is difficult to analyze exactly (but see Ref.~\cite{Ryu-04}).
We also note that the above results apply to quasi-one-dimensional systems with an odd number of channels; by contrast, zero modes of quasi-one-dimensional systems with an even number of channels never exhibit delocalization even in the presence of chiral symmetry~\cite{Brouwer-98, Brouwer-00}.

\section{Symmetry classification}
    \label{asec: symmetry}

We summarize the tenfold internal-symmetry class for noninteracting fermionic systems~\cite{AZ-97, Evers-review, Beenakker-review-sc, Ryu-review}.
We consider a generic noninteracting fermionic system described by the Hermitian Hamiltonian
\begin{align}
    \hat{H} = \sum_{m, n} H_{m, n} \hat{c}_{m}^{\dag} \hat{c}_{n}. 
        \label{aeq: free fermion}
\end{align}
Here, $\hat{c}_{n}$ ($\hat{c}_{n}^{\dag}$) annihilates (creates) a fermion at site $n$, satisfying the canonical anticommutation relations $\{ \hat{c}_{m}, \hat{c}_{n}^{\dag} \} = \delta_{m, n}$.
The indices $n$ describe the lattice sites, as well as possible internal degrees of freedom such as the spin degree of freedom.
The Hermitian matrix $H = \left( H_{m, n} \right)_{m, n}$ is the single-particle Hamiltonian.
While we discuss normal fermionic systems in Eq.~(\ref{aeq: free fermion}) in the following, the discussions can be straightforwardly generalized to Bogoliubov-de Gennes Hamiltonians for superconductors by using the Nambu spinors instead of the complex fermion operators.

We begin with unitary symmetry that does not mix fermion annihilation and creation operators.
We introduce a symmetry transformation by
\begin{align}
    \hat{c}_{m}
    \rightarrow 
    \hat{c}'_{m}
    \coloneqq \hat{\cal U} \hat{c}_{m} \hat{\cal U}^{-1}
    = \sum_{n} U_{m, n} \hat{c}_{n}.
\end{align}
Here, $\hat{\cal U}$ is a unitary operator that acts on the fermionic Fock space, while $U = \left( U_{m, n} \right)_{m, n}$ is a unitary matrix instead of a second-quantized operator.
Because of unitarity of $\hat{\cal U}$, the canonical anticommutation relations are preserved under the symmetry transformation:
\begin{align}
    \left\{ \hat{c}_{m}, \hat{c}_{n}^{\dag} \right\}
    = \hat{\cal U} \left\{ \hat{c}_{m}, \hat{c}_{n}^{\dag} \right\} \hat{\cal U}^{-1}.
\end{align}
Symmetry of the system is described by the invariance of the Hamiltonian $\hat{H}$ under the symmetry operation $\hat{\cal U}$:
\begin{align}
    \hat{\cal U} \hat{H} \hat{\cal U}^{-1} = \hat{H}, 
\end{align}
which is equivalent to 
\begin{align}
    U^{-1} H U = H
\end{align}
for the single-particle Hamiltonian $H$.
The unitary operation $\hat{\cal U}$ is internal when it acts only on the internal degrees of freedom and does not act on the spatial degrees of freedom.
Such internal symmetry is relevant to disordered electron systems and characterizes the universality classes of Anderson localization since only internal symmetry survives in the presence of disorder.
We note that the tenfold symmetry classification~\cite{AZ-97} does not include the unitary symmetry that commutes with single-particle Hamiltonians.
This is because the Hamiltonian is block diagonalized in a trivial manner in the presence of such unitary symmetry.

Time-reversal symmetry is described by the antiunitary operation defined by
\begin{align}
    \hat{\cal T} \hat{c}_{m} \hat{\cal T}^{-1}
    = \sum_{n} T_{m, n} \hat{c}_{n}
\end{align}
and 
\begin{align}
    \forall\,z\in\mathbb{C}
    \quad
    \hat{\cal T}z\hat{\cal T}^{-1} = z^{*}.
\end{align}
Here, $\hat{\cal T}$ is an antiunitary operator that acts on the fermionic Fock space, while $T = \left( T_{m, n} \right)_{m, n}$ is a unitary matrix.
A system respects time-reversal invariance if the Hamiltonian $\hat{H}$ satisfies
\begin{align}
    \hat{\cal T} \hat{H} \hat{\cal T}^{-1}
    = \hat{H}.
\end{align}
In fact, if this relation is satisfied, we have
\begin{align}
    \hat{\cal T} \hat{O} \left( t \right) \hat{\cal T}^{-1}
    = \hat{O} \left( -t \right),
\end{align}
where $\hat{O} \left( t \right) = e^{\ii \hat{H} t} \hat{O} e^{-\ii \hat{H} t}$ is the time-evolved operator of a fermionic operator $\hat{O}$. 
In terms of the single-particle Hamiltonian $H$, time-reversal invariance is equivalent to
\begin{align}
    T^{-1} H^{*} T = H.
\end{align}
Because of antiunitarity of time-reversal symmetry, the symmetry operator and matrix are required to satisfy
\begin{align}
    \hat{\cal T}^{2} = \left( \pm 1 \right)^{\hat{N}},\quad
    T^{*} T = \pm 1
\end{align}
with the number operator $\hat{N} \coloneqq \sum_{n} \hat{c}_{n}^{\dag} \hat{c}_{n}$.
The signs in these equations correspond to the signs of time-reversal symmetry in Table~\ref{tab: AZ}.
For $\hat{\cal T}^2 = -1$, time-reversal symmetry leads to the Kramers degeneracy.
Generally, time-reversal symmetry with $T^{*} T = +1$ enhances Anderson localization~\cite{Gorkov-79, Altshuler-80}, while time-reversal symmetry with $T^{*} T = -1$ suppresses Anderson localization~\cite{Hikami-80}.
The disordered electron system in Eq.~(\ref{eq: Anderson model}) respects time-reversal symmetry with $T = 1$.

Particle-hole symmetry (or equivalently, charge-conjugation symmetry) is described by the unitary operation defined by
\begin{align}
    \hat{\cal C} \hat{c}_{m} \hat{\cal C}^{-1}
    = \sum_{n} C^{*}_{m, n} \hat{c}_{n}^{\dag},
\end{align}
where $\hat{\cal C}$ and $C = \left( C_{m, n} \right)_{m, n}$ are unitary operators and matrices, respectively.
In contrast to time-reversal symmetry, this operation mixes fermion annihilation and creation operators.
It describes the transformation between particles and holes, and flips the sign of the electron charge with respect to the charge neutral point:
\begin{align}
    \hat{\cal C} \hat{Q} \hat{\cal C}^{-1}
    = - \hat{Q}
\end{align}
with $\hat{Q} \coloneqq \hat{N} - N/2$.
The Hamiltonian is particle-hole symmetric if it satisfies
\begin{align}
    \hat{\cal C} \hat{H} \hat{\cal C}^{-1}
    = \hat{H},
\end{align}
which leads to $\mathrm{tr}\,H = 0$ and
\begin{align}
    C^{-1} H^{T} C = - H.
\end{align}
Particle-hole symmetry acts as unitary symmetry on the fermionic Fock space but acts as antiunitary symmetry on the single-particle Hilbert space.
Similarly to time-reversal symmetry, the symmetry operator and matrix are required to satisfy
\begin{align}
    \hat{\cal C}^{2} = \left( \pm 1 \right)^{\hat{N}},\quad
    C^{*} C = \pm 1.
\end{align}
In the presence of particle-hole symmetry, eigenenergy appears in opposite-sign pairs $\left( E, -E \right)$; zero-energy modes are subject to a special constraint.
For $C^{*} C = +1$, zero modes remain to be delocalized even in one-dimensional disordered systems~\cite{Brouwer-00}.

Finally, chiral symmetry (or equivalently, sublattice symmetry) is defined by the antiunitary operation defined by
\begin{align}
    \hat{\cal S} \hat{c}_{m} \hat{\cal S}^{-1}
    = \sum_{n} S_{m, n} \hat{c}_{n}^{\dag},
\end{align}
where $\hat{\cal S}$ is an antiunitary operator on the fermionic Fock space, and $S = \left( S_{m, n} \right)_{m, n}$ is a unitary matrix on the single-particle Hilbert space.
In the simultaneous presence of time-reversal symmetry and particle-hole symmetry, chiral symmetry appears as a combination of the two symmetry.
Even in the absence of time-reversal symmetry and particle-hole symmetry, chiral symmetry is respected, for example, in bipartite hopping models.
The system respects chiral symmetry if the Hamiltonian satisfies 
\begin{align}
    \hat{\cal S} \hat{H} \hat{\cal S}^{-1}
    = \hat{H},
\end{align}
which leads to $\mathrm{tr}\,H = 0$ and
\begin{align}
    S^{-1} H S = - H.
\end{align}
The matrix $S$ can be chosen to be Hermitian and satisfy $S^2 = 1$ without loss of generality.
Similarly to particle-hole symmetry, chiral symmetry imposes a special constraint on zero modes, which results in delocalization even in one-dimensional disordered systems~\cite{Dyson-53, Stone-81A, Stone-81B, Brouwer-98}.
The disordered electron system in Eq.~(\ref{eq: Dyson model}) respects chiral symmetry with $S_{m, n} = \left( -1 \right)^{m} \delta_{m, n}$.

\section{Scattering theory of graphene}
    \label{asec: graphene}

We consider the scattering problem for the two-dimensional Dirac Hamiltonian
\begin{equation}
    H \left( k_x, k_y \right) = \hbar v \left( k_x \sigma_x + k_y \sigma_y \right)
        \label{aeq: graphene}
\end{equation}
with $v > 0$.
This Hamiltonian describes graphene~\cite{graphene-NG-review, graphene-DasSarma-review}, as well as a surface mode of the three-dimensional topological insulator~\cite{Fu-Kane-Mele-07, Moore-Balents-07, Roy-09}.
Let $L_x$ and $L_y$ be the lengths of the system along the $x$ and $y$ directions, respectively.
Along the $x$ direction, the system lies in $0 \leq x \leq L_x$ and is connected with the two leads in $x \leq 0$ and $L_x \leq x$.
The two leads are assumed to be graphene with a large potential $V_0$.
In the limit $\left| V_0 \right| \to \infty$, an infinite number of modes propagate in the leads. 
Along the $y$ direction, on the other hand, we impose a certain boundary condition and use the Fourier representation with the wave number $k_y$.
Such boundary conditions include the smooth edge and metallic armchair edge.
We later take the limit $L_y \to \infty$, for which the effect of the boundary conditions is irrelevant.

\subsection{Derivation of the transmission probability}
    \label{asec: graphene - derivation}

The transmission probability is analytically obtained in a manner similar to Refs.~\cite{Katsnelson-06, Tworzydlo-06}.
We begin with the eigenvalue problem of the two-dimensional Dirac Hamiltonian in Eq.~(\ref{aeq: graphene}):
\begin{align}
    \hbar v \left( - \ii \partial_x \sigma_x - \ii \partial_y \sigma_y \right) \psi \left( x, y \right) + V_0 \psi \left( x, y \right) = E \psi \left( x, y \right).
\end{align}
The eigenenergy is obtained as
\begin{equation}
E = \pm \hbar v \sqrt{k_x^2 + k_y^2} + V_0,
\end{equation}
and the corresponding eigenstate is 
\begin{align}
\psi \left( x, y \right)
= e^{\ii k_x x +\ii k_y y} \begin{pmatrix}
1 \\ \theta \left( k_x, k_y \right)
\end{pmatrix}
\end{align}
with 
\begin{align}
\theta \left( k_x, k_y \right) \coloneqq \pm \frac{k_x + \ii k_y}{\sqrt{k_x^2 + k_y^2}}.
\end{align}
We note that $\theta \left( k_x, k_y \right)$ satisfies
\begin{equation}
    \theta \left( k_x, k_y \right) \theta \left( - k_x, k_y \right) = -1.
        \label{eq: kx - kx}
\end{equation}
In the following, we fix $E$ and define $k_x$ and $k_0$ as
\begin{align}
    k_x &\coloneqq \sqrt{\left( \frac{E}{\hbar v}\right)^2 - k_y^2}, \label{aeq: graphene - kx} \\
    k_0 &\coloneqq \sqrt{\left( \frac{E-V_0}{\hbar v}\right)^2 - k_y^2}.
\end{align}
For $\left| V_0 \right| \to \infty$, we have $k_0 \to \infty$ and $\left| \theta \left( k_0, k_y \right) \right| \to \pm 1$.

\begin{figure}[t]
\centering
\includegraphics[width=86mm]{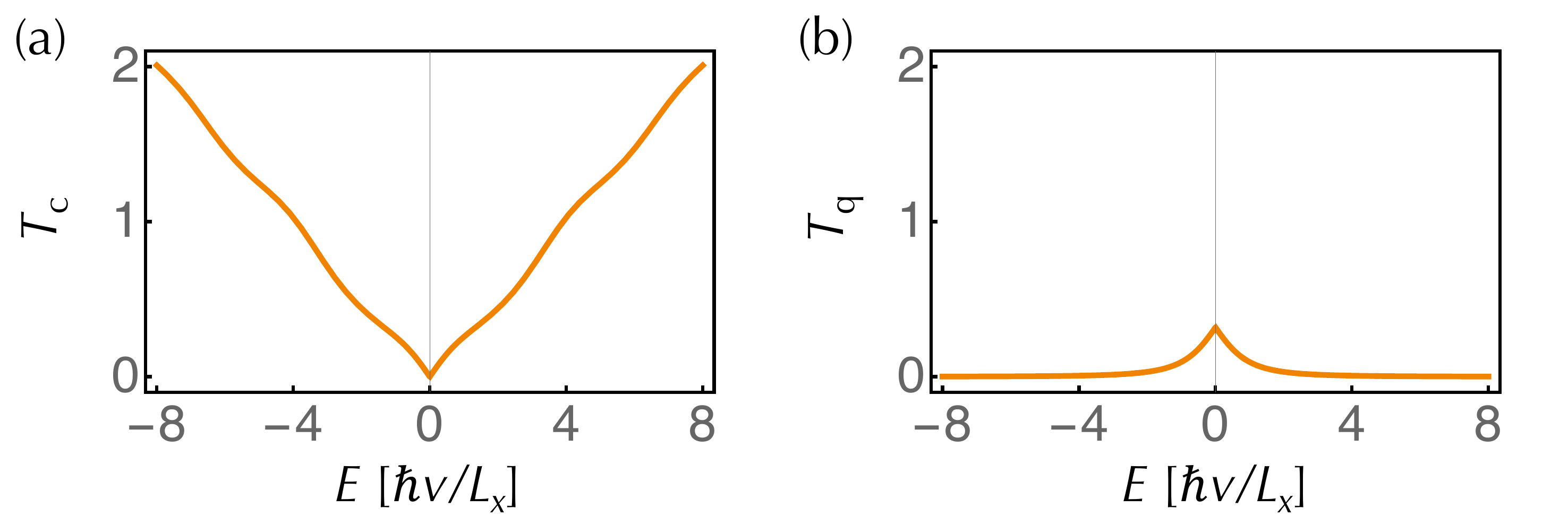} 
\caption{Transmission probability of the two-dimensional Dirac Hamiltonian in Eq.~(\ref{aeq: graphene}) as a function of energy $E$. (a)~Classical transmission probability $T_{\rm c}$. (b)~Quantum transmission (tunneling) probability $T_{\rm q}$.}
	\label{fig: graphene-appendix}
\end{figure}

\begin{widetext}
Now, suppose that the wave function is given as
\begin{align}
\psi \left( x, y \right)
= \left\{ \begin{array}{lc}
e^{\ii k_0 x + \ii k_y y} \begin{pmatrix} 1 \\ \theta \left( k_0, k_y \right)\end{pmatrix}
+ r \left( k_y \right) e^{-\ii k_0 x + \ii k_y y} \begin{pmatrix} 1 \\ \theta \left( -k_0, k_y \right)\end{pmatrix} & \left( x \leq 0 \right); \\
a \left( k_y \right) e^{\ii k_x x + \ii k_y y} \begin{pmatrix} 1 \\ \theta \left( k_x, k_y \right)\end{pmatrix}
+ b \left( k_y \right) e^{-\ii k_x x + \ii k_y y} \begin{pmatrix} 1 \\ \theta \left( -k_x, k_y \right)\end{pmatrix}\quad 
& \left( 0 \leq x \leq L_x \right); \\
t \left( k_y \right) e^{\ii k_0 \left( x-L_x \right) + \ii k_y y} \begin{pmatrix} 1 \\ \theta \left( k_0, k_y \right)\end{pmatrix} & \left( x \geq L_x\right),  \\
\end{array} \right.
\end{align}
for the wave number $k_y$ along the $y$ direction.
The boundary conditions at $x = 0$ reduce to
\begin{align}
&1 + r \left( k_y \right) = a \left( k_y \right) + b \left( k_y \right), \label{eq: x0 - 1} \\
&\theta \left( k_0, k_y \right) + r \left( k_y \right) \theta \left( - k_0, k_y \right)
= a \left( k_y \right) \theta \left( k_x, k_y \right) + b \left( k_y \right) \theta \left( - k_x, k_y \right),        \label{eq: x0 - 2} 
\end{align}
and the boundary conditions at $x=L_x$ reduce to
\begin{align}
&a \left( k_y \right) e^{\ii k_x L_x} + b \left( k_y \right) e^{-\ii k_x L_x} 
= t \left( k_y \right), \label{eq: xL - 1} \\
&a \left( k_y \right) e^{\ii k_x L_x} \theta \left( k_x, k_y \right) + b \left( k_y \right) e^{-\ii k_x L_x} \theta \left( -k_x, k_y \right)
= t \left( k_y \right) \theta \left( k_0, k_y \right).         \label{eq: xL - 2} 
\end{align}
From Eqs.~(\ref{eq: x0 - 1}) and (\ref{eq: x0 - 2}), we have
\begin{align}
a \left( k_y \right) \left[ \theta \left( k_x, k_y \right) - \theta \left( -k_0, k_y \right) \right] + b \left( k_y \right) \left[ \theta \left( -k_x, k_y \right) - \theta \left( -k_0, k_y \right) \right]
= \theta \left( k_0, k_y \right) - \theta \left( -k_0, k_y \right).
\end{align}
In addition, from Eqs.~(\ref{eq: xL - 1}) and (\ref{eq: xL - 2}), we have
\begin{align}
a \left( k_y \right) e^{\ii k_x L_x} \left[ \theta \left( k_x, k_y \right) - \theta \left( k_0, k_y \right) \right]
+ b \left( k_y \right) e^{-\ii k_x L_x} \left[ \theta \left( -k_x, k_y \right) - \theta \left( k_0, k_y \right) \right] = 0.
\end{align}
Combining these equations with Eq.~(\ref{eq: kx - kx}), we have
\begin{align}
a \left( k_y \right) &= \frac{\left( 1 +\theta^2 \left( k_0, k_y \right)\right) \left( 1 + \theta \left( k_x, k_y \right) \theta \left( k_0, k_y \right)\right) e^{-\ii k_x L_x}}{e^{\ii k_x L_x} \left( \theta \left( k_x, k_y \right) - \theta \left( k_0, k_y \right) \right)^2 + e^{-\ii k_x L_x} \left( 1 + \theta \left( k_x, k_y \right) \theta \left( k_0, k_y \right) \right)^2}, \\
b \left( k_y \right) &= \frac{\left( 1 +\theta^2 \left( k_0, k_y \right)\right) \theta \left( k_x, k_y \right) \left( \theta \left( k_x, k_y \right) - \theta \left( k_0, k_y \right)\right) e^{\ii k_x L_x}}{e^{\ii k_x L_x} \left( \theta \left( k_x, k_y \right) - \theta \left( k_0, k_y \right) \right)^2 + e^{-\ii k_x L_x} \left( 1 + \theta \left( k_x, k_y \right) \theta \left( k_0, k_y \right) \right)^2},
\end{align}
and
\begin{align}
t \left( k_y \right) = \frac{\left( 1 +\theta^2 \left( k_0, k_y \right)\right) \left( 1 +\theta^2 \left( k_x, k_y \right)\right)}{e^{\ii k_x L_x} \left( \theta \left( k_x, k_y \right) - \theta \left( k_0, k_y \right) \right)^2 + e^{-\ii k_x L_x} \left( 1 + \theta \left( k_x, k_y \right) \theta \left( k_0, k_y \right) \right)^2}.
\end{align}
In the limit $V_0 \to \infty$, we have $\theta \left( k_0, k_y \right) \to -1$ and hence
\begin{align}
t \left( k_y \right) &\to \frac{2\left( 1 +\theta^2 \left( k_x, k_y \right)\right)}{e^{\ii k_x L_x} \left( 1+ \theta \left( k_x, k_y \right) \right)^2 + e^{-\ii k_x L_x} \left( 1 - \theta \left( k_x, k_y \right) \right)^2} \nonumber \\
&= \frac{1 +\theta^2 \left( k_x, k_y \right)}{\left( 1 +\theta^2 \left( k_x, k_y \right)\right) \cos k_x L_x + 2\ii \theta \left( k_x, k_y \right) \sin k_x L_x}.
\end{align}
Since we have
\begin{align}
\theta \left( k_x, k_y \right)
+ \theta^{-1} \left( k_x, k_y \right)
&= \theta \left( k_x, k_y \right) - \theta \left( -k_x, k_y \right)
= \frac{2k_x}{E/\hbar v},
\end{align}
the transmission amplitude $t \left( k_y \right)$ reduces to
\begin{equation}
t \left( k_y \right) = \frac{k_x}{k_x \cos k_x L_x + \ii \left( E/\hbar v \right) \sin k_x L_x},
\end{equation}
Thus, the transmission probability $T \left( k_y \right)$ is obtained as 
\begin{align}
T \left( k_y \right) = \left| t \left( k_y \right) \right|^2
= \left| \frac{k_x}{k_x \cos k_x L_x + \ii \left( E/\hbar v \right) \sin k_x L_x} \right|^2.
\end{align}
Here, the wave number $k_x$ along the $x$ direction is given as Eq.~(\ref{aeq: graphene - kx}) by the energy $E$ and the wave number $k_y$ along the $y$ direction.

Then, we consider all the modes along the $y$ direction.
In the limit $L_y \to \infty$, we have
\begin{align}
T = \sum_{k_y} T \left( k_y \right)
\to \frac{L_y}{\pi} \int_0^{\infty} dk_y~T \left( k_y \right).
\end{align}
Notably, the transmission probability $T \left( k_y \right)$ behaves differently depending on whether $k_x$ is real-valued or not.
For $0 \leq k_y \leq \left| E\right|/\hbar v$,
we have $k_x \in \mathbb{R}$ and hence
\begin{align}
T \left( k_y \right)
= \frac{1}{\cos^2 k_x L_x + \left( E/\hbar v k_x \right)^2 \sin^2 k_x L_x}.
\end{align}
For $k_y \geq \left| E\right|/\hbar v$, on the other hand,  we have $k_x \in \ii \mathbb{R}$ and hence
\begin{align}
T \left( k_y \right)
= \frac{1}{\cosh^2 \kappa_x L_x + \left( E/\hbar v \kappa_x \right)^2 \sinh^2 \kappa_x L_x}
\end{align}
with $\kappa_x \coloneqq \sqrt{k_y^2 - \left( E/\hbar v\right)^2}$.
Thus, the transmission probability $T$ consists of the following two contributions (i.e., $T = T_{\rm c} + T_{\rm q}$):
\begin{align}
T_{\rm c} &\coloneqq \frac{L_y}{\pi} \int_0^{\left| E \right|/\hbar v} \frac{dk_y}{\cos^2 k_x L_x + \left( E/\hbar v k_x \right)^2 \sin^2 k_x L_x}, \\
T_{\rm q} &\coloneqq \frac{L_y}{\pi} \int_{\left| E \right|/\hbar v}^{\infty} \frac{dk_y}{\cosh^2 \kappa_x L_x + \left( E/\hbar v \kappa_x \right)^2 \sinh^2 \kappa_x L_x}.
\end{align}
Introducing
\begin{equation}
x \coloneqq k_x \left( k_y \right) L_x, \kappa_x \left( k_y\right) L_x,
\end{equation}
we have
\begin{equation}
\frac{dx}{dk_y}
= - \frac{k_y L_x^2}{x}, \frac{k_y L_x^2}{x},
\end{equation}
and hence
\begin{align}
T_{\rm c} &= \frac{L_y}{\pi L_x} \int_{0}^{\left| E \right| L_x /\hbar v} \frac{dx}{\sqrt{\left( EL_x/\hbar vx\right)^2 - 1} \left( \cos^2 x + \left( EL_x/\hbar v x\right)^2 \sin^2 x \right)}
\eqqcolon \frac{L_y}{\pi L_x} \int_{0}^{\left| E \right| L_x /\hbar v} T_{\rm c} \left( x \right) dx,   \label{aeq: graphene - Tc} \\
T_{\rm q} &= \frac{L_y}{\pi L_x} \int_{0}^{\infty} \frac{dx}{\sqrt{\left( EL_x/\hbar vx\right)^2 + 1} \left( \cosh^2 x + \left( EL_x/\hbar v x\right)^2 \sinh^2 x \right)}
\eqqcolon \frac{L_y}{\pi L_x} \int_{0}^{\infty} T_{\rm q} \left( x \right) dx.   \label{aeq: graphene - Tq}
\end{align}
\end{widetext}
Notably, $T_{\rm c}$ describes classical scattering with the real wave numbers $k_x \in \mathbb{R}$. 
As shown in Fig.~\ref{fig: graphene-appendix}\,(a), $T_{\rm c}$ is zero at the Dirac point $E = 0$ and increases away from the Dirac point.
This behavior is compatible with the density of states of graphene, as well as the conductance derived by the Boltzmann equation~\cite{Nomura-06, *Nomura-07, Ando-06}.
By contrast, $T_{\rm q}$ describes quantum tunneling with the imaginary wave numbers $k_x \in \ii\mathbb{R}$.
As shown in Fig.~\ref{fig: graphene-appendix}\,(b), $T_{\rm q}$ gets largest at the Dirac point $E = 0$ and decreases away from the Dirac point.
Thus, while $T_{\rm c}$ dominates the transmission away from the Dirac point, $T_{\rm q}$ dominates the transmission near the Dirac point.

\subsection{Asymptotic analysis}
    \label{asec: graphene - asymptotics}

The integrals in Eqs.~(\ref{aeq: graphene - Tc}) and (\ref{aeq: graphene - Tq}) seem to be unfeasible analytically.
However, their asymptotic behavior is tractable even analytically, which allows us to obtain the linear and nonlinear conductance of the Dirac Hamiltonian.

First, we focus on the behavior for $\left| E L_x /\hbar v \right| \ll 1$.
Since the integrand in Eq.~(\ref{aeq: graphene - Tc}) is expanded as
\begin{equation}
    T_{\rm c} \left( x \right) = \frac{1}{\sqrt{\left( EL_x/\hbar vx\right)^2 - 1}} + \mathcal{O} \left( E^2 \right)
\end{equation}
for $0 \leq x \leq \left| E \right| L_x/\hbar v$, 
we have
\begin{align}
T_{\rm c}
&= \frac{L_y}{\pi L_x} \int_{0}^{\left| E \right| L_x /\hbar v} \left[ \frac{1}{\sqrt{\left( EL_x/\hbar vx\right)^2 - 1}} + \mathcal{O} \left( E^2 \right) \right] dx \nonumber \\
&= \frac{L_y}{\pi L_x}\frac{\left| E \right| L_x}{\hbar v} \int_0^{1} \frac{dx}{\sqrt{x^{-2} - 1}} + \mathcal{O} \left( E^3 \right) \nonumber \\
&= \frac{L_y}{\pi L_x}\frac{\left| E \right| L_x}{\hbar v} + \mathcal{O} \left( E^3 \right).
\end{align}
In addition, the integrand in Eq.~(\ref{aeq: graphene - Tq}) is expanded as
\begin{align}
T_{\rm q} \left( x \right)
&= \frac{1}{\cosh^2 x} - \left( \frac{1}{2\cosh^2 x} + \frac{\sinh^2 x}{\cosh^4 x} \right) \left( \frac{EL_x}{\hbar v x}\right)^2 \nonumber \\
&\qquad\qquad\qquad\qquad\qquad\qquad+ \mathcal{O} \left( E^{4} \right),
\end{align}
which shows the singularity at $x=0$.
To handle this singularity appropriately, we divide the integral range $\left[ 0, \infty \right]$ into $\left[ 0, \left| E \right| L_x/\hbar v \right]$ and $\left[ \left| E \right| L_x/\hbar v, \infty \right]$.
For $x \in \left[ 0, \left| E \right| L_x/\hbar v \right]$, we expand the integrand as
\begin{align}
    T_{\rm q} \left( x \right) = \frac{1}{\left| E \right| L_x/\hbar vx} + \mathcal{O} \left( E^2 \right)
\end{align}
and have
\begin{align}
&\int_{0}^{\left| E \right| L_x/\hbar v} T_{\rm q} \left( x \right) dx \nonumber \\
&\qquad=\int_{0}^{\left| E \right| L_x/\hbar v} \left[ \frac{1}{\left| E \right| L_x/\hbar vx} + \mathcal{O} \left( E^2 \right) \right] dx \nonumber \\
&\qquad= \frac{\left| E \right| L_x}{\hbar v} \int_{0}^{1} xdx + \mathcal{O} \left( E^3 \right) \nonumber \\
&\qquad= \frac{1}{2} \frac{\left| E \right| L_x}{\hbar v} + \mathcal{O} \left( E^3 \right).
    \label{aeq: graphene - smallE - Tq1}
\end{align}
For $x \in \left[ \left| E \right| L_x/\hbar v, \infty \right]$, using 
\begin{align}
f \left( x \right) &\coloneqq \frac{1}{2x^2\cosh^2 x} + \frac{\sinh^2 x}{x^2\cosh^4 x} - \frac{1}{2x^2} \nonumber \\
&= \frac{1}{2} - \frac{4x^2}{3} + \mathcal{O} \left( x^4 \right),
\end{align}
we expand the integrand as
\begin{align}
    T_{\rm q} \left( x \right) = \frac{1}{\cosh^2 x} - \left[ f \left( x \right) + \frac{1}{2x^2} \right] \left( \frac{EL_x}{\hbar v} \right)^2 + \mathcal{O} \left( E^4 \right)
\end{align}
and have
\begin{align}
&\int_{\left| E \right| L_x/\hbar v}^{\infty} T_{\rm q} \left( x \right) dx \nonumber \\
&= \int_{\left| E \right| L_x/\hbar v}^{\infty} \left\{ \frac{1}{\cosh^2 x} - \left[ f\left( x \right) + \frac{1}{2x^2} \right] \left( \frac{EL_x}{\hbar v}\right)^2 \right\} dx \nonumber \\
&\qquad\qquad\qquad\qquad\qquad\qquad\qquad\qquad+ \mathcal{O} \left( E^4 \right) \nonumber \\
&= 1 - \tanh \left( \frac{\left| E \right| L_x}{\hbar v} \right) \nonumber \\
&\quad- \left( \frac{E L_x}{\hbar v} \right)^2 \int_{\left| E \right| L_x/\hbar v}^{\infty} \left[ f \left( x \right) + \frac{1}{2x^2}\right] dx
+ \mathcal{O} \left( E^4 \right).
\end{align}
Here, we have
\begin{align}
\int_{\left| E \right| L_x/\hbar v}^{\infty} \left[ f \left( x \right) + \frac{1}{2x^2}\right] dx
&= \frac{1}{2} \left( \frac{\left| E \right| L_x}{\hbar v} \right)^{-1}
- \frac{c}{2} \nonumber \\
&\qquad\qquad+ \mathcal{O} \left( E \right)
\end{align}
with 
\begin{equation}
    c \coloneqq -2 \int_{0}^{\infty} f \left( x \right) dx
    = 0.201876\cdots,
\end{equation}
which leads to
\begin{align}
&\int_{\left| E \right| L_x/\hbar v}^{\infty} T_{\rm q} \left( x \right) dx \nonumber \\
&\qquad= 1 - \frac{3}{2} \frac{\left| E\right| L_x}{\hbar v} + \frac{c}{2} \left( \frac{E L_x}{\hbar v}\right)^2 
+ \mathcal{O} \left( E^3 \right).
    \label{aeq: graphene - smallE - Tq2}
\end{align}
Combining Eq.~(\ref{aeq: graphene - smallE - Tq1}) with Eq.~(\ref{aeq: graphene - smallE - Tq2}), we have
\begin{align}
T_{\rm q}
&= \frac{L_y}{\pi L_x}\int_{0}^{\infty} T_{\rm q} \left( x \right) dx \nonumber \\
&= \frac{L_y}{\pi L_x} \left[ 1 - \frac{\left| E \right| L_x}{\hbar v}
+ \frac{c}{2} \left( \frac{E L_x}{\hbar v} \right)^2 \right] + \mathcal{O} \left( E^3 \right).
\end{align}
Thus, the total transmission probability for $\left| EL_x/\hbar v \right| \ll 1$ is
\begin{align} 
T &= T_{\rm c} + T_{\rm q} \nonumber \\
&= \frac{L_y}{\pi L_x} \left[ 1 
+ \frac{c}{2}\left( \frac{E L_x}{\hbar v} \right)^2 \right] + \mathcal{O} \left( E^3 \right).
\end{align}
This analytic result is compatible with the numerical results in Figs.~\ref{fig: graphene} and \ref{fig: graphene-appendix}.
Notably, although each of $T_{\rm c}$ and $T_{\rm q}$ contains the nonanalytic term in proportion to $\left| E \right|$, the total transmission probability $T = T_{\rm c} + T_{\rm q}$ is analytic.

Next, we consider the opposite limit $\left| EL_x/\hbar v\right| \to \infty$.
In this limit, the integrand in Eq.~(\ref{aeq: graphene - Tc}) is approximately evaluated as
\begin{align}
    T_{\rm c} \left( x \right) 
    \simeq \frac{1}{\sqrt{\left( EL_x/\hbar vx\right)^2 - 1} \left( 1/2 + \left( EL_x/\hbar v x\right)^2/2 \right)},
\end{align}
and then Eq.~(\ref{aeq: graphene - Tc}) reduces to
\begin{align}
T_{\rm c} 
&\simeq \frac{L_y}{\pi L_x} \frac{\left| E \right| L_x}{\hbar v} \int_0^{1} \frac{2dx}{\sqrt{x^{-2} - 1} \left( 1+x^{-2}\right)} \nonumber \\
&= \frac{L_y}{\pi L_x} \left( 2-\sqrt{2}\,\mathrm{arccoth}\,\sqrt{2}\right) \frac{\left|E \right| L_x}{\hbar v}.
    \label{aeq: graphene - largeE - Tc}
\end{align}
On the other hand, the integrand in Eq.~(\ref{aeq: graphene - Tq}) is evaluated as
\begin{align}
    T_{\rm q} \left( x \right) \simeq \frac{1}{\left| EL_x/\hbar vx\right| \left( EL_x/\hbar v x\right)^2 \sinh^2 x},
\end{align}
and then Eq.~(\ref{aeq: graphene - Tq}) reduces to
\begin{align}
T_{\rm q} 
&\simeq \frac{L_y}{\pi L_x} \left( \frac{\hbar v}{\left| E \right| L_x}\right)^{3} \int_{0}^{\infty} \frac{x^3 dx}{\sinh^2 x} \nonumber \\
&= \frac{L_y}{\pi L_x} \frac{3\zeta \left( 3 \right)}{2} \left( \frac{\hbar v}{\left| E \right| L_x}\right)^{3},
\end{align}
which is much smaller than $T_{\rm c}$ in Eq.~(\ref{aeq: graphene - largeE - Tc}) for $\left| EL_x/\hbar v\right| \to \infty$.
Thus, the total transmission probability is 
\begin{align}
T = T_{\rm c} + T_{\rm q} \simeq \frac{L_y}{L_x} \frac{2-\sqrt{2}\,\mathrm{arccoth}\,\sqrt{2}}{\pi} \frac{\left| E \right| L_x}{\hbar v}.
\end{align}
The numerical results in Figs.~\ref{fig: graphene} and \ref{fig: graphene-appendix} are consistent with this analytic result.
The numerical results include the additional small oscillations around this linear behavior.

\bibliography{Landauer}

\end{document}